
\input phyzzx87
\def\psfig{\relax}
\singlespace
\hoffset=1cm
\voffset=1pc

\chapterminspace=6pc
\sectionminspace=5pc
\unlock
\def\sss{\scriptscriptstyle}
\def\bold#1{\setbox0=\hbox{$#1$}%
     \kern-.025em\copy0\kern-\wd0
     \kern.025em\copy0\kern-\wd0
     \kern-.025em\raise.0233em\box0 }
\def\ifmath#1{\relax\ifmmode #1\else $#1$\fi}
\def\quarter{\ifmath{{\textstyle{1 \over 4}}}}

\def\third{\ifmath{{\textstyle{1 \over 3}}}}
\def\half{\ifmath{{\textstyle{1 \over 2}}}}
\def\eighth{\ifmath{{\textstyle{1 \over 8}}}}
\def\barparen#1{\buildrel {\sss{(-)}} \over #1}%
\def\solid{(---------)}
\def\dashes{($-~-~-~-$)}
\def\dotdot{($\cdot~\cdot~\cdot~\cdot~\cdot~\cdot\,$)}


\def\gev{~{\rm GeV}}
\def\tev{~{\rm TeV}}



\def\etal{{\it et al.}}
\def\etc{{\it etc.}}
\def\ls#1{\ifmath{_{\lower1.5pt\hbox{$\scriptstyle #1$}}}}

%

    \def\fillboxx#1#2{\hbox to #1{\vbox to #2{\vfil}\hfil}
    }

\def\mt{m_t}

\def\hsm{\phi^0}
\def\mhsm{m_{\hsm}}

\def\mh{m_h}
\def\hl{h^0}
\def\mhl{m_{\hl}}
\def\hh{H^0}
\def\mhh{m_{\hh}}
\def\hpm{H^{\pm}}
\def\mhpm{m_{\hpm}}

\def\ifmath#1{\relax\ifmmode #1\else $#1$\fi}
\def\half{\ifmath{{\textstyle{1 \over 2}}}}

\def\quarter{\ifmath{{\textstyle{1 \over 4}}}}
\def\3quarter{{\textstyle{3 \over 4}}}
\def\third{\ifmath{{\textstyle{1 \over 3}}}}
\def\twothirds{{\textstyle{2 \over 3}}}

\def\h{h}
\def\mh{m_{\h}}
\def\lplm{l^+l^-}

\def\tanb{\tan\beta}

\def\mw{m_W}
\def\mz{m\ls Z}
\def\hpm{H^{\pm}}
\def\mhpm{m_{\hpm}}

\def\hthree{H^0_3}

\def\rta{\rightarrow}

\def\snu{\widetilde \nu}

\def\chitil{\widetilde\chi}
\def\mchipa{M_{\tilde \chi^+_1}}
\def\mchipb{M_{\tilde \chi^+_2}}

\def\wp{W^+}
\def\wm{W^-}

\def\hl{h^0}
\def\hh{H^0}
\def\ha{A^0}
\def\mhl{m_{\hl}}
\def\mhh{m_{\hh}}
\def\mha{m_{\ha}}

\def\lam{\lambda}

\def\gam{\gamma}

\def\gam{\gamma}

\def\VEV#1{\langle #1 \rangle}

\def\tanb{\tan\beta}

\def\chitil{\widetilde \chi}
\def\lsp{\chitil^0_1}

\def\gp{g^{\prime}}

\def\wp{W^+}
\def\wm{W^-}
\def\mw{m_W}
\def\mz{m_{Z}}

\def\lam{\lambda}

\def\hthree{A^0}

\def\hpm{H^{\pm}}

\def\mhpm{m_{\hpm}}

\def\epem{e^+e^-}

\def\lplm{l^+ l^-}

\def\rta{\rightarrow}

\def\cca{\chitil^+_{\tilde W^{+}}}

\def\mt{m_t}

\def\M2pm{M^2_{P^\pm}}
\def\m2z{\mz^2}

   \def\chapter#1{\par \penalty-300 
       \gdef\thechapterhead{#1}%
       \gdef\thesectionhead{\relax}%
       \gdef\thesubsecthead{\relax}%
       \gdef\thesubsubsecthead{\relax}%
       \chapterreset
}
\def\section#1{\par \ifnum\lastpenalty=30000\else
   \penalty-200 \vskip12pt plus 2pt minus 0.5pt
   \spacecheck\sectionminspace\fi
   \gl@bal\advance\sectionnumber by 1
   \gl@bal\subsectnumber=0%
   {\pr@tect
   \xdef\sectlabel{\ifcn@@ \chapterlabel.\fi
       \the\sectionstyle{\the\sectionnumber}}%
   \wlog{\string\section\space \sectlabel}}%
   \noindent {\bf \sectlabel~~#1}\par
   \nobreak  \penalty 30000
       \ifcontentson%
          \c@ntentscheck%
          \CONTENTS{S};{#1}%
       \fi%
   }
   \def\FootNoteFonts{\Tenpoint}

   \def\Vfootnote#1{%
      \insert\footins%
      \bgroup%
         \interlinepenalty=\interfootnotelinepenalty%
         \floatingpenalty=20000%
         \singl@true\doubl@false%
         \FootNoteFonts%
         \splittopskip=\ht\strutbox%
         \boxmaxdepth=\dp\strutbox%
         \leftskip=\footindent%
         \rightskip=\z@skip%
         \parindent=0.5%
         \footindent%
         \parfillskip=0pt plus 1fil%
         \spaceskip=\z@skip%
         \xspaceskip=\z@skip%
         \footnotespecial%
         \Textindent{#1}%
         \footstrut%
         \futurelet\next\fo@t%
   }
   \def\eqname#1{%
      \rel@x{\pr@tect%
      \csnamech@ck%
       \ifnum\equanumber<0%
          \xdef#1{{\noexpand\f@m0(\number-\equanumber)}}%
          \immediate\write\csnamewrite{%
            \def\noexpand#1{\noexpand\f@m0 (\number-\equanumber)}}%
          \gl@bal\advance\equanumber by -1%
      \else%
          \gl@bal\advance\equanumber by 1%
          \ifusechapterlabel%
            \xdef#1{{\noexpand\f@m0(\ifcn@@ \chapterlabel.\fi%
               \number\equanumber)}%
            }%
          \else%
             \xdef#1{{\noexpand\f@m0(\ifcn@@%
                 {\the\chapterstyle{\the\chapternumber}}.\fi%
                 \number\equanumber)}}%
          \fi%
          \ifcn@@%
             \ifusechapterlabel
                \immediate\write\csnamewrite{\def\noexpand#1{(%
                  {\chapterlabel}.%
                  \number\equanumber)}%
                }%
             \else
                \immediate\write\csnamewrite{\def\noexpand#1{(%
                  {\the\chapterstyle{\the\chapternumber}}.%
                  \number\equanumber)}%
                }%
             \fi%
          \else%
              \immediate\write\csnamewrite{\def\noexpand#1{(%
                  \number\equanumber)}}%
          \fi%
      \fi}%
      #1%
   }
\lock

\def\ls#1{\ifmath{_{\lower1.5pt\hbox{$\scriptstyle #1$}}}}
\def\lss#1{^{\lower5pt\hbox{$\scriptstyle #1$}}}

\def\notep{\epsilon{\mkern-8.0mu}{/}}
\def\notk{/{\mkern-9.0mu}{k}}
\def\notp{p{\mkern-8.0mu}{/}}

\def\nots{s{\mkern-8.5mu}{/}}
\def\ca{%
    \def\.{\hskip-4pt plus 1pt}
    \def\vr{\vrule height 7pt depth 4pt}
    {\rlap{\raise 7pt \hbox{\vr}}\.\rightarrow}
    }
\def\cca{%
    \def\.{\hskip-4pt plus 1pt}
    \def\vvr{\vrule height 14pt depth 21pt}
    {\rlap{\raise 24pt\hbox{\vvr}}\.\rightarrow}
    }
\def\ccca{%
    \def\.{\hskip-4pt plus 1pt}
    \def\vvr{\vrule height 21pt depth 4pt}
    {\rlap{\raise 7pt \hbox{\vvr}}\.\longrightarrow}
    }

\def\l#1{\lambda_#1}
\def\ifmath#1{\relax\ifmmode #1\else $#1$\fi}
\def\mpl{M_{\rm PL}}
\def\quarter{\ifmath{{\textstyle{1 \over 4}}}}

\def\a{\alpha}
\def\be{{\beta}}

\def\de{{\delta}}

\def\la{\lambda}
\def\si{{\sigma}}


\def\cM{{\cal M}}
\def\cA{{\cal A}}

\def\ymax{y^{\rm max}}
\def\nsd{N_{SD}}
\def\anti{\overline}
\def\gexp{\Gamma_{\rm exp}}
\def\gmax{\Gamma_{\rm res}}
\def\h{H}
\def\mh{m\ls{\h}}
\def\yh{y\ls{\h}}
\def\gammah{\Gamma_{\h}}
\def\vevlam{\VEV{\lami\lamii}}
\def\eepem{E_{\epem}}
\def\egamgam{E_{\gam\gam}}
\def\lami{\lam}
\def\lamii{\lam^{\,\prime}}
\def\lamei{\lam_e}
\def\lameii{\lam_e^{\,\prime}}
\def\pci{P_c}
\def\pcii{P_c^{\,\prime}}

\def\msusy{M_{\rm SUSY}}
\def\lplm{\ell^+\ell^-}

{\Pubnum{SCIPP 93/49 \cr NSF-ITP-94-30}
\date{April 1994}
\titlepage
\baselineskip=11pt
\vskip3cm
\centerline{ \fourteenbf
    Spin Formalism and Applications to New Physics Searches%
\foot{Work supported in part by the U.S. Department of Energy,
grant number DE-FG03-92ER40689 and in part by the National Science
Foundation, NSF grant \#PHY89-04035.}}
\author{Howard E. Haber}
\vskip3pt
\par\kern 5pt
\centerline{\it Santa Cruz Institute for Particle Physics}
\centerline{{\it University of California, Santa Cruz, CA 95064}%
\foot{Permanent address.}}
\par\kern 5pt \centerline{\sl and} \par\kern 5pt
\centerline{\it Institute for Theoretical Physics}
\centerline{{\it University of California, Santa Barbara, CA 93106}%
\phantom{\foot{Imaginary footnote.}}}
\vskip1cm
\centerline{{\bf Abstract}}

An introduction to spin techniques in particle physics is given.
Among the topics covered are: helicity formalism and its applications
to the decay and scattering of spin-1/2 and spin-1 particles,
techniques for evaluating helicity amplitudes (including projection
operator methods and the spinor helicity method), and density matrix
techniques.  The utility of polarization and spin correlations for
untangling new physics beyond the Standard Model at future colliders
such as the LHC and a high energy $e^+e^-$ linear collider
is then considered.  A number of detailed examples are explored
including the search for low-energy supersymmetry, a non-minimal
Higgs boson sector, and new gauge bosons beyond the $W^\pm$ and $Z$.

\vfill
\centerline{To appear in the Proceedings of the
21st SLAC Summer Institute on}
\centerline{Particle Physics: Spin Structure in High Energy Processes}
\centerline{SLAC, Stanford, CA, 26 July---6 August 1993.}
\vfill\endpage}
\hoffset=1cm
\voffset=1pc
{\normalspace
\def\lf{\leaders\hbox to0.9em{\hss.\hss}\hfill}%
\def\hf{\phantom{}}
\centerline{\bf Table of Contents}
\vskip2pc
\line{Introduction\hf\lf3}
\line{Lecture 1:\enskip Spin Formalism and Calculational Techniques\hf\lf5}
\line{\quad{1}.1\enskip Helicity states in quantum mechanics\hf\lf5}
\line{\quad{1}.2\enskip Helicity amplitudes in decay and scattering processes%
       \hf\lf9}
\line{\quad{1}.3\enskip Explicit spin-$\half$ and spin-1 wave
functions\hf\lf11}
\line{\quad{1}.4\enskip Spin projection operator methods\hf\lf15}
\line{\quad{1}.5\enskip The spinor helicity method\hf\lf17}
\line{\quad{1}.6\enskip The Bouchiat-Michel formulae\hf\lf22}
\line{\quad{1}.7\enskip Density matrix techniques in unstable particle
                    production and decay\hf\lf25}
\line{Lecture 2:\enskip Applications to Low-Energy Supersymmetry\hf\lf31}
\line{\quad{2}.1\enskip Raison-d'\^etre
for new physics beyond the Standard Model\hf\lf31}
\line{\quad{2}.2\enskip Introduction to low-energy supersymmetry\hf\lf34}
\line{\quad{2}.3\enskip Polarization and spin analysis as tools for
                    supersymmetry searches\hf\lf41}
\line{Lecture 3:\enskip Applications to Higgs and New Gauge Boson
                        Searches\hf\lf51}
\line{\quad{3}.1\enskip Higgs bosons beyond the minimal Standard Model%
                              \hf\lf51}
\line{\quad{3}.2\enskip Higgs boson production at a $\gamma \gamma$
                    collider\hf\lf53}
\line{\quad{3}.3\enskip Search for CP-violating effects in the Higgs sector%
                             \hf\lf64}
\line{\quad{3}.4\enskip New gauge bosons beyond the $W^\pm $ and $Z$%
                             \hf\lf68}
\line{Coda\hf\lf75}
\line{Acknowledgments\hf\lf76}
\line{References\hf\lf77}
\endpage}
\singlespace
\titlestyle{\bf INTRODUCTION}
\medskip

Every beginning student in particle physics learns how to evaluate
Feynman diagrams and compute decay rates and cross sections.  In
processes involving particles with nonzero spin, one quickly learns
the mantra: {\it average over initial spins and sum over final spins}.
Usually, one learns a number of tricks to perform the spin summations
so that one never has to worry about the explicit forms for the spin
wave functions.  However, in the process of summing over spins, one
loses a great deal of information.  Suppose one is experimentally
studying some new physics phenomenon such as a newly discovered
particle.  In order to maximize the information obtained in the
experiment, it is advantageous not to sum over spins.  For example,
one can study the production of the new particle with polarized beams.
One can also study the polarization of the final state by measuring
decay angle correlations in the decay chain of the new particle.
Clearly, with additional information of this type,
one can potentially learn much more about the properties
of the new particle and its interactions.

\REF\precision{See \eg, D.~Schaile, CERN-PPE/93-213 (1993).}
\REF\cdftop{F.~Abe \etal\ [CDF Collaboration] FERMILAB-PUB-94/097-E
(1994).}
\REF\higgslimit{E. Gross and P. Yepes, {\sl Int. J. Mod. Phys.} {\bf A8}
(1993) 407.}
Today, all particle physics phenomena is successfully described by the
Standard Model.  Precision electroweak measurements at LEP
test the Standard Model to better than one part in
thousand\refmark\precision.  So far, no
convincing deviation from the Standard Model has been found.
Two pieces of the Standard Model remain to be discovered.  One is the
top quark; evidence for its existence has recently been
announced by the CDF collaboration\refmark\cdftop.  The
other is the Higgs boson.  In the Standard Model, electroweak symmetry
breaking is triggered by the dynamics of an elementary complex
doublet of scalar (Higgs) fields.  As a result of these dynamics, the neutral
component of the Higgs field acquires a vacuum expectation value.
Three of the four Higgs degrees of freedom are absorbed and give mass
to the $W^\pm$ and $Z$ gauge bosons.  A fourth neutral scalar degree
of freedom is the Higgs boson.  Present limits from LEP imply that the
Higgs boson mass must be larger than 60~GeV\refmark\higgslimit.

\REF\thooft{S. Weinberg, {\sl Phys. Rev.} {\bf D13} (1976) 974;
{\bf D19} (1979) 1277; L. Susskind, {\sl Phys. Rev.} {\bf D20} (1979)
2619; G. 't Hooft, in {\it Recent Developments in Gauge
Theories,} Proceedings of the NATO Advanced Summer Institute,
Cargese, 1979, edited by G. 't~Hooft \etal\ (Plenum, New York,
1980) p.~135.}
However, most theorists believe that the mechanism of electroweak
symmetry breaking must be more complex.  Theories with elementary
scalar fields suffer from the problem of
{\it naturalness}\refmark\thooft.  Simply
put, the Standard Model cannot be a complete theory of particle
interactions, since it neglects gravity.  However, to a very good
approximation, it is safe to ignore gravitational interactions at all
energy scales below the Planck scale, $\mpl\simeq 10^{19}$~GeV.  The
problem with elementary scalar fields is that the natural value
for its mass is given by
the largest mass scale at which the scalar field can
be viewed as elementary.  By natural, I mean that if one attempts to
set the scalar mass to be light, radiative corrections drive the mass
up to the highest energy
scale.  Thus, to understand why the Higgs scalar mass
(and thus why the scale of electroweak symmetry breaking) is of order
100~GeV rather than $10^{19}$~GeV, one must postulate a remarkable
fine-tuning of parameters in the fundamental Planck scale theory.
Most theorists find this possibility unaesthetic and prefer a less
``miraculous'' explanation of the large hierarchy between the
electroweak scale and the Planck scale.  In attempting
to devise a theory of electroweak symmetry breaking that avoids this
disease, one is led to either assume that the Higgs scalar is not
elementary much beyond the scale of 100~GeV, or else a new symmetry
exists that can protect scalar fields from acquiring large masses.  In
all proposed models of these types, there must be new physics
associated with the dynamics of electroweak symmetry breaking that
exists at or near the scale of electroweak symmetry breaking.

\REF\hehtasi{H.E. Haber,
     in {\it Testing the Standard Model}, Proceedings of the 1990
     Theoretical Advanced Study Institute in Elementary Particle
     Physics, edited by M. Cveti\v c and Paul Langacker
     (World Scientific, Singapore, 1991) pp.~340--475.}
The main goal of the next generation of particle accelerators is to
discover the underlying mechanism of electroweak symmetry breaking%
\refmark\hehtasi.
That is, one must find the Higgs boson and/or new particles associated
with the dynamics of electroweak symmetry breaking that lies at an
energy scale of about 1~TeV or below.  Assuming that new particles or
phenomena are found, it will be crucial to determine
the details  of the new particle mass spectrum and the nature
of their interactions.  Spin will play a major role in this enterprise.
By employing polarized beams and studying the initial and final state
spin correlations,
it will be possible to extract detailed information about
the new particles and their interactions.

The object of these lectures is to describe some of the basic techniques
for making use of spin information to explore new physics beyond the
Standard Model.  In Lecture 1, I review spin formalism in
particle physics and its
practical applications to systems of spin-1/2 and spin-1 particles.  Most
everything in Lecture 1 is well known to practitioners of the field,
but is rarely disseminated in elementary graduate
courses.  I shall highlight the most important results, focusing on
the use of helicity states.  It is interesting to note that although
the material of Lecture 1 is well known, techniques involving the
computation of helicity amplitudes are still evolving.  New and
powerful methods are still being developed today and are permitting
the calculation of some multi-particle processes that could not have
been envisioned even ten years ago.  I will only be able to touch on
some of these new methods briefly here.  In Lecture 2, I will discuss
low-energy supersymmetry as a canonical candidate for physics
beyond the Standard Model.  I will then show how spin and polarization
can be exploited in the detection and elucidation of supersymmetric
particles and their interactions.  In Lecture 3, I focus on two other
examples of the use of spin and polarization for the detection of new
phenomena: Higgs bosons at a $\gamma\gamma$ collider, and the
detection of new gauge bosons beyond the $W^\pm$ and $Z$ at a
future hadron collider.  I end with a few comments on other areas
where spin techniques can play a critical role in the search for new
physics.

 \endpage

\chapter{}

\title{\bf LECTURE {\seventeenbf 1}:    \break
Spin Formalism and Calculational Techniques}
\medskip

\section{\bf Helicity states in quantum mechanics}

Consider a spinning particle in quantum mechanics
with spin vector ${\bold{\vec S}}$, orbital angular momentum vector
${\bold{\vec L}}$ and total angular momentum ${\bold{\vec J}}\equiv
{\bold{\vec L}}+{\bold{\vec S}}$.  In most
elementary textbooks of quantum
mechanics, it is shown how to construct simultaneous eigenstates of
$\bold{\vec J}\lss{\,2}$, $J_z$, $\bold{\vec L}\lss{\,2}$
and $\bold{\vec S}\lss{\,2}$.
Another useful basis is the product basis consisting of
simultaneous eigenstates of
$\bold{\vec L}\lss{\,2}$, $L_z$, $\bold{\vec S}\lss{\,2}$ and $S_z$.
Transformations between these two bases involve the Clebsch-Gordon
coefficients.  However, there is a third basis choice which will prove
far more convenient in processes involving relativistic particles.
Define the helicity operator $\Lambda$ as follows
$$
  \Lambda = {\bold{\vec J\cdot \hat p}} = ({\bold{\vec L}}+{\bold{\vec S}})
           \cdot\bold{\hat p}
  = ({\bold{\vec r\times \vec p}} +{\bold{ \vec S}}){\bold{\cdot \hat p}}
  = {\bold{\vec S\cdot \hat p}}\,,\eqn\helicityop
$$
which
is a scalar operator that commutes
with $\bold{\vec J}$ and $\bold{\vec S}\lss{\,2}$.
We can now build angular momentum states that are
simultaneous eigenstates of $\bold{\vec J}\lss{\,2}$, $J_z$, $\bold{\vec
S}\lss{\,2}$
and $\Lambda$.  These are the helicity states of a single particle.
The helicity states have certain
advantages over, say, simultaneous eigenstates
of $\bold{\vec J}\lss{\,2}$, $J_z$, $\bold{\vec L}\lss{\,2}$ and
$\bold{\vec S}\lss{\,2}$.  In particular, helicity states are (i)
invariant under spatial rotations, (ii)
invariant under boosts along the particle's momentum (as long as the
direction of momentum is not reversed), and (iii)
convenient for describing relativistic scattering of both
massless and massive particles.
\REF\jw{M. Jacob and G.C. Wick, Ann. Phys. (NY) {\bf 7} (1959) 404.}
\REF\qmbooks{See \eg, K. Gottfried, {\it Quantum Mechanics}
(W.A. Benjamin, Inc., New York, 1966), \S 35.3; A. Galindo and
P. Pascual, {\it Quantum Mechanics I} (Springer-Verlag, Berlin, 1990),
\S 5.10.}
\REF\angmom{For a comprehensive treatment of angular momentum theory
in quantum mechanics and a guide to the literature, see
D.A. Varshalovich, A.N. Moskalev and V.K. Khersonskii, {\it Quantum
Theory of Angular Momentum} (World Scientific, Singapore, 1988).}
Let us explicitly construct the single particle helicity
states\refmark{\jw,\qmbooks}.
I shall employ the standard particle physicist's convention of
$\hbar=1$.  We shall need to make use of the $D$ and $d$ rotation
matrices introduced in most quantum mechanics textbooks.  These are
defined as follows.  The unitary rotation operator acting on the quantum
mechanical Hilbert space is\refmark\angmom\
$$U[R(\phi,\theta,\gamma)] \equiv e^{-i\phi J_{\hbox{$\scriptstyle z$}}}
    e^{-i\theta J_{\hbox{$\scriptstyle y$}}} e^{-i\gamma J_{\hbox{$\scriptstyle
z$}}}\,,\eqn\rotop$$
where the rotation $R$ is specified by three Euler angles.  The $D$
and $d$-matrices are then defined by
$$\eqalign{
  D^{(j)}_{mm^\prime} (R)\,\delta_{jj^\prime}
&\equiv \VEV{jm|U[R(\phi,\theta,\gamma)]|
      j^\prime m^\prime}   \crr
       &= e^{-i\phi m} e^{-i\gamma m^\prime}
        d^j_{mm^\prime}(\theta)\,\delta_{jj^\prime}\,.\cr}
\eqn\dmatrices$$
Three important properties that we will need are
$$e^{-i\pi J_{\hbox{$\scriptstyle y$}}}\ket{jm} = (-1)^{j-m}
\ket{j,-m}\,,\eqn\propone$$
$$Y_{\ell m}(\theta,\phi) = \sqrt{{2\ell+1\over 4\pi}}
       D^\ell_{m0} (\phi,\theta,\gamma)^\ast \,,\eqn\spharm
$$
and the orthogonality relation
$$
  \int d\Omega\,D^{(j)}_{m\lambda} (R) D^{(j^\prime)}_{m^\prime\lambda}
     (R)^\ast = {4\pi\over 2j+1}\,\delta_{jj^\prime}\delta_{mm^\prime}
      \,,\eqn\orthogonality
$$
where $d\Omega\equiv d\cos\theta d\phi$ (note that there is no
dependence on the angle $\gamma$).

Let ${\bold{\hat p}}$ be a unit vector pointing in a direction specified
by a polar angle $\theta$ and azimuthal angle $\phi$ with respect to a
fixed $z$-axis.  The three-momentum of
a particle will be denoted by ${\bold{\vec p}}\equiv p {\bold{\hat p}}$.
A plane wave state of helicity $\lambda$ is constructed by starting
with a plane wave eigenstate of $S_z$ (with eigenvalue $\lambda$)
moving in the $z$-direction.  Applying the rotation operator
$U[R(\phi,\theta,-\phi)]$ results in a plane wave moving in the
direction of ${\bold{\vec p}}$.  Note that the third argument of $R$
is purely conventional, since for a particle moving in the
$z$-direction, a rotation about ${\bold{\hat z}}$ has no physical
effect.  The reason for choosing such a convention is that the rotation
$R(\phi,\theta,-\phi)$ is equivalent to a single rotation by an
angle $\theta$ about an axis ${\bold{\hat n}}$, where
$$
{\bold{\hat n}}=(-\sin\phi,\cos\phi,0)\,.\eqn\nhatdef
$$
This result can be proven by verifying that
$$
e^{-i\phi J_{\hbox{$\scriptstyle z$}}}e^{-i\theta J_{\hbox{$\scriptstyle
y$}}}e^{i\phi J_{\hbox{$\scriptstyle z$}}}=
e^{-i\theta(J_{\hbox{$\scriptstyle y$}}
\cos\phi-J_{\hbox{$\scriptstyle x$}}\sin\phi)}\,,\eqn\jidentity
$$
which follows from the commutation relations of the angular
momentum operators.

The helicity plane-wave state is explicitly given by
$$
  \ket{{\bold{\vec p}},\lambda} = U[R(\phi,\theta,-\phi)]
\ket{p{\bold{\hat z}},\lambda}\,,\eqn\planewave
$$
where
$$
  J_z\ket{p{\bold{\hat z}},\lambda} = \lambda\ket{p{\bold{\hat z}},\lambda}\,,
\eqn\jayzee$$
since for ${\bold{\vec p}}=p{\bold{\hat z}}$,
$J_z={\bold{\vec J\cdot \hat p}} = {\bold{\vec S\cdot\hat p}}=S_z$.
Let us expand in a complete set of eigenstates of $\bold{\vec J}\lss{\,2}$ and
$J_z$,
$$
  \ket{p{\bold{\hat z}},\lambda} = \sum^\infty_{j=|\lambda|}
      \ket{p,jm} \VEV{p,jm | p{\bold{\hat z}},\lambda}\,,
\eqn\complete
$$
where the sum is taken over integers (half-integers) for an integer
(half-integer) spin particle.  Next,
apply $U[R]$ to Eq.~\complete\ to obtain $\ket{{\bold{\vec p}},\lambda}$.
Inserting a second complete set of angular momentum states and using
Eq.~\dmatrices, one obtains
$$
  \ket{{\bold{\vec p}},\lambda} = \sum^\infty_{j=|\lambda|}\,
       \sum^j_{m=-j} \ket{p,jm} D^{(j)}_{m\lambda} (R)
      \VEV{p,jm |p{\bold{\hat z}},\lambda}\,.\eqn\expansion
$$
This result is the generalization for spinning particles of the well
known expansion of plane waves in terms of spherical waves.
To see that this is the desired result, multiply Eq.~\expansion\
by $\bra{\bold{\vec r}}$.
For spinless particles, $\lambda=0$, $j=\ell$, and
$$\eqalign{%
  &\VEV{{\bold{\vec r}}\,|\,{\bold{\vec p}}} \sim e^{i{\bold{\vec p\cdot\vec
r}}} \cr
  &\VEV{{\bold{\vec r}}\,|\, p,\ell m} \sim i^\ell
  j_\ell (pr) Y_{\ell m} ({\bold{\hat r}}) \cr
  &\VEV{p,\ell m\,|\,p{\bold{\hat z}}} \hbox{\ is\ a\ constant\ independent\
         of\ }{\bold{\hat p}},{\bold{\hat r}}.  \cr} \eqn\spinless
$$
Eq.~\expansion\ then reduces to the familiar result
$$
  e^{i{\bold{\vec p\cdot \vec r}}} = 4\pi \sum_{\ell m} i^\ell
j_\ell(pr)
 Y_{\ell m}({\bold{\hat r}}) Y^*_{\ell m} ({\bold{\hat p}})\,.\eqn\planewave
$$
The overall factor of $4\pi$ can be obtained by setting $p=0$.

Let us return to Eq.~\expansion.  We can invert this expansion and
evaluate the matrix element $\VEV{p,jm |p{\bold{\hat z}},\lambda}$
by imposing the orthonormality relation
$$
  \VEV{p,jm\lambda | p^\prime,j^\prime m^\prime\lambda^\prime}
     = {1\over p^2}\, \delta(p-p^\prime)\delta_{jj^\prime}
        \delta_{mm^\prime}
              \delta_{\lambda\lambda^\prime}\,.\eqn\orthonormal
$$
Using Eq.~\orthogonality, one ends up with
$$
  \ket{p,jm\lambda} = \sqrt{{2j+1\over 4\pi}} \int d\Omega\,
       D^{(j)}_{m\lambda} (R)^\ast \ket{{\bold{\vec p}},\lambda}\,,
\eqn\helicitystate
$$
where the rotation $R$ is specified by Euler angles $(\phi,\theta,-\phi)$.
Note that I have inserted a helicity label $\lambda$ on the left-hand
side of Eq.~\helicitystate\
since $\ket{p,jm}$ is explicitly an eigenstate of $\Lambda$.
It follows that $\ket{p,jm\lambda}$ are the simultaneous eigenstates of
$\bold{\vec J}\lss{\,2}$, $J_z$, $\bold{\vec S}\lss{\,2}$ and $\Lambda$.  These
are
the one-particle helicity states.

\REF\helicityrefs{J. Werle, {\it Relativistic
Theory of Reactions} (North-Holland, Amsterdam, 1966);
A.D. Martin and T.D. Spearman, {\it Elementary Particle
Physics} (North-Holland, Amsterdam, 1970); S.U. Chung, {\it Spin
Formalisms}, CERN Yellow Report 71-8 (1971); Martin L. Perl,
{\it High Energy Hadron Physics} (John Wiley \& Sons, New York, 1974).}
\REF\pilkuhn{H. Pilkuhn, {\it The Interactions of Hadrons}
(North-Holland, Amsterdam, 1967).}
In scattering processes, the initial state consists of two particles.
It is convenient to work in the center-of-momentum (CM) frame where
${\bold{\vec p}}\equiv {\bold{\vec p_1}} = -{\bold{\vec p_2}}$.
The total helicity of the two-particle system is defined as
$$\Lambda = \Lambda_1+\Lambda_2   =
{\bold{\vec J_1\cdot \hat p_1}} +{\bold{\vec J_2\cdot \hat p_2}}
      = ({\bold{\vec J_1}} -{\bold{\vec J_2}}){\bold{\cdot \hat p}}
\,.\eqn\totalhel
$$
It is also useful to define the relative helicity of the two-particle
system as
$${\bold{\vec J\cdot \hat p}}=({\bold{\vec J_1}}+{\bold{\vec J_2}})
{\bold{\cdot\hat p}}=\Lambda_1-\Lambda_2\,.\eqn\relativehel
$$
We can construct two-particle helicity states\refmark\jw\
in the same way we
obtained the one-particle helicity states above.\foot{%
Good textbook introductions to helicity formalism
in particle physics can be found in Refs.~[\helicityrefs] and
[\pilkuhn].}
First we define
two-particle plane wave states in the CM-frame.  However,
there is a delicate question of phases that must first
be addressed.  As in the derivation of the one-particle helicity
states, one begins from the state moving along the $z$-direction and
rotates to the desired orientation.  But, in the two-particle state in
the CM-frame, if ${\bold{\vec p}}_1=p{\bold{\hat z}}$ then
${\bold{\vec p}}_2=-p{\bold{\hat z}}$.  Thus, we must define the state
$\ket{-p{\bold{\hat z}},\lambda}$.  This can be done in two different
ways: (i) start in the rest frame with a state of helicity
$\lambda$, boost along the positive $z$-direction and then rotate to the
negative $z$-axis, or (ii) start in the rest frame with a state
of helicity $-\lambda$ and boost along the negative $z$-axis.  These
two results yield states that differ by a phase, so a convention is
required.  I shall choose the Jacob-Wick
second particle convention\refmark\jw\
which defines a helicity state of a particle of spin $s$ moving in the
negative $z$-direction to be
$$
  \ket{-p{\bold{\hat z}},\lambda} = (-1)^{s-\lambda} e^{-i\pi
J_{\hbox{$\scriptstyle y$}}}
     \ket{p{\bold{\hat z}},\lambda}\,.\eqn\jacobwick
$$
This definition implies that
$$
  \lim_{p\to0}\,\VEV{p{\bold{\hat z}},-\lambda| -p{\bold{\hat z}},\lambda}
= 1\,,\eqn\phasedef
$$
where Eq.~\propone\ has been used [which explains the origin of the
extra phase $(-1)^{s-\lambda}$ in Eq.~\jacobwick].  The two-particle
plane-wave state is then defined by
$$
  \ket{{\bold{\vec p}}; \lambda_1\lambda_2} \equiv U[R(\phi,\theta,-\phi)]
    \ket{p{\bold{\hat z}},\lambda_1} \otimes \ket{-p{\bold{\hat
    z}},\lambda_2}\,.\eqn\jwplanewave
$$
It follows from Eq.~\relativehel\ that
$$  {\bold{\vec J\cdot\hat p}}\ket{{\bold{\vec p}}; \lambda_1\lambda_2} =
\lambda
       \ket{{\bold{\vec p}}; \lambda_1\lambda_2}\,,\qquad\qquad
       \lambda\equiv \lambda_1-\lambda_2\,.\eqn\twohel
$$
{}From this point on, the analysis follows the derivation of the
one-particle helicity state.  The final result for the two-particle
helicity state [the analogue of Eq.~\helicitystate, with $\lambda\equiv
\lambda_1-\lambda_2$] is
$$
  \ket{p,jm\,\lambda_1\lambda_2} = \sqrt{2j+1\over 4\pi}
     \int d\Omega\, D^{(j)}_{m\lambda} (R)^\ast
     \ket{\bold{\vec p}; \lambda_1,\lambda_2}\,.\eqn\twoparthel
$$
\medskip
\section{Helicity amplitudes in decay and scattering processes}

{}From the expression for the two-particle helicity state given in
Eq.~\twoparthel, it is straightforward to derive formulae for
decay widths and scattering amplitudes in terms of helicity
amplitudes\refmark{\helicityrefs,\pilkuhn}.
Two examples are given below.

First, consider the two-body decay of an unstable particle of
spin $J$.  Let us work in the CM-frame.  Choose a $z$-axis and assume
that the decaying particle is polarized with $J_z$ quantum number
equal to $M$.  The final state particles have spin $s_i$ and helicity
$\lambda_i$ (where $i=1,2$).  The decay angular distribution for
$(J,M) \to (s_1,\l1) + (s_2,\l2)$ is given by
$$
  {d\Gamma\over d\Omega} = {p_f\over 32\pi^2 m^2}
         \bigm|{\cal M}^{JM}_{\l1\l2}(\theta,\phi)\bigm|^2\,,\eqn\decayrate
$$
where $m$ is the mass of the decaying particle,
$p_f$ is the CM-momentum,\foot{Explicitly,
$p_f=\lambda^{1/2}(m^2,m_1^2,m_2^2)/2m$, where $m_i$ ($i=1,2$) are the
masses of the decay products and $\lambda(a,b,c)\equiv
(a+b-c)^2-4ab$.} and
$$
  {\cal M}^{JM}_{\l 1\l 2} (\theta,\phi) =
   \sqrt{{2J+1\over 4\pi}}\ D^J_{M\lambda}(\phi,\theta,-\phi)^\ast\,
         {\cal M}^J_{\l 1\l 2}\,.\eqn\reduceddecay
$$
In the above formula, $\lambda\equiv\lambda_1-\lambda_2$, and
${\cal M}^J_{\l1\l2}$ is a reduced decay
amplitude which is a function of $J$, the outgoing helicities, and the
particle masses, but is independent of the angles $\theta$ and $\phi$.

Second, consider the two-body scattering process.
Again, it is convenient to work in the CM-frame.  Then, the
differential cross section for the scattering of particles of definite
helicities: $a(\l a)+b(\l b)\longrightarrow c(\l c)+d(\l d)$ is given by
$$
  {d\sigma\over d\Omega_{\rm CM}} = {1\over 64\pi^2s}
     \left({p_f\over p_i}\right)
     \bigm|{\cal M}_{\l c\l d;\l a\l b} (s,\theta,\phi) \bigm|^2 \,,
\eqn\dsigdomegacm
$$
where $\sqrt{s}$ is the CM-energy of the process,
$p_i$ ($p_f$) is the initial (final) CM-momentum [explicitly:
$p_i=\lambda^{1/2}(s,m_a^2,m_b^2)/2\sqrt{s}$ and for $p_f$, replace
initial state masses with final state masses], and the helicity
amplitudes for the scattering process are given by
$$
  {\cal M}_{\l c\l d;\l a\l b} (s,\theta,\phi) =
   \sum^\infty_{J=\max\{\l i,\l f\} } (2J+1)\, d^J_{\l i\l f}\, (\theta)
      e^{i(\l i-\l f)\phi}
  {\cal M}^J_{\l c\l d;\l a\l b} (s)\,,\eqn\reducedscattering
$$
summed over integers (half-integers) for integer (half-integer)
$\lambda_i$ and $\lambda_f$, where
$$ \l i\equiv\l a-\l b\,,\qquad\qquad \l f\equiv\l c-\l d\,.\eqn\ldiffs$$
In the above formula, ${\cal M}^J_{\l c\l d;\l a\l b} (s)$ is a
reduced matrix element which is independent of scattering angle.
Note that if all particles are spinless (take $\l i=\l f=0$ and
$J=\ell$, where $\ell$ is the orbital angular momentum), then
Eq.~\reducedscattering\ reduces to the usual partial wave expansion of
nonrelativistic quantum mechanics.  [In verifying this result, use
the fact that $d^\ell_{00}(\theta) = P_\ell(\cos\theta)$, which
follows from Eq.~\spharm.]

If the decay and scattering processes are mediated by parity (P)
and/or time-reversal (T) invariant interactions, then there are
nontrivial constraints on the reduced matrix elements.
Additional restrictions
are obtained if the initial and/or final state particles are
identical.  These constraints are summarized
below\refmark{\jw,\helicityrefs}.

(a) Parity
$$\eqalign{%
 &{\cal M}^J_{\l1\l2} = \eta\eta_1\eta_2 (-1)^{s_1+s_2-J}
 {\cal M}^J_{-\l1,-\l2}\,,\crr
 &{\cal M}^J_{\l c\l d;\l a\l b} (s) = (\eta_c\eta_d/\eta_a\eta_b)
    (-1)^{s_c+s_d-s_a-s_b}
    {\cal M}^J_{-\l c-\l d;-\l a-\l b} (s)\,, \cr  }\eqn\paritycons
$$
where $\eta_i$ is the intrinsic parity of particle $i$.  Note that in the
first formula above, $\eta$ is the intrinsic
parity of the decaying spin-$J$ particle.

(b) Time-reversal
$$
 {\cal M}^J_{\l c\l d;\l a\l b} (s) = {\cal M}^J_{\l a\l b;\l c\l d} (s)\,.
\eqn\timerevcons
$$

(c) Identical particles
$$\eqalign{%
 &{\cal M}^J_{\l1\l2} = (-1)^J
 {\cal M}^J_{\l2\l1}\,,\crr
&{\cal M}^J_{\l c\l d;\l a\l b} (s) = \cases{%
  (-1)^J {\cal M}^J_{\l c\l d;\l b\l a} (s),& $a$, $b$ identical,  \cr
  (-1)^J {\cal M}^J_{\l d\l c;\l a\l b} (s),& $c$, $d$ identical. \cr}}
\eqn\identity
$$
Note that in the case of identical particles, $J$ must be an integer.

In actual computations of decay and scattering processes from Feynman
diagrams, one can compute the helicity amplitudes by
employing explicit representations for the spin wave functions.  Then,
the reduced amplitudes can be identified from
Eqs.~\reduceddecay\ or \reducedscattering.

\section{Explicit spin-$\bold\half$ and spin-1 wave functions}

\REF\carruthers{Peter A. Carruthers, {\it Spin and Isospin in Particle
Physics} (Gordon and Breach, New York, 1971); M.D. Scadron, {\it Advanced
Quantum Theory} (Springer-Verlag, Berlin, 1991).}
In this section we summarize the explicit forms for the spin wave
functions for spin-1/2 and spin-1 particles
(see \eg, Ref.~[\carruthers]).  There are some delicate
phase choices to be made in writing down these explicit forms.
Sometimes, these choices actually matter, so I have been careful to be
consistent in my presentation of the formulae below.

\noindent (1) Spin-$\half$ helicity states

\REF\bjdrell{J.D. Bjorken and S.D. Drell, {\it Relativistic Quantum
Mechanics} (McGraw Hill, New York, 1964).}
The explicit form of the Dirac spinors depends on the choice of
representation for the $\gamma$-matrices.  Consider first the
standard or ``low energy" representation (using the
metric and $\gamma$-matrix
conventions of Bjorken and Drell\refmark\bjdrell):
$$\eqalign{%
 &\gamma^0 = \pmatrix{1&0 \cr 0& -1\cr}\,, \qquad
  \gamma^i = \pmatrix{0& \sigma^i\cr -\sigma^i&0\cr}\,, \qquad
  \gamma\ls{5} = i\gamma^0\gamma^1\gamma^2\gamma^3=
\pmatrix{0& 1\cr 1& 0\cr}\,,\cr}\eqn\gammamatrices
$$
where the $\sigma^i$ are the usual Pauli matrices.
The charge-conjugation matrix is defined by\foot{The reader should be
cautioned that this definition of $C$ differs from that of Ref.~[\bjdrell]
(and other standard texts) by a minus sign.  The choice of sign
is conventional.}
$$
 C = i\gamma^0\gamma^2 = \pmatrix{0& i\sigma^2\cr i\sigma^2&0\cr}
\,.\eqn\chargeconj
$$
The free-particle solutions to the Dirac equation
are four-component spinors
$$
  u(p) = \sqrt{2m} \pmatrix{\cosh\,{\zeta\over 2}\, \chi\cr
\sinh\,{\zeta\over 2}\,{\bold{\vec\sigma\cdot \hat p}}\,\chi\cr}\,,
\eqn\uspinor
$$
where $\chi$ is a two-component spinor, $p=(E\,;\,\bold{\vec p}\,)$ and
$$\eqalign{
\cosh\,{\zeta\over 2} =& \left({E+m\over 2m}\right)^{1/2}\,,\crr
\sinh\,{\zeta\over 2} =& \left({E-m\over2m}\right)^{1/2}\,.\cr}\eqn\czeta
$$
Note that the rapidity $\zeta$ is simply related to the energy and
three-momentum by $E=m\cosh\zeta$ and $p=m\sinh\zeta$.
The overall normalization factor in Eq.~\uspinor\ has been chosen so
that $\bar uu=2m$.  This ensures a smooth $m\to 0$ limit; that is,
for a massless spinor, take $\sqrt{2m}\cosh\zeta/2=\sqrt{2m}\sinh\zeta/2=
\sqrt{E}$ in Eq.~\uspinor.
The four-component antiparticle spinor is defined as
$$
  v(p) = C\bar u^T(p)\,,\eqn\vspinor
$$
where $C$ is the charge conjugation matrix [Eq.~\chargeconj] and
$\bar u\equiv u^\dagger\gamma^0$.

In order to construct the spin-1/2 helicity states, one chooses
the two-component spinors $\chi$ in Eq.~\uspinor\ to be eigenstates
of $\half{\bold{\vec\sigma\cdot\hat p}}$, \ie,
$$
\half {\bold{\vec\sigma\cdot\hat p}}\,\chi\ls\lambda = \lambda\chi\ls
\lambda,     \qquad \lambda = \pm\half\,.\eqn\chilambda
$$
If ${\bold{\hat p}}$ is a unit
vector with polar angle $\theta$ and azimuthal angle $\phi$
with respect to a fixed $z$-axis, then the two-component spinors
are
$$
  \chi\ls{1/2}({\bold{\hat p}}) = \pmatrix{ \cos {\theta\over 2} \cr
                            e^{i\phi} \sin{\theta\over 2} \cr},
  \qquad \chi\ls{-1/2} ({\bold{\hat p}}) = \pmatrix{%
                            -e^{-i\phi} \sin{\theta\over 2} \cr
                              \cos {\theta\over 2} \cr}\,.\eqn\twocomp
$$
One can also construct $\chi\ls\lambda(\bold{\hat p})$ from
$\chi\ls\lambda(\bold{\hat z})$ by employing the spin-1/2 rotation
operator
$$
\chi\ls\lambda(\bold{\hat p}) =
  e^{-i\theta \bold{\hat n\cdot\vec\sigma}/2}\,
  \chi\ls\lambda({\bold{\hat z}})\,,
\eqn\rottwocomp$$
where ${\bold{\hat n}}$ is defined in Eq.~\nhatdef.
Note that the two-component helicity spinors satisfy
$$\chi\ls{-\lambda}({\bold{\hat p}})=-2\lambda i\sigma^2\chi^\ast_\lambda
({\bold{\hat p}})\,.\eqn\twocompprop
$$

The explicit forms for the four-component helicity spinors are
$$\eqalign{%
  &u(p,\lambda) = \sqrt{2m}
 \pmatrix{\cosh\,{\zeta\over 2}\, \chi\ls\lambda({\bold{\hat p}})\cr
 2\lambda\sinh\,{\zeta\over 2}\,\chi\ls\lambda({\bold{\hat p}}) \cr}\ ,\crr
 &v(p,\lambda) =  \sqrt{2m}  \pmatrix{%
  \sinh\,{\zeta\over 2}\,\chi\ls{-\lambda}({\bold{\hat p}}) \cr
  -2\lambda\cosh{\zeta\over 2}\, \chi\ls{-\lambda}({\bold{\hat p}})\cr}\ .\cr}
\eqn\uvspinors$$
In the case of a two-particle helicity state, one must choose
an appropriate definition for $\chi\ls\lambda(-{\bold{\hat p}})$ in
the second particle spinor in order
to be consistent with the Jacob-Wick second particle convention.
First, note that if ${\bold{\hat p}}$ is pointing in the direction
specified by polar and azimuthal angles $(\theta,\phi)$, then
$-{\bold{\hat p}}$ points in the direction specified by $(\pi-\theta,
\phi\pm\pi)$ [the choice of sign is made so that $0\leq\phi\leq 2\pi$].
It follows from Eq.~\twocomp\ that\foot{An equivalent result for
$\chi\ls\lambda(-{\bold{\hat p}})$, which can be obtained from
Eq.~\rottwocomp, is
$$
  \chi\ls\lambda(-{\bold{\hat p}}) =  -2\lambda e^{-2i\lambda\phi}\,i\,
 {\bold{\hat n\cdot\vec\sigma}}\,\xi\ls{-\lambda}
 \chi\ls\lambda({\bold{\hat p}})\,,
$$
where the phase $\xi\ls\lambda$ is defined below Eq.~(1.42).}
$$
\chi\ls{-\lambda}(-{\bold{\hat p}})= \xi\ls\lambda\,\chi\ls{\lambda}
({\bold{\hat p}})\,,
\eqn\backspinor
$$
where the phase $\xi\ls\lambda\equiv(-1)^{s-\lambda}\,e^{-2i\lambda\phi}$
for a particle of spin $s$.  For $s=1/2$, it is convenient to write
$\xi\ls\lambda=2\lambda e^{-2i\lambda\phi}$.
For second particle spinors in the Jacob-Wick convention,
the overall phase of $\chi\ls\lambda(-{\bold{\hat p}})$ is
modified by choosing $\xi\ls\lambda\equiv 1$.

With the explicit forms for the helicity spinors, one can derive a
number of useful relations.  We list three such relations below, which
we will make use of later in this lecture:
$$
v(p,\lambda) = -2\lambda\,\gamma\ls{5} u(p,-\lambda)\,,\eqn\vurelation
$$
and
$$\eqalign{%
u(-p,-\lambda) &= \xi\ls{\lambda}\, \gamma^0 u(p,\lambda)\,,\crr
v(-p,-\lambda) &= \xi\ls{-\lambda}\, \gamma^0 v(p,\lambda)\,,\crr
}\eqn\usefulrelations$$
where $-p \equiv (E\,;\,-{\bold{\vec p}}\,)$ and $\xi\ls\lambda$
is defined below Eq.~\backspinor.

In high energy processes, the chiral or ``high energy'' representation
of the Dirac matrices and spinors is more useful if an
explicit representation is needed. For the record, I note that
$$\eqalign{%
  \gamma^\mu_c &= S\gamma^\mu_s S^\dagger\,,\cr
   \psi_c &= S\psi_s\,, \cr}\eqn\chiralrep
$$
where the subscripts $s$ and $c$ refer to the standard and chiral
representations respectively, $\psi$ is either a $u$ or $v$-spinor and
$$S = {1\over \sqrt 2}\pmatrix{1& -1 \cr 1& 1\cr}\,.\eqn\smatrix
$$
For example,
$$
  u(p) = {1\over \sqrt{2(E+m)}} \pmatrix{(m+E-{\bold{\vec
          p\cdot\vec\sigma}})
        \chi\cr  (m+E+{\bold{\vec p\cdot\vec\sigma}})\chi\cr }
       = \pmatrix{ \sqrt{p\cdot\sigma} \chi\cr
         \sqrt{p\cdot\overline\sigma} \chi\cr}\,,\eqn\chiralspinor
$$
\REF\schroeder{M.E. Peskin and D.V. Schroeder,
{\it An Introduction to Quantum
Field Theory} (Addison-Wesley, Reading, MA), to be published.}
where $\sigma^\mu = (1; {\bold{\vec\sigma}})$ and
$\overline\sigma^\mu = (1;-{\bold{\vec\sigma}})$.
See \eg, Refs.~[\pilkuhn] and [\schroeder] for further details on this
representation.

Although the explicit forms for spinors are occasionally useful in
practical computations, it is often more useful to employ
the helicity projection operators which
are independent of the specific Dirac matrix representation.
For a massive spin-1/2 particle with four-momentum
$p^\mu=(E\,;\,\bold{\vec p}\,)$, the spin four-vector is defined as
$$
  s^\mu = (2\lambda) {1\over m} \left(|{\bold{\vec p}}|\,;\,
E{\bold{\hat p}}
\right)\,,\eqn\spinvec
$$
where $2\lambda=\pm 1$ is twice the spin-1/2 particle helicity.
The spin four-vector satisfies
$$\eqalign{%
  &s\cdot p = 0\,, \cr
  &s\cdot s = -1\,. \cr}\eqn\spinprops
$$
Note that in the rest frame, $s =
2\lambda\,(0\,;\,{\bold{\hat p}})$, while in the
high energy limit (where $E\gg m$),
$s = 2\lambda p/m$.

The helicity spinors satisfy the Dirac equation and are eigenstates
of $\gamma\ls{5}\nots$ with unit eigenvalue.  Explicitly, we have
$$\eqalign{&\notp u(p,\lambda)=mu(p,\lambda)\,,\phantom{-}\qquad\qquad
\gamma\ls{5}\nots\, u(p,\lambda)=u(p,\lambda)\,,\crr
&\notp v(p,\lambda)=-mv(p,\lambda)\,,\qquad\qquad\,
\gamma\ls{5}\nots\, v(p,\lambda)=v(p,\lambda)\,.\cr}\eqn\uvdiraceqs
$$
{}From these results, one can derive the helicity projection operators for
a massive spin-1/2 particle:
$$\eqalign{%
&u(p,\lambda) \bar u(p,\lambda) = \half(1+\gamma\ls{5}\nots)\,(\notp+m)\,,\crr
&v(p,\lambda) \bar v(p,\lambda) = \half(1+\gamma\ls{5}\nots)\,(\notp-m)\,.\cr
}\eqn\projectionops$$

To apply the above formulae to the massless case, recall that
in the $m\to 0$ limit, $s=2\lambda p/m+{\cal O}(m/E)$.  Inserting
this result in Eq.~\uvdiraceqs,
it follows that the massless helicity spinors are eigenstates of
$\gamma\ls{5}$
$$\eqalign{%
  \gamma\ls{5} u(p,\lambda) &= 2\lambda u(p,\lambda)\,, \cr
  \gamma\ls{5} v(p,\lambda) &= -2\lambda v(p,\lambda)\,. \cr}\eqn\mzerospinors
$$
Applying the same limiting procedure to Eq.~\projectionops\
and using the mass-shell condition ($\notp\notp=p^2=m^2$), one obtains
the helicity projection operators for a massless spin-1/2 particle
$$\eqalign{%
&u(p,\lambda) \bar u(p,\lambda) = \half(1+2\lambda\gamma\ls{5})\,\notp\,,\cr
&v(p,\lambda) \bar v(p,\lambda) = \half(1-2\lambda\gamma\ls{5})\,\notp\,.\cr
}\eqn\masslessprojection$$

\endpage
\noindent (2) Spin-1 helicity states

Let $k^\mu=(k^0\,;\,{\bold{\vec k}}\,)$
be the four-momentum of a spin-1 particle moving in the
direction ${\bold{\hat k}}$ which points in a direction specified by
polar and azimuthal angles $\theta$ and $\phi$.  The spin-1 wave
function (or polarization four-vector) for the helicity $\lambda=\pm 1$
states is given by
$$
  \epsilon^\mu(k,\pm1) = \sqrt{\half} e^{\pm i\phi}
    (0\,;\,\mp\cos\theta \cos\phi + i\sin\phi\,,\,
     -i\cos\phi \mp\cos\theta \sin\phi\,,\,\pm\sin\theta)\,.\eqn\spinone
$$
Note that the above result holds both for massless and massive
spin-1 particles.  If the mass $m\neq 0$, one must also list the
polarization four-vector of the longitudinal ($\lambda=0$) state
$$
  \epsilon^\mu(k,0) = \left( {|{\bold{\vec k}}|\over m};\
        {k^0\over m}\ {\bold{\hat k}}\right)\,.\eqn\spinonezero
$$
The spin-1 polarization vectors satisfy
$$\eqalign{%
 &k\cdot \epsilon(k,\lambda) = 0\,, \cr
 &\epsilon(k,\lambda)\cdot \epsilon(k,\lambda^\prime)^\ast
            = -\delta_{\lambda\lambda^\prime}\,. \cr}\eqn\spinoneprops
$$
In addition, the overall phase of $\epsilon$ has been chosen such
that
$$\epsilon(k,-\lambda)=(-1)^\lambda\,\epsilon(k,\lambda)^\ast\,.
\eqn\epsilonphase
$$
For spin-1 particles moving in the $-{\bold{\hat k}}$ direction, one can
check that
$$
\epsilon^\mu(-k,-\lambda)=-\xi\ls\lambda\,g_{\mu\mu}\,\epsilon^\mu(k,
\lambda)\,,\eqn\backspinon
$$
where $-k\equiv(k^0;\,-{\bold{\vec k}})$ and $g_{\mu\nu}={\rm diag}\,
(1,-1,-1,-1)$ [but there is {\it no} implied sum
over $\mu$].  The definition of $\xi\ls\lambda$ is
the one appropritate for $s=1$ [see below Eq.~\backspinor];
\ie, $\xi\ls\lambda=(-1)^{1-\lambda}\,e^{-2i\lambda\phi}$.  As before,
for a second particle in the Jacob-Wick convention, the overall
phase of $\epsilon^\mu(-k,\lambda)$ is modified by choosing
$\xi\ls\lambda=1$.
\medskip
\section{Spin projection operator methods}

\REF\renard{F.M. Renard, {\it Basics of Electron Positron
Collisions} (Editions Fronti\`eres, Gif sur Yvette, France, 1981).}
All particle physics
students learn how to use projection operators to compute
spin-summed and averaged cross-sections in quantum field theory.
These methods also apply when the incoming and/or outgoing particles
are in definite spin states.  Many examples of such computations can
be found in Ref.~[\renard].
Here, I shall illustrate the method in
the computation of the squared matrix element for
$$e^+(p_2,\l2)\,+\, e^-(p_1,\l1)\,\to\,\mu^+(p_4,\l4)\,+\,\mu^-(p_3,\l3)
\,,\eqn\eemumu$$
where each lepton is in a definite helicity state as specified by the
corresponding $\lambda$ (and the $p_i$ are the corresponding
four-momenta).  There is one tree-level Feynman diagram shown below.

\medskip
\vbox{%
\centerline{\psfig{file=eetouu.ps,height=3.75cm}}
}
\medskip

Applying QED Feynman rules, the helicity amplitudes are
$$
  {\cal M}_{\l3\l4;\l1\l2} = {e^2\over s} \bar u(p_3,\l3)
      \gamma^\mu v(p_4,\l4) \bar v(p_2,\l2)\gamma_\mu
      u(p_1,\l1)\,,\eqn\helampmumu
$$
where $s = (p_1+p_2)^2$ is the CM-energy squared.  Squaring the
amplitude, and making use of the helicity projection operators
[Eq.~\projectionops], one obtains
$$\eqalign{|{\cal M}_{\l3\l4;\l1\l2}|^2 &= {e^4\over 16s^2} \Tr\,
        \gamma^\mu (1+\gamma\ls{5} \nots_4)(\notp_4-m_\mu)
         \gamma^\nu(1+\gamma\ls{5}\nots_3)(\notp_3+m_\mu) \crr
&\qquad\qquad \times \Tr\,\gamma_\mu (1+\gamma\ls{5} \nots_1)(\notp_1+m_e)
         \gamma_\nu(1+\gamma\ls{5}\nots_2)(\notp_2-m_e)\,. \cr
}\eqn\ampsquared$$
Although this expression is complicated, it can be worked out by hand.
However, it is instructive to simplify the computation by evaluating
the squared amplitude in the high energy limit.  In this case, I can
neglect all masses and replace $\nots_i$ with $\pm2\lambda$
[where the plus (minus) sign is chosen for (anti-)particles].
Eq.~\ampsquared\ then reduces to
$$\eqalign{%
  |{\cal M}_{\l3\l4;\l1\l2}|^2 &= {e^4\over 16s^2} \Tr\,
                    \gamma^\mu (1-2\l4\gamma\ls{5})\notp_4
         \gamma^\nu(1+2\l3\gamma\ls{5})\notp_3 \cr
   &\qquad\qquad \times \Tr\,\gamma_\mu (1+2\l1\gamma\ls{5})\notp_1
         \gamma_\nu(1-2\l2\gamma\ls{5})\notp_2 \crr
   &= \quarter
 e^4 \left[ (1+\cos^2\theta) (1-4\l1\l2)(1-4\l3\l4)\right.\cr
   &\qquad\qquad  \left.+2\cos\theta(2\l1-2\l2)(2\l3-2\l4) \right]\,.\cr}
\eqn\finalampsq
$$
Keep in mind that the $\lambda_i$ can take on the values $\pm\half$.

Eq.~\finalampsq\ is easy to understand.  First, note that if $\l1=\l2$
and/or if $\l3=\l4$, then ${\cal M}_{\l3\l4;\l1\l2} = 0$.  This is a
consequence of angular momentum conservation,
which is illustrated in the diagram below.

\medskip
\vbox{%
\centerline{\psfig{file=emep.ps,height=1.1cm}}
}
\medskip
\noindent
The long arrows indicate the direction of three-momentum (in the
CM-frame), while the short arrows indicate the direction of helicity
(which is either parallel or antiparallel to the momentum depending on
whether $\lambda$ is positive or negative, respectively).  For
example, in the above configuration, if we choose the positive
$z$-axis to lie along the electron direction, then the total $J_z=0$,
which is in conflict with $J_z=\pm 1$, which is required since
helicity is conserved in QED when fermion masses can be neglected.
By the same reasoning, the
allowed configurations would include the possibilities shown below.

\medskip
\vbox{%
\centerline{\psfig{file=emep2.ps,height=2.5cm}}
}
\medskip
\noindent
For these configurations, the squared amplitude takes the following form
$$
 |{\cal M}_{\l3-\l3;\l1,-\l1}|^2 = e^4\left[ 1+(2\l1)(2\l3)\cos\theta
                                    \right]^2\,.\eqn\helconfig
$$
This amplitude vanishes at $\cos\theta=-1$ [$\cos\theta=1$] if
$\lambda_1=\lambda_3$ [$\lambda_1=-\lambda_3$].  This result
can also be seen to be a consequence
of angular momentum conservation; that is, the amplitude vanishes when
$J_z=\pm 1$ in the initial state but $J_z=\mp 1$ in the final state.

\medskip
\section{The spinor helicity method}

\REF\heltechs{R. Gastmans and T.T. Wu, {\it The Ubiquitous Photon:
Helicity Method for QED and QCD} (Oxford University Press,
Oxford, England, 1990).}
\REF\massivemethods{F.A. Berends, P.H. Daverveldt and R. Kleiss,
{\sl Nucl. Phys.} {\bf B253} (1985) 441;
R. Kleiss and W.J. Stirling, {\sl Nucl. Phys.}
{\bf B262} (1985) 235; C. Mana and M. Martinez, {\sl Nucl. Phys.}
{\bf B287} (1987) 601.}
\REF\xzc{Z. Xu, D.-H. Zhang and L. Chang, Tsinghua University
preprints TUTP-84/3 (1984), TUTP-84/4 (1985); TUTP-84/5a (1985);
{\sl Nucl. Phys.} {\bf B291} (1987) 392.}
\REF\calkul{P. de Causmaecker, R. Gastmans, W. Troost, and T.T. Wu,
{\sl Phys. Lett.} {\bf 105B} (1981) 215;
{\sl Nucl. Phys.} {\bf B206} (1982) 53; F.A. Berends,
P. de Causmaecker, R. Gastmans, R. Kleiss, W. Troost, and T.T. Wu,
{\sl Nucl. Phys.} {\bf B206} (1982) 61; {\bf B239} (1984) 382;
{\bf B239} (1984) 395; {\bf B264} (1986) 243; 265.}
\REF\parke{M.L. Mangano and S.J. Parke, {\sl Phys. Rep.} {\bf 200}
(1991) 301.}
\REF\zbern{See \eg,
Z. Bern, in {\it Recent Directions in Particle Theory},
Proceedings of the 1992 Theoretical Advanced Study Institute in
Elementary Particle Physics, Boulder, CO, 1--26 June, 1992,
edited by J. Harvey and J. Polchinski (World Scientific, Singapore,
1993) pp.~471--535; C.S. Lam, McGill preprint 93-20 (1993).}
For scattering processes with more than two particles in the final
state, the spin projection operator methods quickly become unwieldy.
In this section, I shall give a brief introduction to the spinor
helicity method\refmark{\heltechs-\parke}
which is a powerful technique for computing helicity
amplitudes for multiparticle processes involving massless spin-1/2 and
spin-1 particles.  Although generally applied to tree-level
processes, more general techniques have been developed recently
which are
applicable to one-loop (and multiloop) diagrams\refmark\zbern.
The spinor helicity techniques are ideal for QCD where
light quark masses can almost always be neglected.  Generalizations of
these methods that incorporate massive spin-1/2 and spin-1
particles exist, although the complications that are introduced are
substantial.  The reader can study the papers of Ref.~[\massivemethods]
to learn about these techniques for massive particles; in this section,
I shall restrict my discussion to the massless case.
%

We begin by recalling that massless helicity spinors
are eigenstates of $\gamma\ls{5}$ [see Eq.~\mzerospinors].
Combining this result with Eq.~\vurelation\ yields
$$
  v(p,\lambda) = -2\lambda \gamma\ls{5} u(p,-\lambda)
               = u(p,-\lambda)\,,\eqn\vequalu
$$
and we see that particle and antiparticle massless spinors of opposite
helicity are the same.\foot{Eq.~\vequalu\ motivates the choice of
overall sign in the definition of the
charge conjugation matrix $C$ [Eq.~\chargeconj].}
With this observation, it is clear that
the algebra of massless spinors should simplify significantly.  Here,
I shall describe the spinor-helicity technique of Xu, Zhang and
Chang\refmark\xzc\ (denoted henceforth by XZC),
which is a modification of techniques developed by the CALKUL
collaboration\refmark\calkul.  Following XZC, I shall
introduce a very useful notation for massless spinors
$$\eqalign{%
\ket{p\pm}\equiv u(p,\pm\half) &= v(p, \mp\half)\,,\cr
\bra{p\pm}\equiv \bar u(p,\pm\half) &= \bar v(p, \mp\half)\,.\cr
}\eqn\brackets$$
These massless spinors have the following properties

\pointbegin
Massless Dirac equation
$$
  \notp \ket{p\pm} = \bra{p\pm}\notp = 0\,.\eqn\diraceq
$$

\point
Chirality conditions
$$\eqalign{%
 &(1\pm \gamma\ls{5}) \ket{p\mp} = 0\,, \cr
 &\bra{ p\pm}(1\pm \gamma\ls{5}) = 0\,. \cr}\eqn\chirality
$$

\point
Other massless spinor properties
$$
\bra{p\pm}\gamma^\mu \ket{p\pm} = 2p^\mu\,,
\eqn\propthree$$
$$
\ket{p\pm}\bra{p\pm} = \half(1\pm\gamma\ls{5})\notp\,,
\eqn\propfour$$
from which the following ``completeness''-type relation follows
$$
\ket{p+}\bra{p+} + \ket{p-}\bra{p-} =\notp\,.
\eqn\propfive$$

\point
Vanishing spinor-products
$$\eqalign{%
\VEV{p+|q+} = \VEV{p-|q-} &= 0\qquad {\rm for~any}~p,q\,,\cr
\VEV{p+|p-} &= 0\,.\cr}
\eqn\innerproduct$$

Eq.~\innerproduct\ shows that only a few of the possible
spinor-products are non-zero.  It is therefore
convenient to introduce the following notation for spinor-products:
$$\eqalign{%
  \VEV{pq} \equiv \VEV{p-|q+} = -\VEV{qp}\,, \cr
  [\,pq\,] \equiv \VEV{p+|q-} = -[\,qp\,]\,. \cr}\eqn\spinorproducts
$$
These two quantities are related by
$$
  \VEV{pq}^\ast = -[\,pq\,]\,,\eqn\bracketrelation
$$
where all spinors are assumed to have positive energy.\foot{In
Ref.~[\parke], calculations are performed assuming that all particle
momenta are outgoing.  As a result, energy-momentum conservation
implies that some spinors have negative energy, in which case
Eq.~\bracketrelation\ is replaced by
$\VEV{pq}^\ast = -{\rm sign}(p\cdot q)[\,pq\,]\,.$
}
It follows that
$$
  |\VEV{pq}|^2 = |[\,pq\,]|^2 = 2p\cdot q\,,\eqn\squaredbrackets
$$
which indicates that the spinor-products are roughly the
square roots of dot products.  Other useful relations can be
found in Appendix A of Ref.~[\parke].

Next, XZC introduced
a convenient expression for the massless spin-1
polarization vector.  Let $k$ be the four-momentum of the massless
spin-1 particle.  Let $p$ be a ``reference'' four-vector (usually
taken to be another four-momentum vector in the scattering process of
interest).  The XZC spin-1 polarization vectors are given by
$$
  \epsilon^\mu(k,\pm1) = \pm {1\over\sqrt{2}}\,
     {\bra{p\mp}\gamma^\mu \ket{k\mp} \over
        \VEV{p\mp|k\pm} }\,.\eqn\xzcspinone
$$
The only restriction on $p$ is that it not be parallel to $k$.
One can immediately check that $\epsilon^\mu(k,\lambda)$
[where $\lambda=\pm 1$] so defined
satisfies Eq.~\spinoneprops, which means that it is a valid
polarization four-vector.  Note that
$$\epsilon^\ast(k,\lambda) = \epsilon(k,-\lambda)\,,\eqn\epsone$$
which implies that the choice of phase in the above definition of
$\epsilon(k,\pm 1)$ differs from that of Eq.~\spinone.

To appreciate the significance of the reference four-vector $p$, one
can check that if $p$ is changed then
$\epsilon^\mu$ is shifted by a factor proportional to $k^\mu$.
This does not affect Eq.~\spinoneprops\ since $k^2=0$ for massless
spin-1 particles.  Moreover, this shift does not affect
the final result for any observable (in particular the sum of
amplitudes of any gauge invariant set of Feynman diagrams
remains unchanged).  Thus, the
presence of the arbitrary four-vector $p$ just reflects the gauge
invariance of the theory of massless spin-1 particles.

The following additional properties of
$\epsilon^\mu(k,\lambda)$ defined in Eq.~\xzcspinone\ are noteworthy:
$$p\cdot \epsilon(k,\lambda) = 0\,,\eqn\epstwo$$
$$\sum_\lambda \epsilon_\mu(k,\lambda)\epsilon^\ast_\nu(k,\lambda)
      = -g_{\mu \nu} + {p_\mu k_\nu + p_\nu k_\mu\over p\cdot k}
\,,\eqn\epsthree$$
$$\notep(k) = {\pm\sqrt 2\over \VEV{p\mp|k\pm} }
 \left( \ket{k\mp}\bra{p\mp} + \ket{p\pm}\bra{k\pm}
\right)\,.\eqn\epsfour
$$

We are now ready to apply the XZC technology to a real calculation.
Consider the process:
$$\gamma(k_1,\l1)\,+\,\gamma(k_2,\l2)\,\to\,q(p_1,\l3)\,+\,\overline
q(p_2,\l4)\,,\eqn\gamgamqq
$$
which is relevant as a background to Higgs production in $\gamma\gamma$
collisions (to be discussed in Lecture 3).
In the following computation, I shall
neglect the quark masses.  Two tree-level diagrams contribute to this
process

\medskip
\vbox{%
\centerline{\psfig{file=ggtoqq.ps,height=2.5cm}}
}
\medskip
\noindent
where the separated arrows above indicate the flow of four-momenta.
The corresponding amplitudes are
$$\eqalign{%
  {\cal M}_a &=  -e^2e_q^2 \bar u(p_1)\notep_1\,
       {\notp_1 -\notk_1\over (p_1-k_1)^2}\, \notep_2v(p_2)\,, \crr
  {\cal M}_b &=  -e^2e_q^2 \bar u(p_1)\notep_2\,
       {\notp_1 -\notk_2\over (p_1-k_2)^2}\, \notep_1v(p_2)\,, \cr}
\eqn\ampgamgamqq
$$
where $e_q$ is the quark charge in units of $e$.
It is easy to check that the helicity amplitudes with $\l1=\l2$ and/or
with $\l3=\l4$ are zero.  This leaves four non-zero helicity amplitudes
corresponding to
$$
  (\l3,\l4;\l1,\l2) = (\pm\half,\mp\half;+1,-1),\;
(\pm\half,\mp\half;-1,+1)\,.\eqn\nonzeroamps
$$
Consider first ${\cal M}(+\half,-\half;-1,+1)$.  I shall use the
following notation for the photon polarization four-vectors
$$\eqalign{%
  \notep_1 \equiv \notep(k_1,+1) = \notep(k_1,p_2,+1)\,, \cr
  \notep_2 \equiv \notep(k_2,-1) = \notep(k_2,p_1,-1)\,, \cr}\eqn\polvecs
$$
where the reference momentum chosen in each case is indicated.  Note
that one can choose different reference momentum for different
photons.  The decision on which reference momenta to choose is
somewhat of an art; experience will teach you which choices lead to
the most simplification in a given computation.  Writing the
amplitudes given in Eq.~\ampgamgamqq\ using the XZC bracket notation,
and using the fact that all external particles are massless,
$$\eqalign{%
  {\cal M}_a(+\half,-\half; -1,+1) &= {e^2e_q^2\over 2p_1\cdot k_1}
      \VEV{p_1+|\notep_1 (\notp_1-\notk_1))\notep_2|p_2+}\,, \crr
  {\cal M}_b(+\half,-\half; -1,+1) &= {e^2e_q^2\over 2p_1\cdot k_2}
      \VEV{p_1+|\notep_2 (\notp_1-\notk_2))\notep_1|p_2+}\,.\cr}
\eqn\ampgamgamqqtwo
$$
Next, we employ Eq.~\epsfour\ to write
$$\eqalign{%
     \notep_1 &= {\sqrt 2\over \VEV{p_2-|k_1+} }  \left(
         \ket{k_1-}\bra{p_2-} + \ket{p_2+}\bra{k_1+}\right)\,, \crr
     \notep_2 &={-\sqrt 2\over \VEV{p_1+|k_2-} }  \left(
         \ket{k_2+}\bra{p_1+} + \ket{p_1-}\bra{k_2-}\right)\,. \cr}
\eqn\epsidentities
$$
Using Eq.~\innerproduct, it follows that
${\cal M}_b = 0$ since $\VEV{p_1+|k_2+} = \VEV{p_1+|p_1-} = 0$.
Thus,
$$\eqalign{%
  {\cal M}(+\half,-\half; -1,+1) &= {-e^2e_q^2
          \VEV{p_1+|k_1-}
         \bra{p_2-}\notp_1-\notk_1 \ket{p_1-}
          \VEV{k_2-|p_2+}
\over p_1\cdot k_1
      \VEV{p_2-|k_1+}\VEV{p_1+|k_2-}  } \crr
          &= {e^2e_q^2\VEV{p_1+|k_1-} \VEV{p_2-|k_1+} \VEV{k_1+|p_1-}
         \VEV{k_2-|p_2+} \over
       p_1\cdot k_1 \VEV{p_2-|k_1+} \VEV{p_1+|k_2-}}\,, \cr}\eqn\spinoralgebra
$$
where we have used
$\notp_1\ket{p_1-} = 0$ [Eq.~\diraceq] and $\notk = \ket{k_1+}\bra{k_1+} +
\ket{k_1-}\bra{k_1-}$ [Eq.~\propfive].  Finally, we convert to the
spinor-product notation [Eq.~\spinorproducts].  Writing
$2p_1\cdot k_1 = |[\,p_1k_1\,]|^2$, it follows that
$$\eqalign{%
  {\cal M}\left(+\half,-\half; -1,+1\right) &= -2e^2e_q^2\,
       {[\,k_1p_1\,]\VEV{k_2p_2}\over [\,k_1p_1\,]^\ast[\,p_1k_2\,]}\,,\crr
              &= -2e^2e_q^2\, e^{i\alpha}\,
               {\VEV{k_2p_2}\over [\,p_1k_2\,]}\,, \cr}\eqn\penultimate
$$
where I have noted that $e^{i\alpha}\equiv[\,k_1p_1\,]/[\,k_1p_1\,]^\ast$
is a pure phase.  Thus, we arrive at the final and very simple result
$$
  |{\cal M} \left(+\half,-\half;-1,+1\right)|^2 =
       4e^4e_q^4\, {k_2\cdot p_2\over k_2\cdot p_1}\,.\eqn\finalresult
$$
By parity and identical particles [Eqs.~\paritycons\ and \identity],
the remaining non-zero helicity amplitudes immediately follow.  We
collect the complete set of non-zero helicity amplitudes below
$$\eqalign{%
  |{\cal M}\left(+\half,-\half; -1,+1\right)|^2 &=
  |{\cal M}\left(-\half,+\half; +1,-1\right)|^2 =
           4e^4e_q^4\, {k_2\cdot p_2\over k_2\cdot p_1} \,,  \crr
  |{\cal M}\left(+\half,-\half; +1,-1\right)|^2 &=
  |{\cal M}\left(-\half,+\half; -1,+1\right)|^2 =
    4e^4e_q^4\, {k_1\cdot p_2\over k_1\cdot p_1} \,.  \cr }\eqn\fullset
$$
Finally, it is conventional to introduce the kinematic invariants
$$\eqalign{%
     t &= -2p_1\cdot k_1 = -2p_2\cdot k_2\,, \cr
     u &= -2p_1\cdot k_2 = -2p_2\cdot k_1\,. \cr}\eqn\tandu
$$
{}From Eq.~\fullset, we immediately obtain the squared amplitude for
$\gamma\gamma\to q\overline q$ averaged over initial spins and summed
over final spins and colors ($N_c=3$)
$$
|{\cal M}|^2_{ave} = 2N_c e^4e_q^4
\left( {t\over u} + {u\over t}\right)\,.
\eqn\spinaveraged
$$

\section{\bf The Bouchiat-Michel formulae}

\REF\bailin{D. Bailin, {\it Weak Interactions} (Adam Hilger, Bristol,
England, 1982).}
\REF\michel{C. Bouchiat and L. Michel, {\sl Nucl.\ Phys.} {\bf 5} (1958)
416; L. Michel, {\sl Suppl.\ Nuovo Cim.} {\bf 14} (1959) 95.}
\REF\othermethods{H.W. Fearing and R.R. Silbar, {\sl Phys.\ Rev.}
{\bf D6} (1972) 471; M. Caffo and E. Remiddi, {\sl Helv.\ Phys.\ Acta}
{\bf 55} (1982) 339; G. Passarino, {\sl Phys. Rev.} {\bf D28} (1983)
2867; {\sl Nucl. Phys.} {\bf B237} (1984) 249; K. Hagiwara and
D. Zeppenfeld, {\sl Nucl. Phys.} {\bf B274} (1986) 1;
E. Yehudai, FERMILAB-PUB-92/256-T (1992); A. Ballestrero and E. Maina,
Turin Univ. preprint DFTT-76/93 (1994);
D.A. Dicus and R. Vega, unpublished.}
Instead of trying to generalize the methods of the previous section to
the case of massive fermions, I shall introduce yet another method
for evaluating helicity amplitudes.  This method is well suited for
scattering processes in which the initial state consists of two equal
mass fermions.  First, one introduces three four-vectors
$s^a_\mu$, $a=1,2,3$ such that the $s^a$
and $p/m$ form an orthonormal set of four-vectors.  That is,
$$\eqalign{%
      p\cdot s^a &= 0\,,\cr
      s^a\cdot s^b &= -\delta^{ab}\,,\cr
      s^a_\mu\,s^a_\nu &= -g_{\mu\nu} + {p_\mu p_\nu\over
m^2}\,,\cr}\eqn\orthonormalset
$$
where repeated indices (such as the index $a$ above) are implicitly
summed over unless otherwise stated.
A convenient choice for the $s^a$ is
$$\eqalign{%
s^{1\mu} &= (0\,;\,\cos\theta\cos\phi,\,\cos\theta\sin\phi,\,-\sin\theta)\,,\cr
s^{2\mu} &= (0\,;\, -\sin\phi,\,\cos\phi,\,0)\,, \cr}\eqn\soneandtwo
$$
and
$$
s^{3\mu} =\left({|\bold{\vec p}\,| \over m}\,;\,{E\over m}\,{\bold{\hat
p}} \right)\,,\phantom{,\cos\phi\sin\phi,-\sin\theta,)}
\eqn\sthree
$$
in a coordinate system where
${\bold{\hat p}}= (\sin\theta\cos\phi,\,\sin\theta\sin\phi,\,\cos\theta)$.
Note that $s^3$
is identical to the positive helicity spin vector;
that is, $2\lambda s^3=s$ [see Eq.~\spinvec].  From these explicit
forms, it is easy to derive a number of useful
properties\foot{I do not distinguish between
upper and lower Latin indices.  Thus, $\epsilon^{abc}$ is the
usual Levi-Cevita tensor in three space dimensions with
$\epsilon^{123} = 1$.}
$$\eqalign{%
&\epsilon^{\mu\nu\lambda\sigma}p_\mu s^1_\nu s^2_\lambda s^3_\sigma=m\,,\crr
&s_\mu^a s_\nu^b-s_\nu^a s_\mu^b= {\epsilon^{abc}
\epsilon_{\mu\nu\lambda\sigma}p^\lambda s^{c\sigma}\over m}\,,\crr
&\nots^a\nots^b=-\delta^{ab}+{i\epsilon^{abc}\gamma\ls{5}\,\notp\,\nots^c
\over m}\,,\cr}
\eqn\otherprops
$$
where $\epsilon_{0123}=-\epsilon^{0123}\equiv 1$.
One can check that the helicity spinors satisfy\refmark\bailin\
$$\eqalign{\gamma\ls{5}\nots^a u(p,\lambda^\prime) &=
\tau^a_{\lambda\lambda^\prime}\,u(p,\lambda)\,,\crr
\gamma\ls{5}\nots^a v(p,\lambda^\prime) &=
\tau^a_{\lambda^\prime\lambda}\,v(p,\lambda)\,,\cr}
\eqn\gammafivesa
$$
where the $\tau^a$ are the Pauli matrices,\foot{The first (second) row
and column of the $\tau$-matrices correspond to $\lambda=1/2\ (-1/2)$.
Thus, for example,
$\tau^3_{\lambda\lambda^\prime}=2\lambda\delta_{\lambda
\lambda^\prime}$ (no sum over $\lambda$).}\
and there is an implicit sum over the repeated label $\lambda=\pm\half$.
In Eq.~\gammafivesa, the second formula may be obtained from the first
one by using Eq.~\vurelation\ and noting that $4\lambda\lambda^\prime
\tau^a_{-\lambda,-\lambda^\prime}=-\tau^a_{\lambda^\prime\lambda}$.
Consequently, the helicities $\lambda$ and
$\lambda^\prime$ that label $\tau^a$ appear in the second formula
of Eq.~\gammafivesa\ in reversed order.  Finally, note that for
$a=3$, Eq.~\gammafivesa\ reduces to a result previously obtained
[see Eq.~\uvdiraceqs].

Using Eq.~\gammafivesa, one can derive the following
formulae first introduced by
Bouchiat and Michel\refmark\michel\ for spin-1/2 particles of mass $m$
$$\eqalign{%
  u(p,\lambda^\prime)\bar u(p,\lambda) &= \half
       \left[\delta_{\lambda\lambda^\prime} + \gamma\ls{5} \nots^a
         \tau^a_{\lambda\lambda^\prime}\right](\notp+m)\,,\crr
          v(p,\lambda^\prime)\bar v(p,\lambda) &= \half
       \left[\delta_{\lambda^\prime\lambda} + \gamma\ls{5} \nots^a
     \tau^a_{\lambda^\prime\lambda}\right](\notp-m)\,.\cr}\eqn\bmichel
$$
Note that the Bouchiat-Michel formulae are generalizations of the
helicity projection operators.  The former [Eq.~\bmichel] reduces to
the latter [Eq.~\projectionops] when $\lambda=\lambda^\prime$ after using
$2\lambda s^{3}=s$.
Although the above results apply to the $m\neq 0$ case, the $m=0$
limit of the Bouchiat-Michel formulae can be easily obtained.  Noting
that $s^3=p/m+{\cal O}(m/E)$ in the high energy limit, and using the
mass-shell condition ($p^2=m^2$), it follows that the $m\to 0$
limit is smooth.  The end result is
$$\eqalign{%
  u(p,\lambda^\prime)\bar u(p,\lambda) &= \half
       (1+2\lambda\gamma\ls{5})\,\notp\,\delta_{\lambda\lambda^\prime}
         +\half\gamma\ls{5}[\nots^1\tau^1_{\lambda\lambda^\prime}
         +\nots^2\tau^2_{\lambda\lambda^\prime}]\,\notp\,,\crr
          v(p,\lambda^\prime)\bar v(p,\lambda) &= \half
       (1-2\lambda\gamma\ls{5})\,\notp\,\delta_{\lambda^\prime\lambda}
       +\half\gamma\ls{5}[\nots^1\tau^1_{\lambda^\prime\lambda}
      +\nots^2\tau^2_{\lambda^\prime\lambda}]\,\notp\,.\cr}
\eqn\michelzero
$$
As expected, when $\lambda=\lambda^\prime$, we recover the helicity
projection operators for massless spin-1/2 particles
[Eq.~\masslessprojection].

I shall now illustrate how one can use the Bouchiat-Michel formulae to
evaluate helicity amplitudes involving two equal-mass spin-1/2
particles.  A typical amplitude involving a fermion-antifermion pair
takes the form
$$
  {\cal M}_{\l1\l2} = \bar u(p_1,\l1)\,\Gamma\, v(p_2,\l2)\,,\eqn\typicalamp
$$
where $\Gamma$ can contain Dirac matrices, spin-1 polarization
vectors, \etc\  In the CM-frame, ${\bold{\vec p_1}}= -{\bold{\vec p_2}}
= {\bold{\vec p}}$.  Using the notation
$-p\equiv (E\,;\,{-\bold{\vec p}}\,)$,
$$\eqalign{%
  {\cal M}_{\l1\l2} &= \bar u(p,\l1)\, \Gamma\, v(-p,\l2) \crr
     &= -2\lambda_2 \bar u(p,\l1)\,\Gamma\, \gamma\ls{5} u(-p,-\l2) \crr
     &= -2\lambda_2
        \bar u(p,\l1)\,\Gamma\, \gamma\ls{5} \gamma^0 u(p,\l2) \crr
     &= -\lambda_2
      \Tr\left\{\Gamma\,\gamma\ls{5}\gamma^0
        [\delta_{\l1\l2} +\gamma\ls{5} \tau^a_{\l1\l2}\nots^a]
        (\notp+m)\right\}\,,  \cr}\eqn\manipulation
$$
where Eqs.~\vurelation\ and \usefulrelations\ have
been used (with the Jacob-Wick
second particle convention, \ie, $\xi\ls\lambda=1$)
to manipulate the amplitude into a form where the
Bouchiat-Michel formulae [Eq.~\bmichel] can be applied.
By performing the trace, one completes the direct evaluation of the
helicity amplitude.
As a trivial example, set $\Gamma = 1$ in Eq.~\manipulation.  The
result is
$$\eqalign{%
  \bar u(p,\l1) v(-p,\l2) &= \lambda_2
          m\tau^a_{\l1\l2} \Tr\,(\gamma^0\nots^a)\crr
      &= 4\lambda_2|\,{\bold{\vec p}}\,|  \tau^3_{\l1\l2} \crr
      &= 2|{\bold{\vec p}}\,|  \delta_{\l1\l2}\,. \cr}
\eqn\gammaone
$$

Alternative and related methods for directly evaluating
expressions such as Eq.~\typicalamp\ can
be found in Ref.~[\othermethods].

\medskip
\section{Density matrix techniques in unstable particle production and
decay}

\REF\barut{C. Bourrely, E. Leader and J. Soffer, {\sl Phys. Rep.}
{\bf 59} (1980) 95; T.B. Anders, A.O. Barut, and W. Jachmann,
{\sl Int. J. Mod. Phys.} {\bf A6} (1991) 4223; {\bf A8} (1993) 5383.}
\REF\nieh{F. Bletzacker and H.T. Nieh, {\sl Phys. Rev.} {\bf D14} (1976) 1251.}
\REF\martin{B.R. Martin, E. de Rafael and J. Smith, {\sl Phys. Rev.} {\bf
D2} (1970) 179.}
\REF\tsai{Y.-S. Tsai, {\sl Phys. Rev.} {\bf D4} (1971) 2821).}
\REF\kawasaki{S. Kawasaki, T. Shirafuji and S.Y. Tsai,
{\sl Prog. Theor. Phys.} {\bf 49} (1973) 1656.}
\REF\pham{J.-L. Cortes, M. Gourdin and X.Y. Pham, PAR-LPTHE 79/22 (1979);
M. Gourdin and X.Y. Pham, {\sl Nucl. Phys.} {\bf B164} (1980) 399.}

A typical process in high energy physics starts with a two-body collision
of particles of definite spin.  If the beams are polarized, then the
initial particles are in definite helicity states.  If the beams are
unpolarized, then one must average over initial spins.  Particles that
are produced in the collision are often unstable.  A complete quantum
mechanical description of the scattering process including the decay
of all unstable intermediate states must account for possible
interference effects between the production and decay processes.
In this section, I
shall always work in the narrow width approximation; \ie,
the width of the unstable particle is small compared to its mass.
This approximation is very good in many cases of interest.
In practical terms, it implies that the Breit-Wigner resonance shape
can be approximated by a $\delta$-function in cross-section
calculations.  Explicitly,
$$
{1\over (s-m^2)^2+m^2\Gamma^2}\approx {\pi\over
m\Gamma}\delta(s-m^2)\,.\eqn\narrowwidth
$$
For the scattering process, $A+B\to C_1+C_2+\dots\,$, if
$C_1$ is spinless  and decays via $C_1\to D_1+D_2+\dots\,$,
then in the narrow width approximation, the total cross section is%
\refmark\pilkuhn
$$
\sigma\ls{\rm T}\approx \sigma(A+B\to C_1+C_2+\dots)\,{\rm BR}(C_1
\to D_1+D_2+\dots)\,,\eqn\factorize
$$
where the branching ratio is defined by
$${\rm BR}(C_1\to D_1+D_2+\dots)={\Gamma(C_1\to
D_1+D_2+\dots)\over\Gamma}\,,\eqn\branching
$$
and $\Gamma$ is the total width of particle $C_1$.
Thus, in the narrow width approximation, there is no correlation
between the production and decay of a spinless particle.
On the other hand, if
the unstable particle has non-zero spin, a proper
computation should take into account spin correlations between
the production and decay.  This is most easily done by introducing the
concept of density matrices.  Good textbook introductions to density
matrices in particle physics can be found in Refs.~[\helicityrefs]
and [\pilkuhn].  In addition, see Ref.~[\barut] for a
comprehensive treatment of polarization phenomena using helicity
amplitude techniques.
In this section, I shall illustrate the use of density matrices
in two simple examples.  Further details on the material in
this section can be found in Refs.~[\nieh--\pham].

As a first example, consider the process
$$\eqalign{%
 &A + B \to C_1 + C_2 + \dots   \cr
 &\hphantom{A+B\to C}\ca D_ 1+D_2+\dots  \cr}
\eqn\totalprocess$$
where $C_1$ is a spin-1/2 fermion.
Let ${\cal M}_\lambda$ [${\cal N}_\lambda$] be the matrix element for
production [decay] of $C_1$ with helicity $\lambda$.
We then define the production and decay density matrix elements,
respectively, as follows
$$\eqalign{%
   \rho^P_{\lambda\lambda^\prime} &= \sum {\cal M}_\lambda
        {\cal M}^*_{\lambda^\prime}\,,\crr
   \rho^D_{\lambda\lambda^\prime} &= \sum {\cal N}_\lambda
        {\cal N}^*_{\lambda^\prime}\,,\cr }\eqn\densitymatrices
$$
where the summation sign indicates that one should average over
initial spins and sum over the final spins of all particles excluding
$C_1$.  (More complicated density matrices
can be defined if other initial or final state particles are also
prepared or observed in definite helicity states.)
These matrices are sometimes referred to as helicity density matrices
to emphasize the fact that they are defined with respect to states of
definite helicity (other choices are possible).  Note that the
diagonal elements $\rho_{\lambda\lambda}$ correspond to the usual
squared matrix element for the production or decay of the particle of
helicity $\lambda$.
The total squared matrix element for the process
$A+B \to (D_1+D_2 + \dots ) + C_2 + \dots$ is then given by
$$
  \sum |{\cal M}_{\rm total}|^2 = \Tr(\rho^P\rho^D)
    \equiv \rho^P_{\lambda\lambda^\prime}
\rho^D_{\lambda^\prime\lambda}\,.\eqn\totalcross
$$
The meaning of the summation sign here is similar to its use above.
The nontrivial spin correlations between production and decay are
evident in the above formulae, since the total squared matrix element
contains terms involving products of the off-diagonal density matrix
elements.

Since $C_1$ is assumed to be a spin-1/2 particle,
the most general forms for $\rho^P$ and $\rho^D$ can be deduced
from the Bouchiat-Michel formulae.  In particular, Eq.~\bmichel\
implies that the spin-1/2 density matrices must be linear in
$\delta_{\lambda\lambda^\prime}$ and in
$\tau^a_{\lambda\lambda^\prime} s^a_\mu$.  Thus, we may write\foot{
Note that in the definitions presented above, $\Tr\,\rho$ is equal to
the helicity-averaged squared amplitude.  In the literature, one
often finds that density matrices are normalized such that
$\Tr\,\rho=1$, although I will not follow this convention here.}
$$\eqalign{%
   \rho^P_{\lambda\lambda^\prime} &= A \delta_{\lambda\lambda^\prime}
         + \tau^a_{\lambda\lambda^\prime} s^a_\mu B^\mu\,, \crr
   \rho^D_{\lambda\lambda^\prime} &= C \delta_{\lambda\lambda^\prime}
         + \tau^b_{\lambda\lambda^\prime} s^b_\nu D^\nu\,, \cr}
\eqn\generalform
$$
where $A$, $B$, $C$, and $D$ are functions of the kinematic invariants
of the problem.  These results should be compared with the more
traditional expression
$$
\rho^D=C(I+{\bold{\vec P\cdot\vec\tau}})\,,
\eqn\polarization$$
where $I$ is the $2\times 2$ identity matrix and
$\bold{\vec P}$ is the polarization vector of the decaying spin-1/2
particle.  Comparing the two equations above yields $P^b= s^b_\nu
D^\nu/C$,  which can be inverted [using the second equation of
Eq.~\orthonormalset] to give: $D^\nu= -CP^a s_\nu^a$.

In a parity-conserving process, there are restrictions of the form of
the density matrices.  These can be derived from the parity
constraints on helicity amplitudes [see Eq.~\paritycons].  The end
result is that the density matrix elements must satisfy\refmark\pilkuhn\
$$\rho_{-\lambda,-\lambda^\prime}=(-1)^{\lambda-\lambda^\prime}
\rho_{\lambda\lambda^\prime}\,.\eqn\rhoparity
$$
For the $2\times 2$ spin-1/2 density matrix, this constraint implies
that the two off-diagonal elements are opposite in sign.  Using
Eq.~\polarization, it follows that for parity conserving interactions
$P_x=P_z=0$ while $P_y$ may be non-zero (\ie, the polarization vector is
normal to the production plane), which is a well
known result.  Equivalently, in the notation of Eq.~\generalform,
$B_\mu$ and $D_\mu$ must be proportional to $s^2_\mu$.

Using the general forms for the density matrix elements shown in
Eq.~\generalform, it is easy to compute the total amplitude squared
[Eq.~\totalcross] for the process shown in Eq.~\totalprocess.
Using $\Tr\tau^a = 0$ and $\Tr\tau^a\tau^b = 2\delta^{ab}$,
we obtain [using Eq.~\orthonormalset]
$$
  \Tr\,(\rho^P\rho^D) = 2\left[ AC+B^\mu \left(-g_{\mu\nu} +
         {p_\mu p_\nu\over m^2} \right) D^\nu \right]\,.\eqn\trrhorho
$$
The term in Eq.~\trrhorho\
involving $B$ and $D$ describes quantum mechanical
correlations between the production and decay processes.

As a second example, consider the process
$$\eqalign{%
 &A + B \to C_1 + C_2   \cr
 &\hphantom{A+B\to C_1+C}\ca F_1+F_2+\dots  \cr
\noalign{\vskip-22pt}
 &\hphantom{A+B\to C}\cca D_1+D_2+\dots\cr}
\eqn\fullprocesstwo
$$
where $C_1$ and $C_2$ are both spin-1/2 particles.
Let ${\cal M}_{\lambda\mu}$ be the matrix element for the production
of $C_1$ and $C_2$ with helicities $\lambda$ and $\mu$, respectively.
Let ${\cal N}_\lambda^{(1)}$ [${\cal N}_\lambda^{(2)}$]
be the matrix element for the decay of $C_1$
[$C_2$] with helicity $\lambda$ [$\mu$].
We then define the production and decay density matrix elements,
respectively, as follows:
$$
\eqalign{
\rho^P_{\la\la';\mu\mu'} = & \sum
 {\cal M}_{\la\mu} {\cal M}^*_{\la'\mu'}\,,\crr
\rho^{D_1}_{\la\la'} = & \sum {\cal N}^{(1)}_\la {\cal N}^{(1)*}_{\la'}\,,\crr
\rho^{D_2}_{\mu\mu'} =& \sum {\cal N}^{(2)}_\mu {\cal
N}^{(2)*}_{\mu^\prime}\,, \cr}\eqn\doublerhos
$$
where the summation sign indicates that one should average over
initial spins and sum over final spins of all particles excluding
$C_1$ and $C_2$.
Then, the total squared matrix element for the process
$A+B \to (D_1+D_2+\dots)+(F_1+F_2+\dots)$ is given by
$$
\sum |{\cal M}_{\rm total}|^2 =
\rho^P_{\la\la';\mu\mu'}\rho^{D_1}_{\la'\la} \rho^{D_2}_{\mu'\mu}\,.
\eqn\totalmtwo
$$
Following the same steps as above, we use the Bouchiat-Michel formulae
to write down the general forms for the density matrix elements:
$$
\eqalign{
\rho^P_{\la\la';\mu\mu'} = & A\de_{\la\la'}\de_{\mu\mu'} + \de_{\mu\mu'}
 \tau^a_{\la\la'} s^a_{1\a}B^\a +\de_{\la\la'}\tau^b_{\mu\mu'}s^b_{2\be}C^\be
      + \tau^a_{\la\la'}\tau^b_{\mu\mu'} s^a_{1\a}
      s^b_{2\beta} D^{\a\be}\,,\crr
\rho^{D_1}_{\la\la'} = & E\de_{\la\la'}+\tau^c_{\la\la'} s^c_{1\rho}
F^\rho\,,\crr      \rho^{D_2}_{\mu\mu'} = & G\de_{\mu\mu'} +
\tau^d_{\mu\mu'} s^d_{2\si} H^\si\,. \cr}\eqn\denmatstwo
$$
Restrictions in the case of parity conservation can be obtained as
before.  If parity is conserved in the production process, then $B$
and $C$ are both proportional to $s^2$ (note that
there is no restriction on
$D$).  If parity is conserved in the decay process, then $F$ and $H$
are each proportional to $s^2$.

The total squared amplitude for the process shown
in Eq.~\fullprocesstwo\ is
$$
\eqalign{
&\rho^P_{\la\la';\mu\mu'}\rho^{D_1}_{\la'\la} \rho^{D_2}_{\mu'\mu} =
 4\left[AEG+B^\a \left(-g_{\a\rho}+{p_{1\a}p_{1\rho}\over m_1^2}\right)
 F^\rho G   \right. \crr
&+ C^\beta\!\left(\!-g_{\be\si}+ {p_{2\be}p_{2\si}\over m^2_2}\right)
 H^\si E
+ \left. F^\rho\!\left(\!-g_{\rho\a}+{p_{1\rho}p_{1\a}\over m^2_1}\right)
D^{\a\be}\! \left(\!-g_{\be\si} + {p_{2\be}p_{2\si}\over m^2_2}\right)
H^\si\right]
.\cr}\eqn\doublesigma
$$
Note that all terms in Eq.~\doublesigma\ apart from the first term
proportional to $AEG$ reflect the non-trivial correlations between the
production of $C_1$ and $C_2$ and their subsequent decays.

To gain a better understanding of these results, I shall give another
derivation of Eq.~\doublesigma\ which is more physically motivated,
following the analysis of Ref.~[\tsai].  For simplicity, I shall
consider the case of $B=C=0$.
Let ${\bold{\vec w_1}}$ and ${\bold{\vec w_2}}$ be the polarizations of
$C_1$ and $C_2$ in their
rest frames. For example, if we define an orthonormal set of axes
${\bold{\hat e_i}}$ in three-space, then the components of
${\bold{\vec w_1}}$ are given by
$$w_{1i}= {({\rm number\,of}\,C_1\,{\rm with\,spin\,along} +{\bold{\hat e_i}})-
({\rm number\,of}\,C_1\,{\rm with\,spin\,along} -{\bold{\hat e_i}})\over
({\rm number\,of}\,C_1\,{\rm with\,spin\,along} +{\bold{\hat e_i}})+
({\rm number\,of}\,C_1\,{\rm with\,spin\,along} -{\bold{\hat e_i}})}
\eqn\wonex
$$
and so on.
Now, compute the production cross-section for
$A+B\to C_1({\bold{\vec s_1}}) + C_2({\bold{\vec s_2}})$, \ie, the probability
distribution for producing $C_1$ with its spin in the direction of
${\bold{\vec s_1}}$ in its rest frame, and $C_2$
with its spin in the direction of ${\bold{\vec s_2}}$ in its
rest frame. We take ${\bold{\vec s_1}}$ and
${\bold{\vec s_2}}$ to be unit vectors.
Symbolically, we write
$$
d\si(A+B\to C_1+C_2)=C+D_{ij}s_{1i}s_{2j} \,,\eqn\symbolsigma
$$
where there is an implicit sum over $i,j=1,2,3$.

Next, consider the problem of computing the decay distribution of $C_1$ (or
$C_2$) of arbitrary polarization
${\bold{\vec w}}$. We allow for the possibility of a
mixed state so that the norm of ${\bold{\vec w}}$ need not be unity.
(In general, $|{\bold{\vec w}}|\leq1$.)
For example, the decay of an unpolarized state would correspond to
${\bold{\vec w}}=0$. Now, the decay distributions
for $C_1$ and $C_2$ (in their respective rest frames)
are schematically given by
$$
\eqalign{%
d\Gamma(C_1\to D_1+D_2+\dots)= & A_1+B_1\,{\bold{\vec q_1\cdot\vec w_1}}\,,\cr
d\Gamma(C_2\to F_1+F_2+\dots)=& A_2+B_2\,{\bold{\vec q_2\cdot\vec w_2}}\,,\cr}
\eqn\gammac
$$
where ${\bold{\vec q_1}}$ and ${\bold{\vec q_2}}$ are final state momenta
of one of the decay products of $C_1$ and $C_2$ respectively.
Using Eq.~\symbolsigma,
The number of $C_1$ having spin along the direction ${\bold{\hat e_i}}$ with
the
polarization of $C_2$ in a certain direction ${\bold{\vec s_2}}$
is proportional to
$C+D_{ij}s_{2j}$; whereas the corresponding number of $C_1$ having spin along
the direction $-{\bold{\hat e_i}}$ is $C-D_{ij}s_{2j}$.
Hence, using Eq.~\wonex, one finds
$$
w_{1i}=  {(C+D_{ij}s_{2j})-(C-D_{ij}s_{2j})\over(C+D_{ij}s_{2j})+
   (C-D_{ij}s_{2j})} = {D_{ij}s_{2j}\over C}\,.\eqn\wonei
$$
Inserting this result into the formula for $d\Gamma(C_1\to D_1+D_2+\dots)$
[see Eq.~\gammac], we symbolically have:
$$
d\si\otimes d\Gamma(C_1\to D_1+D_2+\dots)\propto CA_1+B_1(q_1)_i
D_{ij}s_{2j}\,.\eqn\sigmagamma
$$
Similarly, we may compute $w_{2j}$ (holding fixed the
angular distribution of $C_1$).  The result is:
$$
w_{2j} =  {CA_1+B_1(q_1)_iD_{ij}-\left[CA_1-B_1(q_1)_iD_{ij}\right]\over
  CA_1+B_1(q_1)_iD_{ij}+\left[CA_1-B_1(q_1)_iD_{ij}\right]}
=  {B_1(q_1)_iD_{ij}\over CA_1} \,.\eqn\wtwoi
$$
Substituting this into the formula for $d\Gamma(C_2\to F_1+F_2+\dots)$
[see Eq.~\gammac], one ends
up with the combined angular distribution of the decay products of $C_1$ and
$C_2$ at fixed production angle
$$
d\si\otimes d\Gamma(C_1\to D_1+D_2+\dots)\otimes d\Gamma(C_2\to F_1+F_2+\dots)
\propto CA_1A_2 + B_1B_2(q_1)_i(q_2)_jD_{ij}\,.\eqn\combineddist
$$
In Eqs.~\sigmagamma\ and \combineddist, I have omitted the overall
normalization.   But, this factor is easily
obtained by considering the case of $B_1=B_2=0$. Then, the joint probability
distribution is [normalized to our previous calculation; see Eq.~\doublesigma]
equal to $4A_1A_2C$.

Finally, we need to convert the results of Eq.~\combineddist\ into
a covariant form. Recall that ${\bold{\vec q_1}}$ is the momentum
in the $C_1$ rest frame and ${\bold{\vec q_2}}$
is the momentum in the $C_2$ rest frame.
If one defines $({\bold{\vec d_j}})_i\equiv D_{ij}$, then it is
easy to check that the covariant expression
which reduces to ${\bold{\vec q_1\cdot\vec d_j}}$ in
the frame where ${\bold{\vec p_1}}=0$ is
$$
-q_1\cdot d_j+ {q_1\cdot p_1d_j\cdot p_1\over m_1^2}=
q_1^\mu d_j^\nu \left(-g_{\mu\nu}+{p_{1\mu}p_{1\nu}\over m_1^2}\right)
\,,\eqn\covariant
$$
since if $p_1=(m_1\,;0)$, then
$-q_1\cdot d_j+(q_1)_0(p_1)_0\equiv {\bold{\vec q_1\cdot \vec d_j}}$
as desired.
Thus, Eq.~\combineddist\ becomes
$$
d\si\otimes d\Gamma_1\otimes d\Gamma_2\propto CA_1A_2+B_1B_2q_1^\mu
\left(-g_{\mu\nu}+{p_{1\mu}p_{1nu}\over m_1^2}\right) D^{\nu\rho}
\left(-g_{\rho\si}+ {p_{2\rho}p_{2\si}\over m_2^2}\right) q^\si_2\,,
\eqn\recover
$$
which is precisely the expression obtained in Eq.~\doublesigma\ in
the case of $B=C=0$, with appropriate identification of the
corresponding variables.

\endpage

\chapter{}

\def\wt{\widetilde}

\def\gev{{\rm GeV}}
\def\rta{\rightarrow}

\def\china{\widetilde\chi^0_1}
\def\chinb{\widetilde\chi^0_2}

\def\chini{\widetilde\chi^0_i}

\def\chipa{\widetilde\chi^+_1}
\def\chipb{\widetilde\chi^+_2}
\def\chima{\widetilde\chi^-_1}

\def\chipma{\widetilde\chi^\pm_1}

\def\chitil{\widetilde\chi}

\def\mchipa{M_{\widetilde\chi^+_1}}
\def\mchipma{M_{\widetilde\chi^\pm_1}}
\def\mchipb{M_{\widetilde\chi^+_2}}
\def\mchina{M_{\widetilde\chi^0_1}}
\def\mchinb{M_{\widetilde\chi^0_2}}
\def\mchinc{M_{\widetilde\chi^0_3}}
\def\mchind{M_{\widetilde\chi^0_4}}

\def\elp{\tilde e_L^+}
\def\elm{\tilde e_L^-}
\def\erp{\tilde e_R^+}
\def\erm{\tilde e_R^-}
\def\elmp{\tilde e_L^\mp}
\def\elpm{\tilde e_L^\pm}
\def\erpm{\tilde e_R^\pm}

\def\murpm{\tilde\mu_R^\pm}
\def\snu{\wt\nu}
\def\melpm{M_{\tilde e_L^\pm}}
\def\merpm{M_{\tilde e_R^\pm}}

\def\mmurpm{M_{\tilde\mu_R^\pm}}
\def\melllpm{M_{\tilde \ell_L^\pm}}
\def\mellrpm{M_{\tilde \ell_R^\pm}}
\def\msnu{M_{\wt\nu}}
\def\sigler{\sigma\ls{L}(\wt e\ls{R})}
\def\sigrer{\sigma\ls{R}(\wt e\ls{R})}
\def\siglmur{\sigma\ls{L}(\wt\mu\ls{R})}
\def\sigrmur{\sigma\ls{R}(\wt\mu\ls{R})}
\def\siglcpm{\sigma\ls{L}(\chipma)}
\def\sigrcpm{\sigma\ls{R}(\chipma)}
\def\siglelr{\sigma\ls{L}(\elm\erp)}
\def\sigrerl{\sigma\ls{R}(\erm\elp)}
\def\siglerl{\sigma\ls{L}(\erm\elp)}
\def\sigrelr{\sigma\ls{R}(\elm\erp)}
\def\hpm{H^\pm}

\def\hthree{A^0}
\def\mhpm{m\ls{H^\pm}}

\def\mhl{m\ls{h^0}}
\def\mha{m\ls{A^0}}
\def\mz{m\ls Z}
\def\mw{m\ls W}
\def\ifmath#1{\relax\ifmmode #1\else $#1$\fi}
\def\half{\ifmath{{\textstyle{1 \over 2}}}}
\def\quarter{\ifmath{{\textstyle{1 \over 4}}}}

\def\third{\ifmath{{\textstyle{1 \over 3}}}}
\def\twothirds{{\textstyle{2 \over 3}}}

\def\gp{g^{\prime}}
\def\mpl{M_{\rm PL}}

\title{\bf LECTURE {\seventeenbf 2}:       \break
Applications to Low-Energy Supersymmetry}

\section{\bf Raison-d'\^etre for new physics beyond the Standard Model}

\REF\LHC{G. Jarlskorg and D. Rein, editors, {\it Proceedings of the
ECFA Large Hadron
Collider Workshop}, Vols I--III, Aachen, Germany, 4--9 October 1990,
CERN Report 90-10 (1990).}
\REF\LCconfs{R. Orava, P. Eerola and M. Nordberg (editors),
{\it Physics and Experiments with Linear Collers}, Workshop
Proceedings, Saariselk\"a, Finland, 9--14 September, 1991
(World Scientific, Singapore, 1992);
F.A. Harris, S.L. Olsen, S. Pakvasa and X. Tata (editors),
{\it Physics and Experiments with Linear $\epem$ Colliders},
Workshop Proceedings, Waikoloa, Hawaii, 26--30 April, 1993
(World Scientific, Singapore, 1993).}

The search for the origin of electroweak symmetry breaking and new
physics beyond the Standard Model provides the central focus for
particle physics experiments envisioned at the next generation of
colliders.   With the recent demise of the SSC, the
only hadron supercollider now on the drawing boards is the LHC which
will be constructed at CERN.  LHC is a proton-proton supercollider
operating at a
CM-energy of $\sqrt{s}\simeq 14$~TeV, with an instantaneous
luminosity of 10$^{33}$~cm$^{-2}$~sec$^{-1}$, and an eventual
capability to reach luminosities above 10$^{34}$~cm$^{-2}$~sec$^{-1}$%
\refmark\LHC.
Meanwhile, the $e^+e^-$ physics community is vigorously engaged in the
development of an $e^+e^-$ linear collider with CM-energy of
$\sqrt{s}=500$~GeV and luminosity in excess of
10$^{33}$~cm$^{-2}$~sec$^{-1}$, with eventual upgrades to CM energies
of 1~TeV (and perhaps beyond) and luminosities above
10$^{34}$~cm$^{-2}$~sec$^{-1}$\refmark\LCconfs.
There is an active international
collaboration involved in the design of this collider, which has been
recently dubbed the International Linear Collider (ILC).  Its
proponents envision a formal proposal for constructing the ILC to be
ready later in this decade, with completion of construction sometime
in the following decade.

\REF\hhg{%
J.F. Gunion, H.E. Haber, G. Kane, and S. Dawson, {\it The
Higgs Hunter's Guide} (Addison-Wesley Publishing Company,
Redwood City, CA, 1990).}
\REF\technipapers{E. Farhi and L. Susskind, {\sl Phys. Rep.} {\bf 74}
(1981) 277; R.K. Kaul, {\sl Rev. Mod. Phys.} {\bf 55} (1983) 449.}
There are three fundamental goals of the LHC and ILC.  The first goal
is to discover the Higgs boson\refmark\hhg\
(assuming that it has not already been
discovered at LEP-II).  If the Standard Model Higgs boson (or the
Higgs bosons of an extended elementary Higgs sector) is not realized
in nature, then one expects to ascertain
the dynamics responsible for electroweak symmetry breaking.
If elementary Higgs scalars exist, then they are probably light (less than
200~GeV in mass) and weakly-coupled.  An alternative picture is one in
which the dynamics responsible for electroweak symmetry breaking
involves strong forces.  In such an approach, any Higgs-like scalar is
almost certainly composite, heavy (with mass on the order of 1~TeV)
and strongly coupled.  Technicolor is the standard paradigm for such
approaches\refmark\technipapers.

The second goal of the future supercolliders is to elucidate the structure of
the effective low-energy gauge group and associated matter multiplets.
If the Standard Model is the correct description of physics at the
electroweak scale, then the correct low-energy gauge group is
SU(3)$\times$SU(2)$\times$U(1), associated with three generations of
quarks, charged leptons and massless neutrinos.  However, one cannot be
certain at present that this is the whole story.  It is still possible
that physics at the 1~TeV scale contains:
\pointbegin
New gauge bosons beyond the $W^\pm$ and $Z$ (which would indicate that
the Standard Model gauge group must be enlarged).
\point
Massive neutrinos (very small masses for left-handed neutrinos and
large masses of order 1~TeV for right-handed neutrinos).
\point
New quark and lepton generations with the same quantum numbers as
the known generations.\foot{Precision electroweak measurements at LEP
do place some constraints.  Since the $Z$ width implies the existence
of exactly three light neutrinos, any fourth generation neutrino must
have mass greater than $m_Z/2$.  Second, precision measurement of the
$\rho$-parameter (where $\rho=\mw^2/\mz^2\cos^2\theta_W=1$ to better
than $1\%$) places strong constraints on the splittings between masses
of any new fourth generation weak doublet states.}
\point
\REF\roos{J. Maalampi and M. Roos, {\sl Phys. Rep.} {\bf 186} (1990) 53.}
Mirror fermions, whose left and right handed couplings are opposite
relative to those of the Standard Model fermions\refmark\roos.
\point
\REF\hewettrizzo{J.L. Hewett and T.G. Rizzo, {\sl Phys. Rep.} {\bf 183}
(1989) 193.}
Fermions with exotic quantum numbers (\eg, new vector-like $D$-quark
and $E$-lepton which could arise in an E$_6$ model of grand
unification)\refmark\hewettrizzo.

\noindent
The discovery of any one of these particles would dramatically alter
theoretical attempts to extend our understanding of physics beyond the
1~TeV scale.

\REF\gutproton{P. Langacker, {\sl Phys. Rep.} {\bf 72} (1981) 185;
G.G. Ross, \it Grand Unified Theories \rm (Addison-Wesley Publishing
Company, Reading, MA, 1984).}
\REF\suss{E. Gildener, {\sl Phys. Rev.} {\bf B14} (1976) 1667;
S. Weinberg, {\sl Phys. Lett.} {\bf 82B} (1979) 387.}
\REF\susskind{L.~Susskind, {\sl Phys. Rep.} {\bf 104} (1984) 181.}
The third goal of the future supercolliders is to search for new
physics beyond the Standard Model
associated with the dynamics of electroweak symmetry breaking.
Despite the great success of the Standard Model of particle physics,
most theorists strongly believe that the successes of the Standard
Model will not persist to higher energy scales.  This belief arises
from attempts to embed the Standard Model in a more fundamental
theory.  We know that the Standard Model cannot be the ultimate
theory, valid to arbitrarily high energy scales.  Even in the absence
of grand unification of strong and electroweak forces at a very high
energy scale\refmark\gutproton,
it is clear that the Standard Model must be
modified  to incorporate the effects of gravity at the Planck
scale [$\mpl=(c\hbar/G_N)^{1/2}\simeq 10^{19}$~GeV].\foot{The Planck scale
arises as follows.  The gravitational potential energy
of a particle of mass $M$, $G_NM^2/r$
(where $G_N$ is Newton's gravitational constant), evaluated
at a Compton
wavelength, $r=\hbar/Mc$, is of order the rest mass, $Mc^2$, when
$$G_NM^2\left({Mc\over\hbar}\right)\sim Mc^2\,,$$
which implies that $M^2\sim c\hbar/G_N$.  When this happens, the
gravitational energy is large enough to induce pair production, which
means that quantum gravitational effects can no longer be neglected.
Thus, the Planck scale, $\mpl=(c\hbar/G_N)^{1/2}$, represents the energy
scale at which gravity and all other forces of elementary particles
must be incorporated into the same theory.}
In this context, it is a mystery
why the ratio $\mw/\mpl\simeq 10^{-17}$ is so small.  This is called
the hierarchy problem\refmark{\suss,\susskind}.  Moreover,
in the Standard Model, the scale of
the electroweak interactions derives from an elementary scalar field
which acquires a vacuum expectation value of $v=2\mw/g=246$~GeV.
However, if one couples a theory of scalar particles to new physics at
some arbitrarily high scale $\Lambda$, radiative corrections to the
scalar squared-mass are of ${\cal O}(\Lambda^2)$, due to the quadratic
divergence in the scalar self-energy (which indicates quadratic
sensitivity to the largest energy scale in the theory).  Thus, the
``natural'' mass for any scalar particle is $\Lambda$ (which is
presumably equal to $\mpl$).  Of course,
in order to have a successful electroweak theory, the Higgs mass must
be of order the electroweak scale.  The fact that the Higgs mass
cannot be equal to its natural value of $\mpl$ is called the
``naturalness'' problem\refmark\thooft.


\REF\preons{I.A. D'Souza and C.S. Kalman,
{\it Preons} (World Scientific, Singapore, 1992).}
\REF\quaddiv{%
M. Veltman, {\sl Acta Phys. Pol.} {\bf B12} (1981) 437;
R.K. Kaul and P. Majumdar, {\sl Nucl. Phys.} {\bf B199} (1982) 36;
T. Inami, H. Nishino and S. Watamura, {\sl Phys. Lett.} {\bf 117B}
(1982) 197;
N.G. Deshpande, R.J. Johnson and E. Ma, {\sl Phys. Lett.} {\bf 130B}
(1983) 61; {\sl Phys. Rev.} {\bf D29} (1984) 2851.}
Theorists have been hard at work for more than a decade in an attempt
to circumvent the problems raised above.   The proposed solutions
involve removing the quadratic divergences from the theory that are
the root cause of the naturalness problem.  Two
classes of solutions have been proposed.  In one class, the elementary
scalars are removed altogether.  One then must add new fundamental
fermions and new fundamental forces.  For example, in technicolor
models, new fermions $F$ are introduced such that
$\VEV{F\overline F}\neq0$ due to technicolor forces,
which results in the breaking of the electroweak
interactions\refmark\technipapers.
Other models of this class are composite models, where
some (or all) of the
particles that we presently regard as fundamental are
bound states of more fundamental fermions\refmark\preons.
In this class of models,
the physics that is responsible for electroweak symmetry breaking is
strong and its implementation requires non-perturbative techniques.
I believe that it is fair to say that no compelling realistic model of
this type exists at present.  I will say no more about this
approach.  The second
class of models are those in which new particles are introduced to the
Standard Model in such a way that all quadratic divergences exactly
cancel.  Since one retains the Higgs scalars as elementary, the
cancellation of quadratic divergences can only be the result of a new
symmetry\refmark\quaddiv.
This symmetry is called supersymmetry which relates
fermions to bosons.  Because fermion self-energies have no quadratic
divergences, it is possible in a theory with a symmetry that relates
fermions to bosons to guarantee that
no quadratic divergences arise in scalar self-energies.

\REF\topten{H.E. Haber, {\it The Supersymmetric Top-Ten Lists},
SCIPP-93/22 (1993), to appear in the Proceedings of the Workshop on
Recent Advances in the Superworld, Houston Advanced Research Center,
April 14--16, 1993.}
In this lecture I will focus on low-energy supersymmetry as a
compelling model for physics beyond the Standard Model\refmark\topten.
Low-energy supersymmetric models contain a rich phenomenology of new
particles and interactions.  Moreover, polarization and spin
correlations
provide essential tools for disentangling the supersymmetric spectrum
and checking that the underlying interactions have a supersymmetric
origin.  I will first review the ingredients of low-energy
supersymmetry and define the minimal supersymmetric extension of the
Standard Model.  Then, I will discuss some theoretical biases on the
parameters of this model, which are based on additional assumptions
about physics at a very high energy scale (near the Planck scale).
Finally, I will survey how to exploit polarization and utilize spin
information to study supersymmetric phenomenology at future
supercolliders.

\section{\bf Introduction to low-energy supersymmetry}

\REF\nilles{%
H.P.~Nilles, {\sl Phys. Rep.} {\bf 110} (1984) 1;
A.B.~Lahanas and D.V.~Nanopoulos, {\sl Phys. Rep.} {\bf 145} (1987) 1.}
\REF\nath{%
P.~Nath, R.~Arnowitt, and A.~H.~Chamseddine, {\it Applied
$N=1$ Supergravity} (World Scientific, Singapore, 1984).}
\REF\green{%
M.B.~Green, J.S.~Schwarz, and E.~Witten, {\it Superstring
Theory} (Cambridge University Press, Cambridge, 1987).}
\REF\witten{%
E. Witten, {\sl Nucl. Phys.} {\bf B188} (1981) 513;
S. Dimopoulos and H. Georgi, {\sl Nucl. Phys.} {\bf B193} (1981)
150; N. Sakai, {\sl Z. Phys.} {\bf C11} (1981) 153;
R.K. Kaul, {\sl Phys. Lett.} {\bf 109B} (1982) 19.}
\REF\haberkane{%
H.E.~Haber and G.L.~Kane, {\sl Phys. Rep.} {\bf 117} (1985) 75.}
\REF\inoue{%
K.~Inoue, A.~Kakuto, H.~Komatsu, and S.~Takeshita,
{\sl Prog. Theor. Phys.} {\bf 68}, (1982) 927;
[E:~{\bf 70} (1983) 330]; {\bf 71} (1984) 413;
R.~Flores and M.~Sher, {\sl Ann. Phys. (NY)} {\bf 148} (1983) 95.}
\REF\ghi{%
J.F.~Gunion and H.E.~Haber, {\sl Nucl. Phys.} {\bf B272} (1986) 1
[E: {\bf B402} (1993) 567].}
\REF\girardello{%
L.~Girardello and M. Grisaru, {\sl Nucl. Phys.} {\bf B194} (1982) 65.}
\REF\casas{%
For a recent discussion, see
B.~de~Carlos and J.A. Casas, {\sl Phys. Lett.} {\bf B309} (1993) 320.}
\REF\bertolini{%
S.~Bertolini, F.~Borzumati, A.~Masiero, and G.~Ridolfi,
{\sl Nucl. Phys.} {\bf B353} (1991) 591.}
\REF\rparity{%
P.~Fayet, {\sl Phys. Lett.} {\bf 69B} (1977) 489;
G.~Farrar and P.~Fayet, {\sl Phys. Lett.} {\bf 76B} (1978) 575.}
\REF\lsp{%
J. Ellis, J.S. Hagelin, D.V. Nanopoulos, K. Olive, and
M. Srednicki, {\sl Nucl. Phys.} {\bf B238} (1984) 453.}
\REF\rbreak{%
See {\it e.g.}, S. Dimopoulos, R. Esmailzadeh, L.J. Hall, and
G.D. Starkman, {\sl Phys. Rev.} {\bf D41} (1990) 2099;
H. Dreiner and G.G. Ross, {\sl Nucl. Phys.} {\bf B365} (1991) 597.}
\REF\tasisusy{%
H.E.\ Haber, in {\it Recent Directions in Particle Theory},
Proceedings of the 1992 Theoretical Advanced Study Institute in
Particle Physics, edited by J.~Harvey and J.~Polchinski
(World Scientific, Singapore, 1993) pp.~589--686.}
\REF\fischler{%
W. Fischler, S. Paban, and S. Thomas, {\sl Phys. Lett.} {\bf B289}
(1992) 373; S.M. Barr, {\sl Int. J. Mod. Phys.} {\bf A8} (1993) 209.}
\REF\hempfling{%
H.E. Haber and R. Hempfling, {\sl Phys. Rev. Lett.} {\bf 66} (1991)
1815.}
\REF\okada{%
Y. Okada, M. Yamaguchi, and T. Yanagida, {\sl Prog. Theor.}
{\bf 85} (1991) 1; J. Ellis, G. Ridolfi, and F. Zwirner,
{\sl Phys. Lett.} {\bf B257} (1991) 83.}
\REF\seeappenda{See \eg, Appendix A of Ref.~[\ghi].}
\REF\decamp{%
See \eg, D. Decamp \etal\ [ALEPH Collaboration],
{\sl Phys. Rep.} {\bf 216} (1992) 253.}
\REF\rudaz{%
J.~Ellis and S.~Rudaz, {\sl Phys. Lett.} {\bf 128B} (1983) 248.}
\REF\masiero{%
F. Gabbiani and A. Masiero, {\sl Nucl. Phys.} {\bf B322} (1989) 235.}
\REF\dine{%
For a recent discussion, see M.~Dine, R.~Leigh,
and A.~Kagan, {\sl Phys. Rev.} {\bf D48} (1993) 4269.
For an alternate approach, see
Y.~Nir and N.~Seiberg, {\sl Phys. Lett.} {\bf B309} (1993) 337.}
\REF\sugra{%
Recent works include:
 G.G. Ross and R.G. Roberts, {\sl Nucl. Phys.} {\bf B377} (1992) 571;
 S. Kelley, J.L. Lopez, D.V. Nanopoulos, H. Pois,
and K. Yuan, {\sl Nucl. Phys.} {\bf B398} (1993) 3;
 M. Olechowski and S. Pokorski, {\sl Nucl. Phys.} {\bf B404} (1993)
590;  D.J. Casta\~no, E.J. Piard, and P. Ramond, UFIFT-HEP-93-18 (1993);
 V. Barger, M.S. Berger, and P. Ohmann, MAD/PH/801 (1993);
 G.L. Kane, C. Kolda, L. Roszkowski, and J.D. Wells, UM-TH-93-24 (1993).}
\REF\deboer{%
For recent reviews, see R. Arnowitt and P. Nath, CTP-TAMU-52/93 (1993);
W. de~Boer, IEKP-KA/94-01 (1994).}
\REF\susyguts{%
U. Amaldi, W. de~Boer, and H. Furstenau, {\sl Phys. Lett.} {\bf B260}
(1991) 447; U. Amaldi \etal, {\sl Phys. Lett.} {\bf B281} (1992) 374.}
\REF\arason{%
H. Arason \etal, {\sl Phys. Rev.} {\bf D46} (1992) 3945.}
\REF\moresusyguts{%
J. Ellis, S. Kelly, and D.V. Nanopoulos, {\sl Nucl. Phys.} {\bf B373}
(1992) 55; F. Anselmo, L. Cifarelli, A. Peterman, and A. Zichichi,
{\sl Nuovo Cim.} {\bf 105A} (1992) 1817; {\bf 105A} (1992) 581;
{\bf 105A} (1992) 1201;
P. Langacker and N. Polonsky, {\sl Phys. Rev.} {\bf D47} (1993) 4028;
M. Carena, S. Pokorski, and C.E.M. Wagner, {\sl Nucl. Phys.} {\bf B406}
(1993) 59;
V. Barger, M.S. Berger, and P. Ohmann, {\sl Phys. Rev.} {\bf D47}
(1993) 1093.}
\REF\topunified{%
See \eg,
 B. Anantharayan, G. Lazarides, and Q. Shafi, {\sl Phys. Rev.} {\bf D44}
(1991) 1613;  S. Dimopoulos, L.J. Hall, and S. Raby,
{\sl Phys. Rev.} {\bf D45} (1992) 4192;
 L.J. Hall, R. Rattazzi, and U. Sarid, LBL-33997 (1993).}

Super\-symmetry is an attractive theoretical framework that may permit
the consistent unification of particle physics and gravity,
which takes place at an
energy of order the Planck scale\refmark{\nilles-\green}.
However, super\-symmetry is clearly not an exact symmetry of nature,
and therefore must be broken.
In theories of ``low-energy'' super\-symmetry, the effective scale of
super\-symmetry breaking is tied to the electroweak
scale\refmark{\witten,\susskind}.  In this
way, it is hoped that super\-symmetry will ultimately explain the origin
of the large hierarchy between the $W$ and $Z$ masses and the
Planck scale.

The minimal supersymmetric extension of the Standard Model (MSSM)
consists of taking the Standard Model
and adding the corresponding supersymmetric partners\refmark\haberkane.
In addition, the MSSM contains two $Y=\pm 1$ Higgs doublets, which is
the minimal structure for the Higgs sector of
an anomaly-free supersymmetric extension
of the Standard Model that generates mass for both ``up''-type and
``down''-type quarks (and charged leptons)\refmark{\inoue,\ghi}.
Supersymmetric interactions
consistent with (global) $B$--$L$ conservation (where $B$ is
baryon number and $L$ is
lepton number) are included.  Finally, the most general
soft-super\-symmetry-breaking terms are added\refmark\girardello.
If super\-symmetry is relevant
for explaining the scale of electroweak interactions, then the
mass parameters that occur in the soft-super\-symmetry-breaking
terms must be of order 1~TeV or below\refmark\casas.
Some bounds on these parameters exist due to the absence
of super\-symmetry particle production at current accelerators, as well
as the absence of any evidence for virtual supersymmetric particle
exchange in a variety of Standard Model processes\refmark\bertolini.

As a consequence of $B$--$L$ invariance, the MSSM possesses a discrete
$R$-parity invariance, where
${R=(-1)^{3\,(B\hbox{--}L)+2S}}$
for a particle of spin $S$\refmark\rparity.
Note that this formula implies that all the ordinary Standard Model
particles have even $R$-parity, whereas the corresponding
supersymmetric partners have odd $R$-parity.
The conservation of $R$-parity in scattering
and decay processes has a crucial impact on supersymmetric
phenomenology.  For example, starting from an initial state involving
ordinary ($R$-even) particles, it follows that
supersymmetric particles must be
produced in pairs.  In general, these particles are highly unstable
and decay quickly into lighter states.  However, $R$-parity invariance
also implies that
the lightest supersymmetric particle (LSP) is absolutely
stable, and must eventually be produced at the end of a decay chain
of a heavy unstable supersymmetric particle.  In order to be
consistent with cosmological constraints, the LSP is almost certainly
electrically and color neutral\refmark\lsp.
Consequently, the LSP is
weakly-interacting in ordinary matter,
\ie, it behaves like a neutrino
and will escape detectors without being directly observed.
Thus, the canonical signature for ($R$-parity conserving)
supersymmetric theories is missing (transverse)
energy, due to the escape of the LSP.
Some model builders attempt to relax the
assumption of $R$-parity conservation\refmark\rbreak.
Models of this type must
break $B$--$L$ and are therefore strongly constrained, although
such models cannot be completely ruled out at present.
In $R$-parity violating models, the LSP would be unstable, and this
fact (among others) leads to a phenomenology
of broken-$R$-parity models that is very different from that of
the MSSM.

In the MSSM, supersymmetry breaking is induced by the
soft-supersymmetry breaking terms mentioned above.  These terms
parametrize our ignorance of the fundamental mechanism of
supersymmetry breaking.
The parameters of the MSSM are conveniently described by considering
separately the
super\-symmetry-conserving sector and
the super\-symmetry-breaking sector.
A careful discussion of the conventions used in defining the MSSM
parameters can be found in Ref.~[\tasisusy].
Among the parameters of the super\-symmetry conserving
sector are: (i)~gauge couplings: $g_s$, $g$, and $\gp$, corresponding
to the Standard Model gauge group SU(3)$\times$SU(2)$\times$U(1),
respectively; (ii)~Higgs Yukawa couplings: $\lambda_e$, $\lambda_u$, and
$\lambda_d$ (which are $3\times 3$ matrices in flavor space); and
(iii)~a super\-symmetry-conserving Higgs mass parameter $\mu$.  The
super\-symmetry-breaking sector contains the following set of parameters:
(i)~gaugino Majorana masses $M_3$, $M_2$ and $M_1$ associated with
the SU(3), SU(2), and U(1) subgroups of the Standard Model;
(ii)~scalar mass matrices for the squarks and sleptons;
(iii)~Higgs-squark-squark trilinear interaction terms (the so-called
``$A$-parameters'') and corresponding terms involving the sleptons; and
(iv)~three scalar Higgs mass parameters---two
diagonal and one off-diagonal mass terms for the two Higgs doublets.
Explicitly, the tree-level Higgs potential in the MSSM reads
$$\eqalign{%
V_{\rm Higgs} = \ &m^2_{1H} |H_1|^2 + m^2_{2H} |H_2|^2
             - m^2_{12} (H^0_1 H^0_2-H_1^-H_2^+ + {\rm h.c.})\crr
             &+\eighth(g^2+g'^2) \left(|H_1|^2-|H_2|^2\right)^2
             +\half g^2|H^*_1H_2|^2\,,\cr
}\eqn\mssmhiggspot$$
where $H_1$ and $H_2$ are weak SU(2) scalar doublets with hypercharges
$Y=-1$ and $+1$, respectively.\foot{The diagonal Higgs squared mass
parameters actually receive a contribution from the supersymmetry
conserving parameter $\mu$ as well.  That is,
$m^2_{iH}\equiv|\mu|^2+m^2_i\quad (i=1,2)$, where
$m^2_i$ ($i=1,2$) and $m_{12}^2$ are parameters of the supersymmetry
breaking sector.  Conventionally, one writes $m_{12}^2\equiv
B\mu$, which defines the parameter $B$.}
These mass parameters $m^2_{iH}$ ($i=1,2$) and $m^2_{12}$
can be re-expressed in terms of the two
Higgs vacuum expectation values, $v_i\equiv\VEV{H_i^0}$, ($i=1,2$),
and one physical Higgs mass.  Here, $v_1$ ($v_2$) is the vacuum
expectation value of the Higgs field that couples exclusively
to down-type (up-type) quarks and leptons.  Note that $v_1^2+v_2^2=
(246~{\rm GeV})^2$ is fixed by the $W$ mass, while the ratio
$$
\tan \beta = {v_2\over v_1}\eqn\tanbeta
$$
is a free parameter of the model.\foot{In the MSSM,
it is conventional to choose the
phases of the Higgs fields such that $v_1$ and $v_2$ are real and
positive.  Moreover, one can check that the
Higgs sector is automatically CP-conserving (at
tree-level).  Thus, the physical neutral Higgs states are
CP-eigenstates.  The parameter $m_{12}^2$ in Eq.~\mssmhiggspot\ is
directly related to the mass of the CP-odd neutral state, $\ha$, via
$\mha^2=m_{12}^2/\sin\beta\cos\beta$.  Thus, $m_{12}^2$ is real and
positive in the standard convention where $\tan\beta$ is positive.}

The MSSM contains a number of possible new sources of CP violation.
For example, gaugino mass parameters, the $A$-parameters,
and $\mu$ may be
complex.
Some combination of these complex phases
must be less than of order
$10^{-2}$--$10^{-3}$
(for a supersymmetry-breaking scale of 100 GeV) to avoid generating
electric dipole moments for the neutron, electron, and atoms
in conflict with observed data\refmark\fischler.
However, these complex phases have little impact on the direct searches
for supersymmetric particles.  Nevertheless, if supersymmetric
particles are discovered, it will be challenging to attempt to measure
the CP-violating phases by precision measurements of supersymmetric
couplings.

Before describing the supersymmetric particle sector, let us
consider the Higgs sector of the MSSM\refmark\hhg.
There are five physical Higgs particles in this model: a charged Higgs
pair ($\hpm$), two CP-even neutral Higgs bosons (denoted by $h^0$
and $H^0$ where $\mhl\leq\mhh$) and one CP-odd neutral
Higgs boson ($\hthree$).  The properties of the Higgs sector are
determined by the Higgs potential.
The strengths of the Higgs self-interaction terms
are directly related to the gauge couplings by supersymmetry (and are
not affected at tree-level by supersymmetry-breaking).  As a result,
$\tan\beta$ [defined in Eq.~\tanbeta]
and one Higgs mass determine: the Higgs spectrum,
an angle
$\alpha$ [which indicates the amount of
mixing of the original $Y=\pm 1$ Higgs doublet
states in the physical CP-even
scalars], and
the Higgs boson couplings.  When one-loop radiative corrections
are incorporated, additional parameters of the supersymmetric model
enter via virtual loops.  The impact of these corrections
can be significant\refmark{\hempfling,\okada}.
For example, at tree-level, the MSSM predicts
$\mhl\leq m_Z$\refmark{\inoue,\ghi}.  If true,
this would imply that experiments to be performed at
LEP-II operating at its
maximum energy and luminosity would
rule out the MSSM if $h^0$ were not found.
However, this Higgs mass bound need not be respected when
radiative corrections are incorporated.
For example, in Ref.~[\hempfling],
the following upper bound was obtained for
$\mhl$ (assuming $\mha>\mz$)
in the limit of $\mz\ll m_t\ll M_{\widetilde t}$ [where
top-squark ($\widetilde t_L$--$\widetilde t_R$) mixing is neglected]
$$
\mhl^2\lsim\mz^2+ {3g^2\mz^4\over 16
\pi^2\mw^2}
\Biggl\{\ln\left({M_{\widetilde t}^2\over m_t^2}\right)\!\left[
{2m_t^4 -m_t^2\mz^2\over\mz^4}\right]
+{m_t^2\over 3 m_Z^2}\Biggr\}.
\eqn\hlmasslimit
$$
For a top-squark mass of $M_{\widetilde t}=1$~TeV, Eq.~\hlmasslimit\
yields a positive mass shift for $\mhl$
of about 20 GeV for $m_t=150$~GeV, and 40 GeV for $m_t=180$~GeV.
Even if $\tan\beta=1$ (so that $\mhl=0$ at tree-level), there is
a large positive shift in $\mhl^2$
due to radiative corrections of similar size.

Consider next the
supersymmetric particle sector of the MSSM.
The supersymmetric partners of the gauge and Higgs bosons are fermions,
whose names are obtained by appending ``ino'' at the end of the
corresponding Standard Model particle name.  The {\it gluino} is the
color octet Majorana fermion partner of the gluon
with mass $M_{\widetilde g}=|M_3|$.
The supersymmetric partners of the electroweak gauge
and Higgs bosons (the {\it gauginos} and {\it Higgsinos})
can mix.  As a result,
the physical mass eigenstates are model-dependent linear combinations
of these states, called {\it charginos} and {\it neutralinos}, which
are obtained by diagonalizing the corresponding mass matrices.

The chargino mass matrix depends on $M_2$, $\mu$, $\tan\beta$ and
$m_W$.  In the $\wt W^+$--$\wt H^+$ basis, the
chargino mass matrix is\refmark\seeappenda\
$$
  X = \pmatrix{ M_2 &\sqrt 2 m\ls W \sin\beta \cr
       \sqrt 2 m\ls W \cos\beta &\mu \cr}\,.\eqn\charginomatrix
$$
In general, two unitary $2\times 2$ matrices $U$ and $V$ are required
to diagonalize the chargino mass-squared matrix
$$
{\cal M}_{\wt\chi^+}^2=VX^\dagger XV^{-1}=U^\ast XX^\dagger
(U^\ast)^{-1}\,.\eqn\chiplusdiag
$$
The two mass eigenstates are denoted by $\widetilde\chi^+_1$
and $\widetilde\chi^+_2$ with corresponding squared masses
$$
\eqalign{%
M^2_{\chipa,\chipb}=&
\half\left[|\mu|^2+|M_2|^2+2m_W^2\right]
\mp\Biggl\{\left(|\mu|^2+|M_2|^2+2m_W^2\right)^2\cr
&\qquad -4|\mu|^2|M_2|^2
-4m_W^4\sin^2 2\beta
+8m_W^2\sin 2\beta\,{\rm Re}(\mu M_2)
\Biggr\}^{1/2}\,,\cr}
\eqn\chimasses
$$
where the states are ordered such that $\mchipa \leq \mchipb$.
If CP-violating effects are ignored (in which case, $M_2$ and
$\mu$ are real parameters), then one can choose a convention where
$\tan\beta$ and $M_2$ are positive.
(Note that the relative sign of $M_2$ and $\mu$ is meaningful.
The sign of $\mu$ is convention-dependent; the
reader is warned that both sign conventions
appear in the literature.)
The sign convention for $\mu$ fixed by Eq.~\charginomatrix\
is used by the LEP collaborations\refmark\decamp\
in their plots of exclusion contours in the $M_2$ {\it vs.}~$\mu$
plane derived from the non-observation of $Z\rightarrow\chipa\chima$.
The mixing matrix
elements $U_{ij}$ and $V_{ij}$ [see Eq.~\chiplusdiag]
will appear in the chargino Feynman rules.
If CP is conserved, the $U$ and $V$ can be chosen to be
orthogonal matrices.

The neutralino mass matrix depends on $M_1$, $M_2$, $\mu$,
$\tan\beta$, $m_Z$, and the weak mixing angle $\theta_W$.
In the $\wt B$--$\wt W$--$\wt H^0_1$--$\wt H^0_2$ basis,\foot{$\wt
B$ and $\wt W^3$ are superpartners of the U(1)-hypercharge and
SU(2)-weak neutral gauge bosons.}
the neutralino Majorana mass matrix is\refmark\seeappenda\
$$
  Y = \pmatrix{%
    M_1 &0  &-m_Zc_\beta s_W &m_Zs_\beta s_W \cr
    0   &M_2 &m_Z c_\beta c\ls W &-m_Z s_\beta c\ls W \cr
    -m_Z c_\beta s\ls W &m_Z c_\beta c\ls W &0 &-\mu \cr
    m_Z s_\beta s\ls W &-m_Z s_\beta c\ls W &-\mu &0 \cr }\,,
\eqn\neutralinomatrix
$$
where $s_\beta = \sin\beta$, $c_\beta=\cos\beta$, \etc\ 
The $4\times 4$
unitary matrix $Z$  diagonalizes the neutralino mass
matrix
$$
{\cal M}_{\wt\chi^0}=Z^\ast YZ^{-1}\,,\eqn\chizerodiag
$$
where the diagonal elements of ${\cal M}_{\wt\chi^0}$ can be either
positive or negative.
The four mass eigenstates are denoted by $\widetilde\chi^0_i$
($i=1,2,3,4$), with corresponding mass eigenvalues $\eta_i
M_{\wt\chi^0_i}$.
The physical neutralino masses are defined to be
positive, with $\mchina\leq\mchinb\leq\mchinc\leq\mchind$.
The sign of the mass eigenvalue ($\eta_i=\pm1$) is physically
relevant and
corresponds to the CP quantum number of the Majorana neutralino state.
The mixing matrix elements $Z_{ij}$ will appear in the neutralino
Feynman rules.  If CP is conserved, then $Z$ can be chosen to be an
orthogonal matrix.

It is common practice in the literature to reduce the
supersymmetric parameter freedom
by requiring that all three gaugino mass parameters
are equal at some grand unification scale.  Then,
at the electroweak scale, the gaugino mass parameters can be expressed in
terms of one of them (say, $M_2$). The other
two gaugino mass parameters are given by
$$
M_3=(g_s^2/g^2)M_2\qquad
M_1=(5g^{\prime\,2}/3g^2)M_2\,.
\eqn\eqmass
$$
Having made this assumption, the chargino and neutralino masses and
mixing angles depend only on three unknown parameters: the gluino mass,
$\mu$, and $\tan\beta$.

The supersymmetric partners of the quarks and leptons are spin-zero
bosons:  the {\it squarks}, charged {\it sleptons}, and {\it sneutrinos}.
For a given fermion
$f$, there are two supersymmetric partners $\widetilde
f_L$ and $\widetilde f_R$ which are scalar partners of the
corresponding left and
right-handed fermion.  (There is no $\widetilde\nu_R$.)  However, in
general,
$\widetilde f_L$ and $\widetilde f_R$ are not mass-eigenstates since there
is $\widetilde f_L$-$\widetilde f_R$ mixing which is proportional in
strength to the corresponding element of the scalar
mass-squared-matrix\refmark\rudaz
$$
M_{LR}^2=\cases{m_d(A_d-\mu\tan\beta),&for ``down''-type $f$,\cr
       m_u(A_u-\mu\cot\beta),&for ``up''-type $f$,\cr}
\eqn\lrmass
$$
where $m_d$ ($m_u$) is the mass of the appropriate
``down'' (``up'') type quark or lepton.
Here, $A_d$ and $A_u$ are (unknown)
soft-super\-symmetry-breaking $A$--parameters and $\mu$ and $\tan\beta$
have been defined earlier.
The signs of the $A$~parameters are also
convention dependent; see Ref.~[\tasisusy].
Due to the appearance
of the {\it fermion} mass in Eq.~\lrmass,
 one expects $M_{LR}$ to be small
compared to the diagonal squark and slepton masses, with the possible
exception of the top-squark, since $m_t$ is large,
and the bottom squark if $\tan\beta\gg1$.
The (diagonal) $L$ and $R$-type
squark and slepton masses are given by\refmark\nath.
$$
\eqalign{%
  M^2_{\widetilde u_L}
        &= M^2_{\wt Q}+m_u^2+m_Z^2\cos 2\beta(\half-
\twothirds \sin^2\theta_W)\,,\cr
  M^2_{\widetilde u_R}
        &=M^2_{\wt U}+m_u^2+\twothirds m_Z^2\cos 2\beta
\sin^2\theta_W\,,\cr
  M^2_{\widetilde d_L}
        &=M^2_{\wt Q}+m_d^2-m_Z^2\cos 2\beta(\half-\third
\sin^2\theta_W)\,,\cr
   M^2_{\widetilde d_R}
        &=M^2_{\wt D}+m_d^2-\third m_Z^2\cos 2\beta
\sin^2\theta_W\,,\cr
   M^2_{\widetilde \nu}\;
        &=M^2_{\wt L}+\half m_Z^2\cos 2\beta\,,\cr
   M^2_{\widetilde e_L}
        &=M^2_{\wt L}+m_e^2-m_Z^2\cos 2\beta(\half-
\sin^2\theta_W)\,,\cr
   M^2_{\widetilde e_R}
        &=M^2_{\wt E}+m_e^2-m_Z^2\cos 2\beta
        \sin^2\theta_W\,.\cr}
\eqn\diagonalmasses
$$
The soft-super\-symmetry-breaking parameters: $M_{\wt Q}$,
$M_{\wt U}$, $M_{\wt D}$, $M_{\wt L}$, and $M_{\wt E}$ are
unknown parameters.  In the equations above, the notation of first
generation fermions has been used and generational indices have been
suppressed.  Further
complications such as intergenerational mixing are possible, although
there are some constraints from the nonobservation of flavor-changing
neutral currents (FCNC)\refmark\masiero.

One way to guarantee the absence of significant FCNC's mediated
by virtual supersymmetric particle exchange is to posit that the
diagonal soft-supersymmetry-breaking scalar mass matrices are
proportional to the unit matrix (in flavor space) at some energy
scale (normally taken to be the Planck scale)\refmark\dine.
Renormalization
group evolution is used to determine the
low-energy values for the scalar
mass parameters listed above.  This assumption substantially reduces
the MSSM parameter freedom.  For example, supersymmetric grand unified
models with universal scalar masses at the Planck scale typically
give\refmark\sugra\
${M_{\wt L} \approx M_{\wt E} < M_{\wt Q} \approx}$
${M_{\wt U}
\approx M_{\wt D}}$
with the squark masses somewhere between a factor of 1--3 larger
than the slepton masses (neglecting generational
distinctions).  More specifically,
the first two generations are thought to be nearly degenerate in
mass, while
$M_{\wt Q_3}$ and $M_{\wt U_3}$ are typically reduced by a factor of 1--3
from the other soft-super\-symmetry-breaking scalar masses because
of renormalization effects due to the heavy top quark mass.
As a result, four flavors of squarks
(with two squark eigenstates per flavor)
 and $\wt b_R$ will be nearly mass-degenerate and somewhat heavier than
six flavors of approximately mass-degenerate
sleptons (with two per flavor for the charged
sleptons and one per flavor for the sneutrinos).  On the other hand,
the $\wt b_L$ mass and the diagonal $\widetilde t_L$ and $\widetilde t_R$
masses are reduced compared to the common squark mass of the
first two generations.  In addition, third generation squark masses are
sensitive to the strength of the respective
$\wt q\ls L$--$\wt q\ls R$ mixing as discussed below Eq.~\lrmass.

Two additional theoretical frameworks are often introduced to
further reduce the MSSM parameter
freedom\refmark{\nilles,\nath,\deboer}.
The first is that of grand unified
theories (GUTs) and the desert hypothesis (\ie, no new physics
between the TeV-scale and the GUT-scale).  In the absence of low-energy
supersymmetry, the simplest models of this type fail because the three
SU(3)$\times$SU(2)$\times$U(1) gauge couplings fail to unify at a common
scale\refmark{\susyguts,\arason}.
Remarkably, in the case of the MSSM (with a
supersymmetry-breaking scale of order 1~TeV or below), the three gauge
couplings do unify at a common energy scale of order $10^{16}$~GeV
(with only very mild assumptions about the
GUT-scale theory)\refmark{\susyguts,\moresusyguts}.
Unification constraints on the Higgs-fermion Yukawa couplings may
also exist but are more GUT-model dependent\refmark\topunified.
The second theoretical
framework is that of minimal supergravity theory, which can impose
nontrivial constraints on the soft-supersymmetry breaking parameters.
Referring to the parameter list given above Eq.~\mssmhiggspot, the
Planck-scale values of the soft-supersymmetry-breaking parameters in
the simplest supergravity models take the following form:
(i) a universal gaugino mass $m_{1/2}$ [assuming
grand unification; Eq.~\eqmass\ is a consequence
of this assumption]; (ii) a universal diagonal scalar mass parameter
$m_0$ [whose consequences were described in the preceding paragraph];
(iii) a universal $A$-parameter, $A_0$; and (iv) three scalar Higgs
mass parameters [{\it cf}. Eq.~\mssmhiggspot]---two common diagonal
squared-masses given by $|\mu_0|^2+m_0^2$
and an off-diagonal squared-mass given by $B_0\mu_0$
(which defines the Planck-scale supersymmetry-breaking
parameter $B_0$), where $\mu_0$ is
the Planck-scale value of the $\mu$-parameter.  As before,
renormalization group evolution is used to compute the low-energy values
of the supersymmetry-breaking parameters, and determines the
supersymmetric particle spectrum.  Moreover, in this approach,
electroweak symmetry breaking is induced radiatively if one of the
Higgs diagonal squared-masses is forced negative by the evolution.
This occurs in models with a large Higgs-top quark Yukawa
coupling ({\it i.e.}, large $m_t$).  As a result, the two Higgs
vacuum expectation values (or equivalently, $m_Z$ and $\tan\beta$)
can be expressed as a function of the Planck-scale supergravity
parameters.  The simplest procedure\refmark\sugra\
is to remove $\mu_0$ and $B_0$
in favor of $m_Z$ and $\tan\beta$ (the
sign of $\mu_0$ is not fixed in this
process). In this case, the MSSM spectrum
and its interactions are determined by $m_0$, $A_0$, $m_{1/2}$,
$\tan\beta$, and the
sign of $\mu_0$  (in addition to the parameters of the Standard Model).
Combining both grand unification and the minimal supergravity
approach yield the most constrained version of the MSSM.

\section{Polarization and spin analysis as tools for supersymmetry
searches}

\REF\sqgl{P.R. Harrison and C.H. Llewellyn Smith, {\sl Nucl.\
Phys.} {\bf B213} (1983) 223 [E: {\bf B223} (1983 542];
S. Dawson, E. Eichten and C. Quigg, {\sl Phys.\ Rev.} {\bf D31}
(11985) 1581.}
First, let us briefly
consider supersymmetry searches at hadron colliders.
The supersymmetric particles with the largest production
cross sections are the squarks and gluinos\refmark\sqgl.
These particles have
non-trivial color quantum numbers and are produced in
gluon-gluon and quark-antiquark collisions.  (Gluinos and squarks can
also be produced in association via gluon-quark and gluon-antiquark
collisions.)
Since gluinos are color octets, their production cross-section
is larger than that of the color-triplet squarks.  However,
because there are twelve squark types (six flavors, with two mass
eigenstates per flavor), with at least eight types rather close in
mass (as discussed in the previous section), the total cross section
for the production of squarks of all types is competitive with the
gluino cross sections.

One can also directly produce sleptons, neutralinos and charginos at hadron
colliders via the Drell-Yan mechanism (virtual $s$-channel gauge
boson exchange in $q\bar q$ annihilation).  However, these processes
are mediated by the electroweak interactions, so their cross sections
are substantially smaller than those of squarks and gluinos.  On the
plus side, the sleptons, neutralinos and charginos states are expected
to be lighter than the squarks and gluinos.  Nevertheless, gluinos and
squarks remain the most likely supersymmetric candidates for discovery
at the LHC.

\REF\xtata{X. Tata, in {\it The Standard Model and Beyond}, Proceedings
of the 9th Symposium in Theoretical Physics, Mt. Sorak, Korea,
20--25 August 1990, edited by J.E. Kim (World Scientific, Singapore,
1991) pp.~304--378.}
\REF\chjr{N.S. Craigie, K. Hidaka, M. Jacob and F.M. Renard,
{\sl Phys.\ Rep.} {\bf 99} (1983) 69.}
\REF\bsrt{C. Bourrely, J. Soffer, R.M. Renard and P. Taxil,
{\sl Phys.\ Rep.} {\bf 177} (1989) 319.}
The phenomenology of squarks and gluinos at hadron colliders is well
known\refmark\xtata.
Can one exploit polarization and spin analysis to untangle
such signals if they are discovered?
For a review of spin effects at supercollider energies, see
Refs.~[\chjr] and [\bsrt].
Here I will quote one attempt to answer this question
if polarized beams were available at a future hadron collider.
Suppose the initial proton beams have helicity $\lambda$ and
$\lambda^\prime$, respectively.  Consider the process
$p(\lambda)+p(\lambda^\prime)\rta~{\rm jet}+X$, and denote some
differential cross section for this process by
$d\sigma_{\lambda\lambda^\prime}$.  Then, one can define the
double helicity asymmetry
$$A_{LL}={d\sigma_{++}-d\sigma_{+-}\over d\sigma_{++}+d\sigma_{+-}}\,.
\eqn\asymmetryll
$$
In the parton model, this asymmetry is obtained from
$$A_{LL}d\sigma=\sum_{ij}\,{1\over1+\delta_{ij}}\,\int\,dx_a dx_b
\left[\Delta f_i(x_a)\Delta f_j(x_b)+(i\leftrightarrow j)\right]
\hat a\ls{LL}^{ij}d\hat\sigma_{ij}\,,
\eqn\partonall
$$
where $\Delta f\equiv f_+ - f_-$, and $f_{2\lambda}$ is the parton
distribution in a polarized proton of helicity $\lambda$.  The sum
over $i$ and $j$ in Eq.~\partonall\ is taken over all possible
elementary scattering processes $i+j\rightarrow k+\ell$, in which the
observed jet originates from one of the final state partons.

\REF\kandlchandr{N.S. Craigie, K. Hidaka and P. Ratcliffe,
{\sl Phys.\ Lett.} {\bf 129B} (1983) 310.}
Note that $A_{LL}$ is generally nonzero, even in parity conserving
interactions.  (In contrast,
single helicity asymmetries are nonzero only in parity
non-conserving processes.)  One can work out expressions for the
elementary partonic cross sections ($d\hat\sigma_{ij}$) and the
partonic double helicity asymmetries ($\hat a\ls{LL}^{ij}$), and
derive predictions for $A_{LL}$ based on the partonic subprocesses of
QCD in the Standard Model.  Predictions for $A_{LL}$ (at zero
rapidity) as a function of transverse momentum tend to be small and
positive, of order a few percent\refmark\kandlchandr.
Consider now the contributions to
$A_{LL}$ from supersymmetric particle production.  Due to helicity
conservation, in the limit of zero mass squarks and gluinos, the
production of $\wt q$ and $\wt g$ has the property that $\hat
a\ls{LL}=-1$.  This result is diluted somewhat when the squark and
gluino masses are taken into account.  Nevertheless, it continues to
be true that jet events of a supersymmetric origin would have
$A_{LL}<0$.  Whether this is an observable effect under realistic
experimental conditions remains to be proven.

For the rest of this lecture, I will consider the search for
supersymmetry at $e^+e^-$ colliders.  In particular, I will exhibit
some of the power of the ILC for disentangling the supersymmetric
particle interactions and testing model assumptions in the MSSM.  The
advantages of an $e^+e^-$ collider for a detailed study of the
supersymmetric spectrum (over the corresponding search for
supersymmetry at a hadron supercollider) are:

\pointbegin
In an
$e^+e^-$ collider, Standard Model backgrounds to new physics signals
tend to be of the same order of magnitude in cross section as the
signals themselves.   This is true because all tree-level
cross sections of processes produced at $e^+e^-$ colliders are
electroweak in strength, so all two-body tree-level cross sections
are roughly a unit of $R$ (where one unit of $R$ is equal to the
cross section for $e^+e^-\rta\gamma^\ast\rta\mu^+\mu^-$).

\point
The beam energy constraint can be used (assuming that beamstrahlung
effects are negligible).  That is, the CM-energy of the final state is
known to be equal to the CM-energy of the $e^+e^-$ collider.  In
contrast, at hadron colliders, the CM-energy of the final state
partons is generally not known unless it can be directly measured.

\point
$W^\pm$ and $Z$ bosons can be detected in their two-jet decay modes.
This is very difficult in hadron colliders, where QCD backgrounds
are severe.

\point
Polarized beams can provide a powerful tool for studying new physics
and rejecting Standard Model background.  The SLC has demonstrated the
feasibility of polarized beams at a linear $e^+e^-$ collider.
Although it is possible in principle to polarize the beams at a hadron
collider, the interest in doing so has been limited. (Perhaps these
lectures will encourage more study of the feasibility
of polarized beams at future hadron supercolliders and their
potential in detecting and elucidating new physics.)

\point
Complex final states are more easily managed at $e^+e^-$ colliders due
to the relative cleanliness of the environment (\eg, smaller
multiplicities, less gluon radiation, \etc).

\point
The production rates for uncolored and colored particles with
electroweak quantum numbers are similar (of order one unit of $R$).

The main disadvantages of $e^+e^-$ linear colliders as compared
to hadron supercolliders are:

\pointbegin
The CM-energy of future hadron supercolliders are significantly larger
than any $e^+e^-$ linear collider that will exist during the same era.
Of course, at a hadron collider, the new physics reach is determined
by the CM-energy of the relevant parton-parton interaction, which is
of order 10\% of the $pp$ CM-energy.  Even so, the physics
reach of the LHC is substantially larger than that of a 500~GeV ILC.
Since the goal of the next generation of colliders is to uncover the
mechanism of electroweak symmetry breaking, the LHC is therefore
essential for maximizing the probability for the discovery of new
physics at the TeV energy scale.

\point
It is very difficult to directly produce colored particles that are
singlets under the electroweak gauge group (such as the gluino) at an
$e^+e^-$ collider.\foot{Gluinos can be more easily studied
at $e^+e^-$ colliders
if they are lighter than squarks, assuming that the production of
squark pairs is kinematically allowed.  In this case, the dominant
squark decay, $\wt q\rta q\wt g$, provides the gluino source.}
In contrast, such particles are produced with
significant cross sections at hadron colliders (via gluon-gluon
scattering), and under most circumstances are easily observed.

In summary, hadron colliders and $e^+e^-$ colliders are complementary.
The LHC is likely to be the discovery machine for new physics beyond
the Standard Model.  Sorting out the nature of the new physics will
primarily be the job of the ILC.

\REF\jlc{T. Tsukamoto, K. Fujii, H. Murayama, M. Yamaguchi, and Y. Okada,
KEK Preprint 93-146 (1993).}
Polarized beams at the ILC provide an effective tool in studies of
supersymmetric particle production.  A comprehensive analysis
by Tsukamoto \etal\refmark\jlc\ demonstrates that one can make precision
measurements of the MSSM parameters and test various theories for
these parameters at the ILC.  I will briefly describe some of their
work here; for a more detailed
description of their methods and strategies, see Ref.~[\jlc].

First, based on the theoretical remarks at the end of section 2.2,
one expects that the lightest states of the MSSM are the sleptons,
charginos and neutralinos.  The lightest supersymmetric particle is
assumed to be $\china$.  However, since $\china$ behaves like a
neutrino in particle detectors, one cannot detect $e^+e^-\rta
\china\china$.  Consider next $e^+e^-\rta\china\chinb$, where
$\chinb\rta\china+f\bar f$ (and $f$ can be either a quark or lepton).
Note that the $\china\chinb$ final state is produced via $t$-channel
selectron-exchange or via $s$-channel $Z$-exchange.  There is no
$s$-channel virtual photon exchange; as a result, the cross section for
this process tends to be less than that for charged supersymmetric
particle production.  Thus, one should first focus on charged
slepton production $e^+e^-\rta \wt\ell^+\ls{L,R}\wt\ell^-\ls{L,R}
\; (\ell=e,\mu,\tau)$,
which is mediated at tree-level by
\medskip
\vbox{%
\centerline{\psfig{file=slep.ps,height=2.75cm}}
}
\medskip
\noindent
and on pair-production of the lightest charginos
$e^+e^-\rta\chipa\chima$
which is mediated at tree-level by the following graphs.
\medskip
\vbox{%
\centerline{\psfig{file=charg.ps,height=2.75cm}}
}
\medskip

I assume that the initial electron beam is polarized.  There are two
main advantages of employing a polarized electron beam.  First, it
provides another handle for determining the supersymmetric parameters.
For example, by controlling the electron polarization, one can affect
the relative strengths of the competing Feynman diagrams depicted
above. That permits the isolation of neutralino couplings in slepton
production and sneutrino couplings in chargino production.  In
particular, note that $\elp\erm$ (or $\elm\erp$) production occurs
only via $\chini$-exchange, since the $\gamma$ and $Z$ couplings to
slepton pairs is diagonal at tree-level.  Furthermore,  only the $\wt
B$ component of the $\china$ contributes to $\wt e\ls{R}^\pm$ production;
clearly this requires an incoming $e\ls{R}^-$ beam.  In $\chipa\chima$
production, $\snu$ exchange is absent in the case of an incoming
$e\ls{R}^-$ beam, since the $\snu$ is the superpartner of the neutrino
which possesses only left-handed couplings.  Moreover, the
$e\ls{R}^-$ beam enhances the higgsino components of the produced
$\chipma$.  Second, controlling the polarization enhances the ability
to separate Standard Model backgrounds from the above signals.  For
example, one of the main Standard Model backgrounds to the processes
considered above is $e^+e^-\rta W^+W^-$.  This background can be
significantly reduced by employing an $e\ls{R}^-$ beam, because
(i) $W^+W^-$ production via $s$-channel exchange
is suppressed and (ii) the
$t$-channel neutrino exchange contribution to $W^+W^-$ production
(which is particularly large in the forward direction due to the
exchange of a massless particle) is completely absent.

\FIG\jlcii{}
\FIG\jlciii{}
\pageinsert
   \tenpoint \baselineskip=12pt   \narrower
\centerline{\psfig{file=ssi02.ps,height=5.5cm,angle=90}\quad}
\vskip12pt\noindent
{\bf Fig.~\jlcii.}\enskip
Total cross sections for slepton production:
(a) $\epem\to \tilde e^+_R\tilde e^-_R,
\;  \epem\to \tilde e^\pm_L\tilde e^\mp_R,$
and $\epem\to\tilde e^+_L \tilde e^-_L$,
(b) $\epem\to \tilde\mu^+_R\tilde\mu^-_R$ and $\epem\to
\tilde\mu^+_L\tilde\mu^-_L$, where dashed, solid, and
dotted lines correspond to electron beam polarizations of $-1$, 0,
and $+1$, respectively.  The cross sections were evaluated
at the lowest order, without including initial state radiation
nor beam effects. Taken from Ref.~[\jlc].
\vskip4pc
\centerline{\psfig{file=ssi03.ps,height=6.0cm,angle=90}}
\vskip12pt\noindent
{\bf Fig.~\jlciii.}\enskip
Acoplanarity angle distributions for final-state leptons from
right-handed slepton-pair productions (solid) at $\sqrt s = 350$ GeV
with an integrated luminosity of 20 fb$^{-1}$ after including
the initial state radiation and the beamstrahlung effects:
(a) $\epem\to\tilde e^+_R\tilde e^-_R$ with $P_{\em}=0$,
(b) $\epem\to\tilde\mu^+_R\tilde\mu^-_R$ with $P_{\em}=0$, and
(c) $\bar\mu^+_R\bar\mu^-_R$ with $P_{\em} = +0.95$.  The
dashed lines indicate the $W^+W^-$ background, while the
dotted line represents that from the $e^\pm\skew2{\barparen\nu_{\hskip-3pt e}}{W^\mp}$ process. Taken from Ref.~[\jlc].
\vfill
\endinsert

Tsukamoto \etal\refmark\jlc\ describe the following strategy for
the supersymmetric particle search at the ILC using
a polarized electron beam.  Suppose that the supersymmetric
particle spectrum satisfies
$\mchina<\mellrpm<\mchinb,\mchipa<\melllpm,\msnu$.
For example, the cross sections for slepton pair production for a
representative set of parameters for three
choices of electron beam polarization is shown in Fig.~\jlcii.
Then, in the first stage of the supersymmetric particle search,
$\erpm$ and $\murpm$ will be discovered.\foot{Discovery of the
$\wt\tau\ls{R}^\pm$ is somewhat more involved and will be neglected in
the following discussion.}  A detailed experimental analysis will then
produce measurements of $\merpm$, $\mmurpm$, $\mchina$,
$\sigler$, $\sigrer$, $\siglmur$, $\sigrmur$, and the slepton angular
distributions.  The notation here should be obvious; for example,
$\sigler=\sigma(e^+e^-\ls{L}\rta\erp\erm)$, \etc\  The $\china$ will
also be discovered by virtue of the decay
$\wt\ell\ls{R}\rta\ell\china$ ($\ell=e,\mu$).  The ability to separate
the supersymmetric signal from Standard Model background can be
enhanced with polarization.  For example, one potentially important
background to slepton pair production is
$e^+e^-\rta W^+W^-\rta\ell^+\ell^-$+ missing energy, which yields the
same type of final state as the signal events.  However, this
background can be suppressed by employing a right-handed electron
beam, since $W$-bosons couple only to $e^-\ls{L}$.  The power of
polarization in the background suppression is illustrated in
Fig.~\jlciii.  When the full analysis is performed, it is found that
the supersymmetric particle masses can be measured
rather accurately, typically to within a few GeV.  With the
information of the masses and cross sections, one can already deduce
important information about the supersymmetric spectrum.  Assuming the
relation of gaugino mass parameters given in Eq.~\eqmass, a
measurement of $\mchina$ is sufficient to provide an upper bound on
the mass of the lightest chargino.  One finds that for
$\mchipma\geq\mw$, the light chargino mass is bounded by
$\mchipma\leq(M_2/M_1)\mchina\simeq 2\mchina$, where Eq.~\eqmass\ was
applied in the last step.  One also tests the universality of
scalar masses by comparing the values of $\merpm$ and $\mmurpm$.
Finally, adding in the cross section measurements and angular
distributions allows one to check the assignment of slepton quantum
numbers and provides some constraints on the parameters of the
neutralino mass matrix.

Proceeding in the scenario under consideration,
in the second stage of the supersymmetric particle search the next
particle to be discovered is $\chipma$.  Experimental observables
include: $\mchipma$, $\siglcpm$, $\sigrcpm$, and the chargino angular
distribution.  The chargino mass search is very similar to the search
for a new heavy charged lepton.  Assuming that squarks and the charged
Higgs boson are much
heavier than $\mw$, the dominant decay of the chargino is
$\chipma\rta\china W^\pm$, where the final state $W$ is either real or
virtual.  According to the analysis of Ref.~[\jlc],
the chargino and neutralino masses can be measured to an
accuracy of about $5\%$.   Moreover, by comparing the relative
chargino production rates using left-handed and right-handed electron
beams, one can distinguish between the higgsino and gaugino components of
$\chipma$, and separate out the sneutrino-exchange contribution.
For example, gauginos and the sneutrino couple only to left-handed
electrons, while higgsinos couple to both $e^-\ls{L}$ and $e^-\ls{R}$.
Thus, one can perform a global fit using the measured masses of
$\chipma$ and $\china$, $\sigrer$ [from stage 1 of the search],
and the chargino pair cross sections $\siglcpm$ and $\sigrcpm$
to reconstruct the sneutrino mass and the four unknown parameters, $M_1$,
$M_2$, $\mu$ and $\tanb$ of the neutralino mass matrix.  One can now
begin to test models of the supersymmetry breaking parameters.  For
example, having deduced $M_1$ and $M_2$, one can directly test the
unification of gaugino mass parameters [Eq.~\eqmass].

\FIG\jlcxvii{}
\REF\bhl{R.M. Barnett, H.E. Haber, and K.S. Lackner, {\sl Phys. Rev.
Lett.} {\bf 51} (1983) 176; {\sl Phys. Rev.} {\bf D29} (1984) 1381.}

\topinsert
   \tenpoint \baselineskip=12pt   \narrower
\centerline{\psfig{file=ssi17.ps,height=8cm,angle=90}\quad}
\vskip12pt\noindent
{\bf Fig.~\jlcxvii.}\enskip
The scatter plot of $e^+$ energies at $\sqrt s = 400$ GeV with an
integrated luminosity of 20 fb$^{-1}$ and an electron beam
polarization of $P_{\em} = +0.95$ for $\epem \to
\tilde e^\pm_L\tilde e^\mp_R$ and $\epem\to\tilde e^+_R\tilde e^-_R$,
including the initial state radiation and the beamstrahlung effects.
Taken from Ref.~[\jlc].
\endinsert

Finally, in the third stage of the supersymmetric particle search,
$\elpm$ is discovered in associated production of $\elmp\erpm$.%
\foot{One may also be able to discover the sneutrino in
$e^+e^-\rta\snu\overline{\snu}$, which is mediated by $s$-channel
$Z$-decay and $t$-channel chargino exchange.  See \eg, Ref.~[\bhl].}
Note that for the associated production, there is no $s$-channel gauge
boson exchange contribution, since the $\gamma$ and $Z$ couple
diagonally to slepton pairs at tree-level.  Thus, only the $t$-channel
neutralino exchange contributes in this case.
If the electron beam is polarized, then there are four possible
cross sections to consider:
$$\eqalign{%
\siglelr&\equiv\sigma(e^+e^-\ls{L}\rta\elm\erp)\,,\cr
\sigrerl&\equiv\sigma(e^+e^-\ls{R}\rta\erm\elp)\,,\cr
\siglerl&\equiv\sigma(e^+e^-\ls{L}\rta\erm\elp)\,,\cr
\sigrelr&\equiv\sigma(e^+e^-\ls{R}\rta\elm\erp)\,.\cr}\eqn\associated
$$
In fact, $\siglerl=\sigrelr=0$, since chirality is conserved at
the $\ell\wt\ell\chitil^0$ vertex.  That is, in the
$e\wt e\chitil^0$ interaction, $e^-\ls{L}$ [$e^-\ls{R}$] couples
exclusively to $\elm$ [$\erm$].  The fact that one can speak of a
chirality for the scalar sleptons is a deep consequence of
supersymmetry which associates  scalar particles with each
left and right-handed fermion component.  As a result, the
experimental observables in the stage 3 analysis are: $\melpm$, $\siglelr$,
$\sigrerl$ and the angular distribution of the final state sleptons.
However, one can test experimentally the absence of $\siglerl$ and
$\sigrelr$ directly with polarized beams.   In Fig.~\jlcxvii,
the results of a Monte Carlo simulation of Ref.~[\jlc] are displayed.
This figure illustrates a case in which the initial electron beam
is almost purely
right-handed electron beam.  The final state sleptons are assumed to
decay via $\wt\ell\rta\ell\china$, so the experimentally observed
process is $e^+e^-\ls{R}\rta e^-e^+ +$ missing energy.  Since
$\melpm>\merpm$, one expects that in the associated slepton
production, the
final state positron energy should be larger than the corresponding
electron energy.\foot{In addition, one also expects events arising
from $e^+e^-\rta\erm\erp$ which should be symmetric
about $E_{e^-}=E_{e^+}$.}
This is indeed the case, as shown in Fig.~\jlcxvii.
The few events seen with $E_{e^-}>E_{e^+}$ correspond to $\siglelr$
since the plot assumes that the initial electron beam is not $100\%$
polarized.  Such a plot if observed in a real experiment
would constitute strong evidence for the
absence of $\sigrelr$ and support the notion of the association of
chirality for sleptons.  With the measurement of $\melpm$ and the
associated slepton pair cross sections, one can make additional checks
of parameters already obtained in stages 1 and 2 above and
test additional model predictions.  For example, in models based
on minimal supergravity, one obtains relations among slepton masses of
the same generation.  These relations can be checked and provide
additional probes of the structure of the theory at the grand unification
scale.
\REF\feng{J.L. Feng and D.E. Finnell, {\sl Phys. Rev.} {\bf D49} (1994)
2369.}

\FIG\jlfi{
The number of squark pairs of the first two generations produced at a
500 GeV $e^+e^-$  collider with polarized $e^-$ beams,
unpolarized $e^+$ beams,  and an integrated luminosity of 20
${\rm\,fb}^{-1}$ for each $e^-$ beam polarization. The four helicity
combinations plotted are $e^-_{L,R}e^+ \rightarrow
\wt q\ls{L,R}\skew6\bar{\tilde q}$.  Taken from Ref.~[\feng].}
\FIG\jlfv{
The distribution of $m\ls {\wt q}^{\rm min}$, the minimum allowed
squark mass for a given event, in region 1 at the point $(\mu , M_2)$ =
$(-500 {\rm\,GeV}, 300 {\rm\,GeV})$. The distribution for $e^-_L$
($e^-_R$) polarized beams is given by the solid (dashed) histogram and
is sharply peaked at the actual $\wt q\ls L$ ($\wt q\ls R$) mass of
220 (210) GeV. The integrated luminosity assumed is 10 ${\rm\,fb}^{-1}$
per polarization, and the bin size is 5 GeV.  Taken from Ref.~[\feng].}

If the energy of the ILC is sufficiently high, one may be able to
reach the squark threshold.  The phenomenology of squark pair
production at a future $e^+e^-$ linear collider has been recently
treated in detail in Ref.~[\feng].  As in the case of $\wt\mu^+\wt\mu^-$
production, only $s$-channel $\gamma$ and $Z$-exchange diagrams contribute.
Again, polarizing the electron beam can play a
critical role in separating out the $\wt q\ls{L}\skew2\overline{\wt q}\ls{L}$
and $\wt q\ls{R}\skew2\overline{\wt q}\ls{R}$ final states.  The
cross sections are very sensitive to the electron helicity, as shown
in Fig.~\jlfi.  In particular, there is a strong tendency for
left [right] handed electrons to produce $\wt q\ls{L}\skew2\overline{\wt
q}\ls{L}$
[$\wt q\ls{R}\skew2\overline{\wt q}\ls{R}$] final states.  Thus, by controlling
the polarization of the electron beam, one can separately determine
the masses of $\wt q\ls{L}$ and $\wt q\ls{R}$.  A Monte-Carlo
simulation of Ref.~[\feng], shown in Fig.~\jlfv,
suggests that a mass difference of 10~GeV
between $\wt q\ls{L}$ and $\wt q\ls{R}$ could be detected.

Clearly, polarization is a valuable tool for dissecting the properties
of supersymmetric particles.  One still must check that the observed
sleptons and squarks have spin-zero, while the charginos and
neutralinos have spin-1/2.  Threshold behavior of cross sections
provides some indication, although this requires that one study
the production rates as a function of the
CM-energy of the collider.  The spin of the final state particles can
also be determined by careful measurement of the distribution of
final state decay products.  These methods are well-known and have
been used often in the past to determine the quantum numbers of
hadronic resonances.  To describe these methods in detail would take
us beyond the scope of these lectures.  Instead, I will refer you
to some of the old textbooks of the field
(see \eg, Refs.~[\helicityrefs,\pilkuhn]),
which perhaps should be
better known to the current young generation of particle physicists!
\vfill\endpage

\vbox{%
   \tenpoint \baselineskip=12pt   \narrower
\centerline{\psfig{file=ssi2-1.ps,height=7cm}\qquad\quad}
\vskip12pt\noindent
{\bf Fig.~\jlfi.}\enskip
The number of squark pairs of the first two generations produced at a
500 GeV $e^+e^-$ collider with polarized $e^-$ beams,
unpolarized $e^+$ beams,  and an integrated luminosity of 20~%
fb$^{-1}$ for each $e^-$ beam polarization. The four helicity
combinations plotted are $e^-_{L,R}\,e^+ \rightarrow
\wt q\ls{L,R}\,\skew2\overline{\wt q}$.  Taken from Ref.~[\feng].
\vskip2pc
\centerline{\psfig{file=ssi2-5.ps,height=7.5cm}\qquad}
\vskip12pt\noindent
{\bf Fig.~\jlfv.}\enskip
The distribution of $m_{\tilde q}^{\rm min}$, the minimum allowed
squark mass for a given event, in region 1 at the point $(\mu, M_2)$ =
$(-500 {\rm\,GeV}, 300 {\rm\,GeV})$. The distribution for $e^-_L$
($e^-_R$) polarized beams is given by the solid (dashed) histogram and
is sharply peaked at the actual $\wt q\ls L$ ($\wt q\ls R$)
mass of
220 (210) GeV. The integrated luminosity assumed is 10~fb$^{-1}$
per polarization, and the bin size is 5 GeV.  Taken from Ref.~[\feng]
}
\endpage

\chapter{}

\title{\bf LECTURE {\seventeenbf 3}:  \break
Applications to Higgs and New Gauge Boson Searches}

In this lecture, I will discuss a few more examples of utilizing
polarization and spin analysis in new physics
searches at future colliders.
This lecture is not meant to be
comprehensive.  Instead, I have selected just a few examples from the
search for Higgs bosons at a future $e^+e^-$ linear collider (ILC) and
new gauge bosons beyond the $W^\pm$ and $Z$ at
a future hadron supercollider (such as the LHC).
Some other areas of investigation will be mentioned briefly
at the end of these lectures.

\section{\bf Higgs bosons beyond the minimal Standard Model}

In the Standard Model, the electroweak gauge symmetry is
broken when the neutral component of a complex Higgs doublet (with
hypercharge $Y=1$) acquires a vacuum expectation value.  The
scalar spectrum
of the theory then contains one neutral CP-even Higgs field;
the other scalar degrees of freedom are Goldstone bosons which are
absorbed by the $W^\pm$ and $Z$ (thereby generating the gauge boson
masses).  In this Lecture, the Standard Model Higgs boson will be
denoted by $\hsm$ in order to distinguish it from other CP-even neutral
Higgs scalars that may appear in non-minimal extensions of the Standard
Model.  Although the Standard Model is very well tested at LEP,
there is no direct experimental information on
the underlying dynamics that is responsible for electroweak symmetry
breaking.  As remarked in the introduction to
these lectures, the study of
electroweak symmetry breaking is the main goal
of the next generation of colliders.

\REF\moretwohiggs{For an extensive review and guide to the literature,
see chapter 4 of Ref.~[\hhg].}
If new physics exists beyond the Standard Model, then the minimal
Higgs structure described above will almost certainly be supplanted by
a more complicated electroweak symmetry breaking sector.  What new
features is this sector likely to possess?  If Higgs bosons are
elementary scalars (on the scale of TeV physics), then one should
consider the possibility of an extended Higgs sector.  Here, I shall
consider the simplest of the extended Higgs sectors---the two-Higgs
doublet model\refmark\moretwohiggs.

Consider a model with two complex Higgs doublets with hypercharges
$Y=-1$ and $Y=+1$, respectively.  For simplicity, I shall assume that
the Higgs sector conserves CP.  Moreover, I will arrange
the Higgs-fermion couplings such that the Higgs doublet with $Y=-1$
[$Y=+1$] couples exclusively to down-type [up-type] fermions.  (This
insures that there are no Higgs-mediated tree-level flavor changing
neutral currents.)  Of the eight scalar degrees of freedom, three
Goldstone bosons are absorbed by the gauge bosons, leaving five
physical Higgs states: two CP-even neutral states $\hl$ and $\hh$, a
CP-odd neutral state $\ha$, and a charged Higgs pair $\hpm$.
In addition, the model has two additional parameters: the ratio of
Higgs vacuum expectation values $\tanb=v_2/v_1$ and the CP-even Higgs
mixing angle $\alpha$ (the latter arises after diagonalizing the
CP-even neutral Higgs mass matrix in the basis of $Y=\pm 1$ states).
The
reader has surely noticed that this is precisely the Higgs spectrum of
the MSSM.  However,  the MSSM is a highly constrained
two-Higgs doublet model.  Whereas the non-supersymmetric two-Higgs
doublet model described above depends on six parameters ($\alpha$,
$\tanb$, and four Higgs masses), the Higgs sector of the MSSM (at
tree-level) is fixed by two parameters (usually taken to be $\tanb$
and $\mha$).  Both versions of the two-Higgs doublet model have been
studied extensively in the literature\refmark\moretwohiggs.

\REF\hnht{H.E. Haber and Y. Nir,
{\sl Nucl. Phys.} {\bf B335} (1990) 363; H.E. Haber and S. Thomas,
SCIPP-94/08 (1994).}
In discussing the prospects for Higgs discovery at future colliders,
one must consider two separate aspects: (i) the discovery of a light
CP-even scalar ($\hl$) and (ii) the discovery of evidence for a non-minimal
Higgs sector.  It may appear that the discovery of $\hl$ would
immediately address point (ii) as well, since if $\hl$ derives from a
non-minimal Higgs sector, then it seems reasonable to assume that its
properties will differ from $\hsm$ of the Standard
Model. However,  under a few
very mild assumptions, one can show that in a two-Higgs doublet model
where $\mhl$, $\mz\ll\mha$, $\mhh$, $\mhpm$, the properties of
$\hl$ approach those of the Standard Model \foot{This limit is called
the {\it decoupling} limit.  One can show that in most cases, the
heavier Higgs states are nearly degenerate, with the mass degeneracy
broken by terms of ${\cal O}(\mz)$.}\refmark\hnht.
Thus, the discovery of
$\hl$ is likely to shed no light on the possible existence of a
non-minimal Higgs sector.

\REF\bargersusyii{V. Barger, M.S. Berger and A.L. Stange,
{\sl Phys. Rev.} {\bf D45} (1992) 4128;
V. Barger, K. Cheung, R.J.N. Phillips and A.L. Stange,
{\sl Phys. Rev.} {\bf D46} (1992) 4914.}
\REF\hhzz{R. Bork, J.F. Gunion, H.E. Haber and A. Seiden,
{\sl Phys. Rev.} {\bf D46} (1992) 2040.}
\REF\azz{J.F. Gunion, H.E. Haber, and C. Kao,
{\sl Phys. Rev.} {\bf D46} (1992) 2907.}
\REF\gunorr{J.F. Gunion and L.H. Orr, {\sl Phys. Rev.} {\bf D46}
(1992) 2052.}
\REF\kzsecond{
Z. Kunszt and F. Zwirner, {\sl Nucl. Phys.} {\bf B385} (1992) 3.}
\REF\bkbtd{H. Baer, C. Kao, M. Bisset, X. Tata and D. Dicus,
{\sl Phys. Rev.} {\bf D46} (1992) 1067; {\bf D47} (1993) 1062.}
\REF\rubbiasc{A. Rubbia, in {\it Proceedings of the 1992 Workshop
on High Energy Physics with Colliding Beams}, Volume 3, ``Electroweak
Symmetry Breaking at Colliding Beam Facilities,'' December 11-12, 1992,
Santa Cruz, CA, edited by J. Rogers, SLAC-Report-428 (1993) p.~800--836.}

Present LEP bounds based on the search for
the lightest CP-even Higgs boson imply that $\mhl\geq 60$~GeV
(assuming $\hl=\hsm$).  At LEP-II with $\sqrt{s}\simeq 190$~GeV
running at design luminosity, it should be possible to
discover the $\hl$ or set a bound of $\mhl\gsim\mz$\refmark\higgslimit.
As remarked in
Lecture 2, in the MSSM the tree-level bound of $\mhl\leq\mz$ is
significantly modified by radiative corrections.  For $\mt=180$~GeV,
the bound reads $\mhl\lsim 130$~GeV, which implies that
if the MSSM were correct, then there is a
significant possibility that LEP will not discover the Higgs boson.
In this circumstance,
we will be forced to wait until the 21st century to get our
first glimpse of the Higgs sector.  The LHC will be able to probe a
considerable Higgs mass range\refmark{\bargersusyii-\rubbiasc}.
If $140\lsim\mhl\lsim 600$~GeV, then
the LHC will discover the $\hl$ via the gold-plated signature:
$$gg\rta\hl\rta ZZ\rta\ell^+\ell^-\ell^+\ell^-\,,\eqn\goldplate
$$
under the assumption that the $\hl ZZ$ coupling is equal to that of
the minimal Higgs coupling in the Standard Model.  [Note that for
$\mhl<2\mz$, at least
one of the $Z$-bosons in Eq.~\goldplate\ is off-shell.]
For Higgs mass values in the ``intermediate mass regime'' of
$\mz\lsim\mhl\lsim 140$~GeV, the Higgs search at a hadron collider is
much more difficult.  The dominant signal, $gg\rta\hl\rta b\bar b$, is
completely swamped by the QCD background ($q\bar q$,
$gg\to b\bar b$). Other
signatures have been proposed, such as
$$\eqalign{%
gg\rta&\,\hl\rta\gamma\gamma\,,\cr
gg\rta&\,t\bar t\hl\,,\phantom{\rta W\hl}
\qquad(\hl\rta b\bar b\ ~{\rm or}\ ~\hl\rta\gamma\gamma)\,,\cr
q\bar q^\prime\rta&\,W^\ast\rta W\hl\,,
\qquad(\hl\rta\gamma\gamma)\,.\cr}
\eqn\higgsigs
$$
LHC detectors are being designed with some of these Higgs search
channels in mind\refmark\rubbiasc.
One would hope to be able to detect a signal in at
least two channels in order to have confidence that a Higgs signal was
being observed.

What about the prospects for detection of the other states of the
non-minimal Higgs sector?  At LEP-II, $\hpm$ can be detected via
$e^+e^-\rta H^+H^-$, and $\ha$ can be detected via $e^+e^-\rta\hl\ha$.
Of course, if $\ha$, $\hpm$ and $\hh$ are substantially heavier than
$\hl$ then no evidence of the extended Higgs search will emerge prior
to the era of the supercolliders.  Moreover, LHC will be hard-pressed
to find such states.  Both $\hpm$ and $\ha$ do not couple to gauge
boson pairs, so that the signatures of these states at a
hadron supercollider are notoriously
difficult to separate from Standard Model backgrounds.

\REF\barg{V. Barger, K. Cheung, B.A. Kniehl, and R.J.N. Phillips,
{\sl Phys. Rev.} {\bf D46} (1992) 3725.}
\REF\janotnlc{P. Janot, in
{\it Physics and Experiments with Linear $\epem$ Colliders},
Workshop Proceedings, Waikoloa, Hawaii, 26--30 April, 1993,
edited by F.A. Harris, S.L. Olsen, S. Pakvasa and X. Tata,
(World Scientific, Singapore, 1993) pp.~192--217.}
\REF\habersummary{H.E. Haber, in
{\it Physics and Experiments
with Linear Colliders}, Proceedings of the Linear Collider Workshop,
Saariselk\"a, Finland, 9--14 September, 1991, edited by R. Orava,
P. Eerola and M. Nordberg (World Scientific, Singapore, 1992)
pp.~235--275; J.F. Gunion, in
{\it Physics and Experiments with Linear $\epem$ Colliders},
Workshop Proceedings, Waikoloa, Hawaii, 26--30 April, 1993,
edited by F.A. Harris, S.L. Olsen, S. Pakvasa and X. Tata,
(World Scientific, Singapore, 1993) pp.~166--191.}

\REF\yamada{A. Yamada, {\sl Mod. Phys. Lett.} {\bf A7} (1992) 2789.}

We therefore turn to the prospects of Higgs detection at the ILC.
An $e^+e^-$ linear collider with $\sqrt s \geq 300$ GeV
will be able to fully explore the
intermediate mass Higgs regime.  The two primary mechanisms for Higgs
production at the ILC
are: (i) $e^+e^-\rta Z\hl$ via $s$-channel $Z$-exchange
(the same mechanism responsible for
Higgs production at LEP-II energies); and (ii)
$e^+e^-\rta\nu\bar\nu\hl$ via $W^+W^-$
fusion (this latter mechanism becomes
increasingly important as $\sqrt{s}$ becomes larger).   An ILC with
CM-energy
$300\lsim\sqrt{s}\lsim 500$~GeV and an integrated luminosity of 10 to
20~fb$^{-1}$ would have a discovery reach of $\mhl\lsim 0.7\sqrt{s}$,
enough to cover completely the intermediate mass regime\refmark\barg.
The LEP-II search for non-minimal Higgs states mentioned above also
applies at higher energies.  The various signatures appear to be
detectable at the ILC if the processes are kinematically allowed (and
not too close to threshold)\refmark{\janotnlc--\yamada}.

\section{\bf Higgs boson production at a $\bold{\gamma\gamma}$ collider}

\REF\telnovi{H.F. Ginzburg, G.L. Kotkin, V.G. Serbo and V.I. Telnov,
{\sl Nucl. Inst. and Methods} {\bf 205} (1983) 47.}
\REF\telnovii{H.F. Ginzburg, G.L. Kotkin, S.L. Panfil,
V.G. Serbo and V.I. Telnov,
{\sl Nucl. Inst. and Methods} {\bf 219} (1984) 5.}
\REF\barklow{T.L. Barklow,
in {\it Research Directions for the Decade}, Proceedings of the 1990
Summer Study on High Energy Physics, Snowmass, CO, June 25--July
13, 1990, edited by E.L. Berger (World Scientific, Singapore,
1992), pp.~440--450.}
\REF\caldwell{D.L. Borden, D.A. Bauer and D.O. Caldwell,
{\sl Phys. Rev.} {\bf D48} (1993) 4018.}
All $\epem$ colliders are also $\gam\gam$ colliders.  However,
the $\gam\gam$ luminosity resulting from the
Weizs\"acker-Williams spectrum of photons falls rapidly
as a function of the $\gamma\gamma$ invariant mass.
The ILC provides a more promising alternative for directly studying
$\gamma\gamma$ collisions.
By Compton backscattering of laser photons off
the ILC electron and positron beams, one can produce high luminosity
$\gamma\gamma$ collisions with a wide spectrum of $\gamma\gamma$
CM-energy ($\egamgam$)\refmark{\telnovi--\caldwell}.
In comparison with the
Weizs\"acker-Williams spectrum and luminosity, the $\gamma\gamma$
collider mode produces a significantly harder $\gamma\gamma$
spectrum with a substantially higher luminosity at large $\egamgam$.
In addition, a high degree of circular polarization for each of the colliding
photons can be achieved by polarizing the incoming electron
and positron beams and the laser beams\refmark\telnovii. Both the
$\gam\gam$ luminosity spectrum and subprocess cross
sections are strongly influenced by the polarizations
of the colliding photons.

\REF\russians{A.I. Vainshtein, M.B. Voloshin, V.I. Zakharov and
M. Shifman, {\sl Yad. Fiz.}, {\bf 30} (1979) 1368
[{\sl Sov. J. Nucl. Phys.} {\bf 30} (1979) 711].}
Higgs bosons can be produced in $\gamma\gamma$ collisions via
a one-loop diagram, in which all charged particles of the theory
whose masses derive from the Higgs mechanism can appear inside the
loop\refmark\russians.
(The relevant formulae are
conveniently summarized in section 2.1 of Ref.~[\hhg].)
Thus, the detection of Higgs bosons in the $\gamma\gamma$ collider
mode at the ILC can provide fundamental information about the
particle spectrum and mass generation mechanism of the theory.

\FIG\widthratios{}
\REF\gamgamgunhab{J.F. Gunion and H.E. Haber,
{\sl Phys. Rev.} {\bf D48} (1993) 5109.}
To illustrate the sensitivity of the $\gamma\gamma$-Higgs coupling
to new particles that can appear in the loop, Gunion and I computed
$\Gamma({\rm Higgs}\rta\gamma\gamma)$ in a variety of model
scenarios\refmark\gamgamgunhab.
Our results are shown in Fig.~\widthratios.  First, we
compared the Standard Model to a model
with one extra heavy generation of quarks and leptons.
In Fig.~\widthratios, the results are exhibited
in the case of a mass-degenerate heavy fourth
generation quark doublet of mass 500~GeV and a heavy charged lepton of
mass 300~GeV.  These mass values were chosen so that none of the fourth
generation fermions could be pair produced at an
$\epem$ linear collider with $\sqrt{s}=500$~GeV (ILC-500).
Note that even a crude measurement of
the $\gamma\gamma$-Higgs coupling
would be sufficient to distinguish between the three and four generation
Standard Model (except in a small Higgs mass region where the ratio
of $\gamma\gamma$ couplings is accidentally near 1).

\midinsert
   \tenpoint \baselineskip=12pt   \narrower
\centerline{\psfig{file=ggh1.ps,angle=90,height=7.5cm}}
\vskip6pt\noindent
{\bf Fig.~\widthratios}\enskip
The ratio of $\Gamma({\rm Higgs}\rta \gam\gam)$
computed in two different models, for a number of model choices.
For the Standard Model (SM) $\hsm$: the ratio of $\Gamma(
\hsm\rta\gam\gam)$ as computed in the 4-generation {\it vs.} the
3-generation SM, as a function of $\mhsm$ (solid curve).   The
extra generation (EG) of fermions includes a 300~GeV charged lepton
and a mass-degenerate 500~GeV quark doublet.
For the MSSM $\hl$: the ratio
$\Gamma(\hl\rta\gam\gam)/\Gamma(\hsm\rta\gam\gam)$
as a function of $\mhl=\mhsm$ (dotted curve), with $\mha=400~\gev$.
Squarks and charginos have been taken to be
as light as possible without being observable at ILC-500 (see text).
For the MSSM $\hh$, two curves are shown.   The dot-dashed curve is
$\Gamma(\hh\rta\gamma\gamma)$ in a model
with light charginos ($M_2=-\mu=150~\gev$)
divided by the corresponding width with heavy charginos
($M_2=-\mu=1$~TeV), keeping the squarks and sleptons heavy
(with masses of order 1~TeV);  the dashed curve is
$\Gamma(\hh\rta\gamma\gamma)$ in a model with
light squarks and sleptons (see text)
divided by the corresponding width computed with heavy squarks and
sleptons, keeping the charginos heavy as before.
For the latter two curves,
the ratio of widths is plotted as a function of $\mhh$, for
$\tanb=2$. Taken from Ref.~[\gamgamgunhab].
\endinsert

Next, consider the case of the MSSM.
Suppose the lightest CP-even Higgs boson has
been discovered, but no experimental evidence for either the heavier
Higgs bosons or any supersymmetric particles has been found at the ILC.
Could a measurement of the $\hl\gamma\gamma$ coupling provide indirect
evidence for physics beyond the Standard Model?  Unfortunately, the
answer is no.
If the MSSM parameters are chosen such that all new particles beyond the
Standard Model are too heavy to be produced at the ILC-500,
then the deviation of
$\Gamma(\hl\rta\gamma\gamma)$ from the corresponding Standard Model
value is less than 15\%.  Because of decoupling, as the
supersymmetry breaking scale and the scale of the heavier Higgs bosons
become large, all couplings of the $\hl$ approach their Standard Model
values. This is illustrated in Fig.~\widthratios\
where we plot the ratio $\Gamma(\hl\rta\gam\gam)/\Gamma(
\hsm\rta\gam\gam)$, as a function of the Higgs mass,
for chargino mass parameters $M_2=-\mu=300~\gev$
and a common soft-supersymmetry-breaking diagonal mass of 300 GeV
for all squarks and sleptons, with all off-diagonal squark and
slepton masses set to zero.
Even with the MSSM parameters chosen such that the supersymmetric
partners lie just beyond the reach of ILC-500, the ratio of
$\gamma\gamma$ decay widths is still close to 1 (in
Fig.~\widthratios, the plotted ratio lies between 0.89 and 0.94).

On the other hand, suppose that some of the other Higgs bosons
of the MSSM ($\hpm$, $\hh$ and/or $\ha$)
are light enough to be produced and studied at the ILC.
In this case, a measurement of the $\gamma\gamma$ couplings of
$\hh$ and $\ha$ can provide useful information on the spectrum
of charged supersymmetric particles (even if the latter are too heavy to be
directly produced at the ILC).\foot{The widths of $\hh$ and $\ha$
into $\gamma\gamma$ will almost always differ from the
width of the Standard Model
Higgs boson and vary as a function of the MSSM parameters.}
Figure \widthratios\ provides two
examples of the sensitivity of the $\hh\gamma\gamma$ couplings to
supersymmetric particle masses.  Suppose that the masses
of all supersymmetric particles appearing in the loop (charginos,
squarks, and sleptons) are 1~TeV in mass.  Consider then
two different
scenarios: (i) light charginos and heavy squarks and sleptons (with
chargino parameters $M_2=-\mu= 150$~GeV), and (ii) light squarks and
sleptons and heavy charginos (with a common soft-supersymmetry breaking
diagonal mass of 150~GeV for all squarks and sleptons, and with
all off-diagonal squark and slepton masses set to zero).\foot{In the
light squark and slepton scenario (with $\tan\beta=2$),
all sleptons and squarks, with the
exception of the top-squark are roughly degenerate in mass, ranging
between 141 and 157~GeV, while the two top-squark masses are 208 and
210~GeV, respectively.  Thus, there are two distinct thresholds
for squark (and slepton) pair production, which account for the
two dips on the corresponding curve in Fig.~\widthratios.}
The ratio of $H^0\rta\gamma\gamma$ widths
(relative to the case of 1~TeV supersymmetric
particle masses) in these two scenarios is depicted in Fig.~\widthratios\
and demonstrates the sensitivity of the $\hh\gamma\gamma$ coupling to
the details of the supersymmetric spectrum.

The connection between the $\h\gam\gam$ coupling and the Higgs
production rate in $\gam\gam$ collisions is most conveniently
expressed in terms of the $\h\rta\gam\gam$ decay width.\foot{
Henceforth, the symbol $\h$ will be used to denote any neutral CP-even
or CP-odd Higgs boson.}\
For a given value of the $\gam\gam$ CM-energy $\egamgam$ we have:
$$\sigma(\gam\gam\rta\h\rta X)={8\pi \Gamma(\h\rta\gam\gam)
\Gamma(\h\rta X)
 \over (\egamgam^2-\mh^2)^2 +\gammah^2\mh^2}
(1+\lami\lamii)
 \,,\eqn\sigmaform$$
where $\gammah\equiv\Gamma(\h\rta {\rm all})$
is the total decay width of $\h$, and $\lami$ and $\lamii$ ($=\pm1$)
are the helicities of the two colliding photons.

The final states $X$ of greatest interest are $\wp\wm$, $ZZ$ and
$Q\overline Q$ (with $Q=b$ or $t$).  In this section, I shall focus on
$X=Q\overline Q$; see Refs.~[\caldwell] and [\gamgamgunhab]
for the analysis of other
possible final state signals (and their corresponding backgrounds).
In order to compute the expected number of Higgs bosons, Eq.~\sigmaform\
must be folded together with the appropriate
$\gam\gam$ luminosity.
The number of Higgs bosons produced and detected is given by:
$$N(\gam\gam\rta \h\rta Q\overline Q)=
\int_{y_-}^{y_+} dy\,{d{\cal L}_{\gam\gam}\over d y}
\sigma_{\gam\gam\rta\h\rta Q\overline Q}
(\egamgam=y\eepem)\,,\eqn\integratedsigma$$
where $d{\cal L}_{\gam\gam}/d y$ is the differential $\gam\gam$
luminosity as a function of $y\equiv \egamgam/\eepem$ and
$y_{\pm}=(\mh\pm\gmax/2)/\eepem$. Here, $\gmax$ is a resolution factor
which is chosen to maximize the significance of the $\gamma\gamma\rta
H\rta Q\overline Q$ signal over the $\gamma\gamma\rta Q\overline Q$
continuum background.  In Ref.~[\gamgamgunhab], Gunion and I adopted a
strategy of integrating over a region of $E_{\gamma\gamma}$ of size
$\gmax\equiv {\rm max}\{\gexp,\gammah\}$, where $\gexp$ is the
experimental $Q\overline Q$ mass resolution and $\gammah$ is the total
Higgs width.  If one defines
$${d{\cal L}_{\gam\gam} \over d y}\equiv F(y) {\cal L}_{\epem}
\,, \eqn\fdef$$
then $F(y)$ and the average value of
$\lami\lamii$ at $y$ [denoted by $\vevlam_y$]
are obtained after convoluting over the possible energies and polarizations
of the colliding photons that yield a fixed value of $y$.  Both quantities
will depend upon the experimental arrangement for creating
the back-scattered laser beams, the polarization of the incoming
electron and positron, and the polarizations of the two initial
laser beams\refmark{\telnovi--\caldwell}.
In the present discussion, it is a good approximation to
assume that both $F(y)$ and $\vevlam_y$ are constant over the region of
integration in Eq.~\integratedsigma\ where the Higgs cross section is dominant.

It is a simple exercise to compute the number of detected Higgs
events directly from Eq.~\integratedsigma.
If $F(y)$ and $\vevlam_y$ are slowly varying,
$$\eqalign{
N(\gam\gam\rta\h\rta Q\overline Q)=&{8\pi BR(\h\rta X)\over \eepem
 \mh^2} \tan^{-1}\left({\gmax\over
\gammah}\right) \Gamma(\h\rta\gam\gam) \crr
&\times F(\yh) (1+\vevlam_{\yh})
{\cal L}_{\epem}\,,\cr}\eqn\nevents$$
where $\yh\equiv \mh/\eepem$.  Note that
in the limit where $\gexp\gg\gammah$ the inverse tangent approaches
$\pi/2$, and $N$ is independent of $\gmax$.

As already noted, $F(y)$ and $\vevlam_y$ can be adjusted by
appropriately choosing the experimental arrangement and polarizations
of the electron, positron and two initial laser beams.
The basic formalism for computing these quantities appears in
Ref.~[\telnovii] and a large number of specific cases were examined
in Ref.~[\caldwell].  For the reader's convenience, the most important
points are reviewed here.  Define $\lamei$ ($\lameii$) and $\pci$
($\pcii$) to be the helicity and circular polarization of the electron
and corresponding laser beam responsible for producing photon 1 (photon 2).
It is useful to consider the extreme cases of $2\lamei \pci=\pm1$,
\ie, maximal helicity for the incoming electron and full circular
polarization for the initial laser photon. For $2\lamei \pci=-1$
the energy spectrum of photon 1 is peaked just below the highest allowed
photon energy, whereas for $2\lamei \pci=+1$ one finds a rather
flat spectrum over a broad range of photon energy falling sharply
to zero as one approaches the maximum possible energy.
Meanwhile, the helicity of the back-scattered photon,
$\lami$, approaches $+\pci,-\pci$ for photon energy
equal to zero or the maximum allowed, respectively.
In the case of $2\lamei \pci=1$, $\lami=+\pci$ over almost the entire
photon energy range; only very near to the maximum allowed energy
does $\lami$ change sign and approach $-\pci$.  In contrast,
in the case of $2\lamei \pci=-1$, associated with a peaked
energy spectrum, $\lami$ slowly switches sign in the middle
of the allowed energy range.  Note that it is unlikely that
the ({\it a priori} unknown) Higgs boson mass will be
approximately equal to the full $\epem $energy.  Thus,
the flat energy spectrum obtained for $2\lamei \pci=+1$
will generally be preferred for Higgs boson searches.  In addition,
one sees that this choice will imply a relatively constant (and large)
value for $|\lami|$ over most of the energy range of interest.

The functions $F(y)$ and $\vevlam_y$ are obtained by convoluting
together the spectra and polarizations for the individual photons
1 and 2.  In order to maximize the Higgs cross section, one sees from
Eq.~\nevents\ that it is desirable to have as large a value for $F(\yh)$
as possible {\it and} to have $\vevlam_{\yh}\sim +1$.
Moreover, we shall see below that $Q\overline Q$ backgrounds are
proportional to $1-\vevlam_{\yh}$ for
$\mh\gg 2m_Q$, and will be suppressed if $\vevlam_{\yh}\sim +1$.
Typical behaviors for $F(y)$ and $\vevlam_y$ are illustrated in
Ref.~[\caldwell] and can be summarized as follows.
First, at $\epem$ energies of the order of $500~\gev$,
the kinematic limit for $\egamgam$ is roughly $0.8\eepem$; \ie,
$\ymax\sim 0.8$ for the typical machine design.
Second, one finds that large $F(y)$ and $\vevlam_y\sim +1$ can be
{\it simultaneously} achieved for all $y$ between $\sim 0.1$ and $\ymax$.
For Higgs searches below about $0.7\eepem$, it is most useful to
employ the broad spectrum for $F(y)$ that results from choosing
$2\lamei \pci$ and $2 \lameii \pcii$ both as close to $+1$ as possible.
For these choices and typical machine design parameters at $\eepem=500~\gev$,
$F(y)\gsim 1$ for $50~\gev\lsim\egamgam\lsim 350~\gev$, \ie, for $y$ values
between 0.1 and 0.7. If, in addition to choosing
$2\lamei \pci=+1$, $2\lameii \pcii=+1$, we also have $\pci\pcii\sim+1$
then $\vevlam_y$ is near $+1$ for this entire range, $y\lsim 0.7$.
For Higgs searches between $y\sim 0.7$ and $y\sim\ymax$, $F(y)\gsim1$
and $\vevlam_y\sim+1$ can again be simultaneously achieved by the
alternative choices of $2\lamei\pci=2\lameii\pcii=-1$, $\pci\pcii=+1$.
Thus, there is a fortunate conspiracy in which polarization choices
can be made that
yield large $\gam\gam$ luminosity (for $0.1\lsim y\lsim \ymax$)
and also maximize
the Higgs cross section while tending to suppress backgrounds.

\REF\eboli{O.J.P. Eboli, M.C. Gonzalez-Garcia, F. Halzen and
D. Zeppenfeld, {\sl Phys. Rev.} {\bf D48} (1993) 1430.}
In order to establish the significance of the Higgs signal, one
must consider the background from
$\gam\gam\rta Q\overline Q$\refmark{\caldwell,\gamgamgunhab,\eboli}.
The differential cross section for $\gamma(\lambda)+
\gamma(\lambda^\prime)\rta Q\overline Q$ is
$${d\sigma\over dz}= {4\pi\alpha^2 e_Q^4N_c\beta
\left[1-\beta^4+\half (\lami\lamii-1)(1+\beta^2z^2)(1-2\beta^2+\beta^2z^2)
\right] \over s
(1-\beta^2 z^2)^2}\,,\eqn\gamgamqqsig$$
where $z\equiv\cos\theta$ is the cosine of the angle
of the outgoing heavy quark, $Q$, relative to the beam direction
in the $\gamma\gamma$ CM-frame,
$e_Q$ is the quark charge (in units of $e$), $m_Q$ is the quark mass,
$\beta\equiv (1-4m_Q^2/s)^{1/2}$, and $s\equiv \egamgam^2$.
For $m_Q=0$, one can easily check that Eq.~\gamgamqqsig\ reproduces the
results of my Lecture 1 computation [see Eq.~\spinaveraged].
If $\egamgam\gg 2m_Q$ (\ie, $\beta\rta 1$)
then the cross section for this
background subprocess is strongly suppressed for $\lami\lamii=+1$,
the choice which maximizes the Higgs boson cross section as discussed above.
Furthermore,
for $\beta\rta 1$, the cross section is strongly forward-backward peaked
as a function of $z$, whereas the $\h\rta
Q\overline Q$ decay is isotropic.  Thus, to maximize the significance of
the Higgs boson signal in the $b\overline b$ channel it is desirable
to arrange for a value of $\vevlam_{\yh}$ as near to $+1$ as possible,
and to integrate signal and background over a region $|z|<z_0$
away from $|z|=1$.  Even in the $t\overline t$ channel this same procedure
provides some improvement in the statistical significance of the Higgs
signal, despite the fact that $2m_t$ is only somewhat smaller than $\mh$.

Integrating Eq.~\gamgamqqsig\ over the region $|z|\leq z_0<1$ yields
$$
\eqalign{
 \sigma_{Q\overline Q}(s,z_0)=
{4\pi\alpha^2e_Q^4N_c\over s}
\Biggl\{&-\beta z_0\left[1+{(1-\beta^2)^2\over (1-\beta^2z_0^2)}\right]
+\half\left[ 3-\beta^4\right]\log {1+\beta z_0\over 1-\beta z_0}\crr
&+{\lami\lamii}\beta z_0\left[1+{2(1-\beta^2)\over 1-\beta^2z_0^2}
-{1\over \beta z_0}\log {1+\beta z_0\over 1-\beta z_0}\right]
\Biggr\}\,.\cr}\eqn\integratedbkgnd$$
The effectiveness of using polarization and angular cuts in reducing
the $\gamma\gamma\rta Q\overline Q$ background is clear.  For example,
if both $\beta$ and $\lambda\lambda^\prime$ are near 1, then
Eq.~\integratedbkgnd\ yields
$$\eqalign{
\sigma_{Q\overline Q}(s,z_0)\simeq
{4\pi\alpha^2e_Q^4N_c\over s}
\Biggl\{&(1-\beta^2)\left[{2z_0\over
1-z_0^2}+\ln\left({1+z_0\over 1-z_0}\right)\right]\crr
&+(\lambda\lambda^\prime
-1)\left[z_0-\ln\left({1+z_0\over 1-z_0}\right)\right]\Biggr\}\,.\cr}
\eqn\sigapprox$$
That is, for $\lambda\lambda^\prime=1$, the cross-section is suppressed
by a factor of $1-\beta^2$, as expected from our previous analysis.
The effectiveness of the
cut on the decay angle  in reducing the $Q\overline Q$ background
can be seen in the following example.
For $Q=b$ and $\sqrt s=\egamgam=300~\gev$, choosing $z_0=0.5$ reduces
the $Q\overline Q$ cross section by a factor of 14, while retaining
50\% of the Higgs events.  This yields a net gain by a factor of $7$
in Higgs signal over background.
We can compute the net number of background events by
multiplying the cross section of Eq.~\integratedbkgnd\ by the
integral of $d{\cal L}_{\gam\gam}/dy$ [Eq.~\fdef] over the interval
$\Delta y=\gmax/\eepem$, with $\gmax$ defined below Eq.~\integratedsigma\
$$
N(\gam\gam\rta Q\overline Q)= {\gmax\over \eepem}
F(\yh) {\cal L}_{\epem} \sigma_{Q\overline Q}(\mh^2,z_0)\,.
\eqn\nbkgnd$$

For numerical estimates, Gunion and I took
$F(\yh){\cal L}_{\epem}=2\times
10^{33}~{\rm cm}^{-2}~{\rm sec}^{-1}$
\ie, an integrated luminosity of $20~{\rm fb}^{-1}$
for a standard collider year, and we chose
$\vevlam_{\yh}=0.8$. As discussed previously, these choices for $F(\yh)$
and $\vevlam_{\yh}$ are both well within the range
of possibility for $\yh\lsim 0.8$ (\eg, $\egamgam\lsim 400~\gev$
at $\eepem=500~\gev$).
Detector resolution was assumed to be
such that $\gexp=5~\gev$ is possible for the observation of
$\h\rta Q\overline Q$.
When considering $Q\anti Q$ final states, we employed the background
cross section given in Eq.~\integratedbkgnd. Since
the Higgs decays to $Q\anti Q$ are uniform in $z=\cos\theta$,
the effect of the angular cut on the number of Higgs events in a
$Q\anti Q$ final state is simply to multiply the total event rate by a
factor of $z_0$. In our numerical work, we
found that the value $z_0=0.85$ was the most effective in reducing
the $Q\overline Q$ background.

\FIG\smfig{Number of events per year for
the Standard Model Higgs boson ($\hsm\rta b\bar b$,
$\hsm\rta t\bar t$, $\hsm\rta ZZ$) and for
the $Q\anti Q$ backgrounds ($\gam\gam\rta b\bar b$
and $\gam\gam\rta t\bar t$).
In computing event rates for the $Q\overline Q$ ($Q=b,t$) final states,
an angular cut of $|\cos\theta|\leq z_0=0.85$ is imposed
and $\gmax$ is as specified in Eq.~\gmaxdef\ [with $\gexp=5~~\gev$];
$\gmax=\infty$ has been taken for the $ZZ$ final state.
An integrated luminosity of $20~{\rm fb}^{-1}$ has been assumed,
and we have adopted the average value $\vevlam=0.8$.
Results for $\mt=150~\gev$ and $\mt=200~\gev$ are displayed.}

In Fig.~\smfig\ we present results for the
Standard Model Higgs boson, $\hsm$.
Results for two choices of
top quark mass: $m_t=150~\gev$ and $\mt=200~\gev$ are shown.
It is apparent that where $\hsm\rta b\bar b$
decays are dominant, $\hsm$ masses down to roughly $40~\gev$
can be probed.  In contrast, the $t\bar t$ mode
never has a sufficiently large branching ratio (due to the dominance of the
vector boson pair decay modes of the $\hsm$ above the $W^+W^-$
threshold) to be useful, given
the large size of the $\gam\gam \rta t\anti t$ background.
The importance of polarization for enhancing the statistical significance
of the Higgs signal in the $b\anti b$ channel
is illustrated
\FIG\polbb{Minimum and maximum Higgs masses for which the number of
standard deviations of the Higgs signal in the $b\anti b$ channel
$\nsd\geq 5$ and the number of signal Higgs events $S\geq 10$,
as a function of $\vevlam$. Both Standard Model and MSSM Higgs boson
signals are exhibited, with
$\mt=150~\gev$ and $\tanb=2$.   Not shown explicitly is
$\mha^{\rm max}$ which is always very close to $2\mt=300~\gev$.  The
$\hl\rta b\anti b$ signal is statistically significant for
all $\mhl$ masses from $\mhl^{\rm min}$ up to its theoretical limit.}
in Fig.~\polbb, which depicts the maximum and minimum masses for
which the number of Higgs signal events ($S$) and background events
($B$) are such that $S>10$ and the number of standard deviations
[$\nsd\equiv S/\sqrt B$] is greater than 5.
{}From Fig.~\polbb, it is clear that being able to achieve large $\vevlam$
greatly expands the mass range over which the $\hsm$ can be detected
in the $b\anti b$ channel.  Unfortunately,
detection of the $\hsm$ in the $t\anti t$ channel is
not possible even with perfect polarization.

\pageinsert
   \tenpoint \baselineskip=12pt   \narrower
\centerline{\psfig{file=ggh4.ps,angle=90,height=7.5cm}}
\vskip6pt\noindent
{\bf Fig.~\smfig.}\enskip
{Number of events per year for
the Standard Model Higgs boson ($\hsm\rta b\bar b$,
$\hsm\rta t\bar t$, $\hsm\rta ZZ$) and for
the $Q\anti Q$ backgrounds ($\gam\gam\rta b\bar b$
and $\gam\gam\rta t\bar t$).
In computing event rates for the $Q\overline Q$ ($Q=b,t$) final states,
an angular cut of $|\cos\theta|\leq z_0=0.85$ is imposed
and a $Q\overline Q$ mass resolution of 5 GeV has been assumed.
Taken from Ref.~[\gamgamgunhab].}
\vskip2pc
\centerline{\psfig{file=ggh2.ps,angle=90,height=7.5cm}}
\vskip6pt\noindent
{\bf Fig.~\polbb.}\enskip
{Minimum and maximum Higgs masses for which the number of
standard deviations of the Higgs signal in the $b\anti b$ channel
$\nsd\geq 5$ and the number of signal Higgs events $S\geq 10$,
as a function of $\vevlam$. Both Standard Model and MSSM Higgs boson
signals are exhibited, with
$\mt=150~\gev$ and $\tanb=2$.   Not shown explicitly is
$\mha^{\rm max}$ which is always very close to $2\mt=300~\gev$.  The
$\hl\rta b\anti b$ signal is statistically significant for
all $\mhl$ masses from $\mhl^{\rm min}$ up to its theoretical limit.
Taken from Ref.~[\gamgamgunhab].}
\endinsert

\FIG\hhfig{As in Fig.~\smfig, but for the heavy
CP-even Higgs boson of the MSSM, $\hh$.  The exotic $\hh\rta\hl\hl$
decay mode event rate is also shown.
(It should be noted that in the region $165\lsim\mhh\lsim 220~\gev$,
the decay $\hh\rta\hl\hl$ is kinematically forbidden for the MSSM
parameters chosen.)
Supersymmetric partners are assumed to be sufficiently heavy that
they do not influence the $\hh\gam\gam$ coupling or $\hh$ decays.
Radiative corrections to the MSSM Higgs sector are incorporated with
$\msusy=1\tev$ and squark mixing neglected. Results for $\mt=150~\gev$
and $\mt=200~\gev$ are displayed for both $\tanb=2$ and $\tanb=20$.}
\FIG\hafig{As in Fig.~\hhfig, but for the CP-odd Higgs
boson of the MSSM, $\ha$. The exotic $\ha\rta Z\hl$ decay mode
event rate is shown.
The $\ha$ has no tree-level couplings to vector boson pairs,
so that the $ZZ$ final state is not present.}
\FIG\poltt{Statistical significance (\ie, number of standard deviations
$\nsd$) of the $\ha\rta t\anti t$ and $\hh\rta t\anti t$ Higgs signals
at $\mha,\mhh=500~\gev$ as a function of $\vevlam$, for
$\mt=150~\gev$ and $\tanb=2$.
$\gmax$ and $z_0$ are chosen as in Fig.~\smfig.}

\pageinsert
   \tenpoint \baselineskip=12pt   \narrower
\centerline{\psfig{file=ggh8.ps,angle=90,height=7.5cm}}
\vskip6pt\noindent
{\bf Fig.~\hhfig.}\enskip
{As in Fig.~\smfig, but for the heavy
CP-even Higgs boson of the MSSM, $\hh$.  The exotic $\hh\rta\hl\hl$
decay mode event rate is also shown.
(It should be noted that in the region $165\lsim\mhh\lsim 220~\gev$,
the decay $\hh\rta\hl\hl$ is kinematically forbidden for the MSSM
parameters chosen.)
Supersymmetric partners are assumed to be sufficiently heavy that
they do not influence the $\hh\gam\gam$ coupling or $\hh$ decays.
Radiative corrections to the MSSM Higgs sector are incorporated with
$\msusy=1\tev$ and squark mixing neglected. Results for $\mt=150~\gev$
and $\mt=200~\gev$ are displayed for both $\tanb=2$ and $\tanb=20$.}
\vskip2pc
\centerline{\psfig{file=ggh6.ps,angle=90,height=7.5cm}}
\vskip6pt\noindent
{\bf Fig.~\hafig.}\enskip
{As in Fig.~\hhfig, but for the CP-odd Higgs
boson of the MSSM, $\ha$. The exotic $\ha\rta Z\hl$ decay mode
event rate is shown.
The $\ha$ has no tree-level couplings to vector boson pairs,
so that the $ZZ$ final state is not present.}
\endinsert

In the case of the Higgs bosons of the MSSM,
results will be presented for one
moderate value of $\tanb=2$ and one large
value of $\tanb=20$, and two choices of top quark mass as in the
previous analysis.
The masses of all the supersymmetric
partners, in particular, the charginos, neutralinos, squarks, and sleptons
are assumed to be large. This means not only
that they will not enter into the Higgs decay modes, but also,
in the case of the charged supersymmetric particles, their
contributions to the $\gam\gam$ couplings at one-loop will be small.
Results for the heavier CP-even state $\hh$ and for the
CP-odd $\ha$ are given in Figs.~\hhfig\ and \hafig, respectively.
Sensitivity of the $\mt=150~\gev$, $\tanb=2$ results for the $\hh$, $\ha$ and
$\hl$ to the degree of polarization that can be achieved
is illustrated in Figs.~\polbb\ and \poltt, for the $b\anti b$
and $t\anti t$ channels, respectively. Figure~\polbb\ shows
the minimal and maximal observable masses,
$\mhh^{\rm max,min}$, $\mha^{\rm min}$, and $\mhl^{\rm min}$, as defined using
the
criteria stated earlier for $\hsm$.  Not explicitly plotted are:
$\mha^{\rm max}$, which is always very close to $2\mt=300~\gev$; and $\mhl^{\rm
max}$
which is equal to the theoretical upper limit for $\mhl$
within the MSSM [see Eq.~\hlmasslimit].
That is, the $\hl\rta b\anti b$ channel satisfies the criteria for
observability
for all $\mhl$ masses above $\mhl^{\rm min}$.
Note how close $\mhl^{\rm min}$ is to the corresponding Standard Model Higgs
result [see Fig.~\smfig]; this is
simply a reflection of how similar the $\hl$ is to the $\hsm$ of the same
mass once $\mha$ is large enough. Clearly, the range of masses for which
the $\ha$ and $\hl$ are detectable is greatly diminished if a high
degree of polarization cannot be achieved.
For the $t\anti t$ channel,
detection of the $\hh$ and $\ha$ for Higgs masses above $2\mt$
should generally be possible for $\vevlam=0.8$.
Figure \poltt\ shows
the value of $\nsd$ achieved in the $t\anti t$ channel
at $\mha,\mhh=500~\gev$ as a function of $\vevlam$.  These results
clearly indicate that in spite of the large $t$-quark mass,
polarization can still play a critical role
in the observability of the $\ha$ and
$\hh$ in the $t\anti t$ channel.

\midinsert
   \tenpoint \baselineskip=12pt   \narrower
\centerline{\psfig{file=ggh3.ps,angle=90,height=7cm}}
\vskip6pt\noindent
{\bf Fig.~\poltt.}\enskip
{Statistical significance (\ie, number of standard deviations
$\nsd$) of the $\ha\rta t\anti t$ and $\hh\rta t\anti t$ Higgs signals
at $\mha,\mhh=500~\gev$ as a function of $\vevlam$, for
$\mt=150~\gev$ and $\tanb=2$.
$\gmax$ and $z_0$ are chosen as in Fig.~\smfig.}
\endinsert

In summary,
detection of the Standard Model $\hsm$ should be possible for all Higgs masses
between the present day LEP lower limit up to the $\gam\gam$ collider
kinematic limit (roughly $\egamgam^{\rm max}\simeq 0.8E_{\epem}$).
That is, the kinematic reach of the ILC for $\hsm$ detection is somewhat
increased (relative to the limit of $\mhsm\lsim 0.7\eepem$
achievable in conventional $\epem$
collisions) using the $\gamma\gamma$ collider mode option,
while at the same time the
$\gam\gam$ coupling of the $\hsm$ is determined.  In the case of
the MSSM Higgs bosons, the $\gam\gam$ collider mode at the ILC
proves to be an enormously powerful tool.  For moderate $\tanb$,
detection of the $\hh$ and $\ha$ will be possible for all masses
up to roughly the $\gam\gam$ collider kinematical limit, often
in more than one final state decay mode. This represents a significant
increase in Higgs mass reach
as compared to a Higgs  mass limit of roughly $0.4E_{\epem}$
obtained by using the ILC in its search for
$\epem\rta \hl\ha,\hh\ha$.   For large $\tanb$,
a four-fold increase in $\gam\gam$ luminosity beyond
that assumed in this section would allow $\hh$ detection for all masses
up to the $\egamgam$ kinematic limit.
In reaching these optimistic conclusions, the ability to achieve substantial
polarization for the colliding photon beams has been assumed.  We have
illustrated the fact that the mass ranges for which the Standard Model and
MSSM Higgs bosons can be detected deteriorate significantly as
the degree of polarization decreases, especially in the
$b\anti b$ and $t\anti t$
channels. Every effort should be made to achieve the highest possible
polarization for the colliding photons.

\section{Search for CP-violating effects in the Higgs sector}

In the Standard Model with one Higgs weak doublet, the physical Higgs
boson is a CP-even scalar with CP-conserving tree-level interactions.
In the MSSM, the Higgs sector consists of two scalar weak doublets.
But, due to the supersymmetric constraints which restrict the form of
the scalar potential, the physical scalar mass eigenstates are states of
definite CP, with tree-level CP-conserving interactions.\foot{Soft-%
supersymmetry breaking terms do not affect this conclusion [see
Eq.~\mssmhiggspot].}
However, in non-supersymmetric models with a non-minimal Higgs sector,
and in non-minimal supersymmetric models, the scalar sector typically
contains sources of CP-violation.  It is possible that Higgs-sector
CP-violation effects could be detected in low-energy processes (such
as in the interactions of $K$ and $B$-mesons, or the neutron electric
dipole moment).  However, the absence of such low-energy effects
does not necessarily imply that CP-violating Higgs effects
must be absent.
It could turn out that Higgs induced CP-violating
effects can only be seen directly in
high-energy processes in which the Higgs bosons are produced.  One
consequence of CP-violation in the Higgs self-interactions is that
the neutral scalar Higgs mass eigenstates are no longer eigenstates of CP.
In other words, the neutral Higgs mass eigenstates are mixtures of
CP$=\pm 1$ scalar states.

\REF\grz{B. Grzadkowski and J.F. Gunion, {\sl Phys. Lett.} {\bf B294}
(1992) 361.}
\REF\grzii{J.F. Gunion, T.C. Yuan and B. Grzadkowski   {\sl Phys. Rev. Lett.}
{\bf 71} (1993) 488 [E: {\bf 71} (1993) 2681]; J.F. Gunion, UCD-92-26
(1992).}
In this section, I shall discuss a technique suggested by
Gunion and collaborators\refmark{\grz,\grzii}
for detecting the existence of CP-violating Higgs interactions by
searching for scalar states with mixed CP quantum
numbers.  One might think that if the Higgs boson were discovered,
its CP-quantum numbers could be determined from its production
characteristics and/or the energy and
angular distribution of its decay products.  Such methods will
fail if the Higgs boson is detected through its interactions with massive
gauge bosons.  For example, at $e^+e^-$ colliders, the Higgs boson is
typically produced either by bremsstrahlung off the $Z$-boson or via
$W^+W^-$ fusion.  In addition, in many cases,
$W^+W^-$ and $Z^0Z^0$ are the
dominant decay modes of the heavy Higgs boson.
However, at tree-level, the $ZZ$ or $W^+W^-$ couples only to the CP-even
part of a CP-mixed scalar state.\foot{The coupling of a CP-odd scalar to
two gauge bosons takes the form $\epsilon^{\mu\nu\alpha\beta}
F^a_{\mu\nu} F^a_{\alpha\beta}$ which can only arise in one-loop (or
higher-loop) radiative corrections.}

The $\gamma\gamma$ collider provides an efficient environment for the detection
of mixed CP properties (if present) of the neutral Higgs bosons
since $\gamma\gamma$ does not possess tree-level couplings to
the neutral Higgs states.  At one-loop, $\gamma\gamma$
couples to both the CP$=\pm1$ pieces of the scalar state.
The corresponding matrix elements are proportional to
$$
\eqalign{%
{\cal M}(\gamma\gamma \hl)&\sim
{\bold{\hat\epsilon\cdot\hat\epsilon^\prime}}\,,\cr
{\cal M}(\gamma\gamma
\ha)&\sim{\bold{\hat\epsilon\times\hat\epsilon^\prime}}\,.\cr}\eqn\gamgamcp
$$
where $\bold{\hat\epsilon}$ is the space components of the spin-1
polarization vector [Eq.~\spinone].
The availability of polarized bosons is particularly useful, and provides a
number of CP-violating asymmetries:
$$
\eqalign{%
\cA_1 & \equiv {|\cM_{++}|^2-|\cM_{--}|^2
\over|\cM_{++}|^2+|\cM_{--}|^2}\,,\crr
\cA_2 & \equiv {2\,{\rm Im}\,(\cM_{++}\cM^*_{--})
\over|\cM_{++}|^2+|\cM_{--}|^2}\,, \crr
\cA_3 & \equiv {2\,{\rm Re}\,(\cM_{++}\cM^*_{--})
\over|\cM_{++}|^2+|\cM_{--}|^2}\,,\crr}\eqn\gamgamasymmetries
$$
where $\cM_{\lambda\lambda'}$ is the amplitude for
$\gamma(\lambda)+\gamma({\lambda'})\to\h$.  Each asymmetry can lie
between $-1$ and $1$.
$\cA_1\neq0$, $\cA_2\neq0$, and/or $|\cA_3|<1$ would all constitute
signals of CP-violation.

\FIG\grzi{}
Grzadkowski and Gunion\refmark\grz\
have computed the three asymmetries above for a
scalar Higgs field arising from the most general CP-nonconserving two-%
Higgs doublet model.  They repeat the asymmetry calculation many times,
searching over the model parameter space for the maximum CP-violating
signal.  This is accomplished by choosing a particular value of $m_t$,
$\tan\beta$ and Higgs mass, and varying at random the Higgs mixing
angles, \ie, the parameters that
specify the Higgs mass eigenstates (relative to the interaction
eigenstate basis).  In Fig.~\grzi, the asymmetries defined in
Eq.~\gamgamasymmetries\ are plotted as a function of the Higgs mass for
$\mt=150$~GeV and $\tan\beta=2$.  Each asymmetry plotted is the maximum
asymmetry achievable as the Higgs mixing angles are varied,
for the specified Higgs mass.  The maximal CP-violating asymmetries
can be large, and the statistical significance of the signal scales
with the square root of the number of events.  Under plausible operating
conditions for a $\gamma\gamma$ collider operating at a CM energy of
500 GeV, with an integrated luminosity of $20~{\rm fb}^{-1}$, the
authors of Ref.~[\grz] show that the asymmetry $\cA_1$ is the most
promising for the detection of a CP-violating signal.

\topinsert
   \tenpoint \baselineskip=12pt   \narrower
\centerline{\psfig{file=gun1.ps,height=7cm}}
\vskip6pt\noindent
{\bf Fig.~\grzi.}\enskip
The values for $|\cA_1|$ \solid\ and $|\cA_2|$ \dashes\
and $(1-|\cA_3|)$ \dotdot\ which yield the largest
standard scenario statistical significances, as a function
of $\mhsm$, for $\tanb=2$ and $\mt=150~\gev$. Extrema
are obtained for 150,000 random choices of Higgs mixing angles
subject to the requirement that there be at least
80 events in the $b\anti b$ decay channel of the $\hsm$,
or 20 events in the $ZZ$ (one $Z\rta \lplm$) channel, or 80
events in the $t\anti t$ channel when the colliding photon polarizations
are averaged over.  Taken from Ref.~[\grz].
\endinsert

\FIG\statsig{}
At hadron supercolliders, the dominant mechanism for Higgs production
is gluon-gluon fusion to the Higgs boson through a heavy quark loop.
Since the same diagram gives rise to the Higgs coupling to
$\gamma\gamma$ (where the gluons are replaced by photons), one can
also search for a CP-violating asymmetry at hadron supercolliders
with polarized beams.  Of
course, in this case, the theoretical prediction for the asymmetry
and its experimental detection are not as clean
(compared to the case of the $\gamma\gamma$ collider discussed above).
For example,
one does not polarize the gluons directly; rather one must polarize the
initial proton beams, and attempt to determine the resulting gluon
polarization, based on a model (making the best possible use of
polarized deep inelastic scattering data to determine the polarized
gluon distribution function).  In order to detect a CP-violating
signal, it is sufficient
to polarize one of the proton beams. The CP-violating
asymmetry of interest is
$$
\cA\equiv {\sigma_+-\sigma_-\over\sigma_++\sigma_-}\,,\eqn\gluonasymmetry
$$
where $\sigma_\lambda$ is the Higgs production cross-section for the
case of one proton beam with helicity $\lambda$.
To evaluate ${\cal A}$, one
must compute the asymmetry at the partonic level, and then fold in the
helicity-dependent gluon distribution function.
Let $\Delta g(x)\equiv g_+(x)-g_-(x)$, where $g_+(x)$ [$g_-(x)$] is
the probability of finding a gluon with positive [negative] helicity
inside a proton of {\it positive} helicity.\foot{Note that $g_+(x)$
[$g_-(x)$] is also
the probability of finding a gluon with negative [positive] helicity
inside a proton of {\it negative} helicity.}
The unpolarized gluon distribution is given by
$g(x)\equiv g_+(x)+g_-(x)$.
If ${\cal M}_{\lambda\lambda^\prime}$
denotes the amplitude for
$g(\lambda)+g(\lambda^\prime)\rta H$,
then\foot{Recall that for $\gamma\gamma\rta H$ (and $gg\rta H$),
${\cal M}_{\lambda\lambda^\prime}$ vanishes unless
$\lambda=\lambda^\prime$ [see Eq.~\sigmaform].}
\REF\qui{E. Berger and J. Qui, {\sl Phys. Rev.} {\bf D40} (1989) 778.}
\topinsert
   \tenpoint \baselineskip=12pt   \narrower
\centerline{\psfig{file=guntop.ps,height=7cm}\hskip.75cm}
\vskip6pt\noindent
{\bf Fig.~\statsig.}\enskip
Maximal statistical significance achieved for an
asymmetry signal as a function of $\mhsm$ at the SSC with
$L=10$~fb$^{-1}$.  The branching ratio for $\hsm\rta ZZ
\rta\ell^+\ell^-X$ is included.  A model for the polarized
gluon distribution function is used where $\Delta g(x)=g(x)$ for
$x>0.2$ and $\Delta g(x)=5xg(x)$ for $x<0.2$\refmark\qui.
This yields $\Delta g\equiv\int g(x)dx\simeq 3$ over the Higgs mass range
depicted above (where $g(x)\equiv g(x,Q)$ is evaluated at $Q=\mhsm$).
Taken from Ref.~[\grzii].
\endinsert
$$
\cA={\int_{m_H^2/s}^1\,dx\,g\left({m_H^2/sx}\right)\Delta g(x)
\left[|{\cal M}_{++}|^2-|{\cal M}_{--}|^2\right]\over
\int_{m_H^2/s}^1\,dx\,g\left({m_H^2/sx}\right)g(x)
\left[|{\cal M}_{++}|^2+|{\cal M}_{--}|^2\right]}\,.\eqn\cagg
$$
Note that $\cA$ vanishes if $H$ is an eigenstate of
CP.  This can be understood from Eq.~\paritycons\ which implies that
$|{\cal M}_{++}|=|{\cal M}_{--}|$
if P is conserved.  However, for neutral scalars, eigenstates of P
are necessarily eigenstates of CP, since there are no scalar
interactions with fermion pairs or vector boson pairs that conserve P
without conserving CP.
Thus, the detection of a nonzero value for $\cA$ would be an indication
of CP-violation.
The distribution function difference $\Delta g(x)$ is not very well
constrained by data, and thus introduces model dependence
into the calculation.  As a result, the evaluation of the
statistical significance of the signal is more uncertain than in the
previous case of $\gamma\gamma\to H$.
Figure~\statsig, taken from Ref.~[\grzii], depicts a typical result.
Although the detection of a CP-violating
signal of this kind presents a real challenge for experimentalists
working at a hadron supercollider, Fig.~\statsig\ does suggest that a
polarized proton beam at hadron supercolliders can play a useful role
in probing the details of new physics.

\section{\bf New gauge bosons beyond the $\bold{W^\pm}$ and $\bold Z$}

\REF\cvetic{P. Langacker, R. W. Robinett and J. L. Rosner,
{\sl Phys. Rev.} {\bf D30} (1984) 1470;
F. del Aguila, M. Cvetic and P. Langacker, {\sl Phys.
Rev.} {\bf D48} (1993) 969; M. Cvetic, P. Langacker, and J. Liu,
{\sl Phys. Rev.} {\bf D49} (1994) 2405.}
In this section, I present the final example of these
lectures.  One of the key questions that one hopes to answer with the
next generation of supercolliders is:  what is the correct electroweak
gauge group of the low-energy effective TeV-scale theory?  Is it
SU(2)$_L\times$U(1) or is the gauge group larger, implying the
existence of new gauge bosons beyond the $W^\pm$ and $Z$ whose masses
are of ${\cal O}$ (1 TeV).  To answer this question with full
confidence requires a collider that can probe deep into the TeV energy
region.  Clearly (unless we are very lucky), the ILC will not be
sufficient for this task.
If new $W^\prime$ and/or $Z^\prime$ gauge bosons
are discovered, it will be essential to measure their
couplings to fermions.  At a hadron supercollider, there are a
variety of techniques that can be employed for
this purpose\refmark\cvetic.
Here, I shall briefly mention two techniques.

\REF\EHLQ{E. Eichten, I. Hinchliffe, K. Lane and C. Quigg,
{\sl Rev.\ Mod.\ Phys.} {\bf 56} (1984) 579 [E: {\bf 58} (1986) 1065].}
If one of the proton beams is polarized, then one can consider the
single helicity asymmetry
$$
A_L = {d\sigma_- -d\sigma_+\over d\sigma_- +d\sigma_+}\,,\eqn\asubl
$$
where $d\sigma_\lambda$ is the differential cross-section (integrated
either partially or completely over the available phase space)
for $p(\lambda)+p\rta$ heavy gauge boson.  This is a similar
asymmetry to the one defined in the previous section in the case of
Higgs production.  However,
in this case, $A_L\neq 0$ implies only parity nonconservation, since
vector boson interactions can violate P without violating CP.
The production mechanism for new gauge bosons at a hadron collider
is $q\bar q$ fusion. For very
heavy gauge bosons, the momentum fractions of the
initial partons are not particularly small.\foot{Recall that in the
parton model computation, $x_1 x_2 s=m_V^2$,
where $s$ is the CM energy squared of the supercollider and $x_1$
and $x_2$ are the momentum fractions of the quark and antiquark,
respectively\refmark\EHLQ.}
Hence, one would
expect that the polarization of a valence parton would
reflect the polarization of the corresponding proton.  Thus, it seems
plausible that in new gauge boson production,
large single helicity asymmetries should be observed.
This is illustrated by a calculation of
Ref.~[\bsrt], which demonstrates that values for $A_L$ as large as
$25\%$ are possible in the production of a neutral heavy gauge boson
of a few TeV in mass.  Moreover, the obtained value of $A_L$ is a
sensitive function of the vector boson mass and its quantum numbers.
By changing the assumptions of the $Z^\prime$ quantum numbers, one
can find dramatic shifts in $A_L$ by an order of magnitude.
Finally, if both proton beams are polarized, then additional
observables can be used to probe the new gauge boson couplings to
fermion pairs.  For example, Ref.~[\bsrt] shows that in the
production of left-handed and/or right-handed charged gauge bosons,
$\sigma_{--}/\sigma_{++}$ may deviate substantially from 1.

\REF\summaryofresults{H. Haber, in
{\it Proceedings of the 1984 Summer Study on the Design
and Utilization of the Superconducting Super Collider},
Snowmass, Colorado, June 23-July 13, 1984, edited by R. Donaldson
and J.G. Morfin (Fermilab, Batavia, IL, 1985) pp.~125--143;
and pp.~157--159.}
\REF\hagiwara{B.K. Bullock, K. Hagiwara and A.D. Martin,
{\sl Nucl. Phys.} {\bf B395} (1993) 499.}
If no polarized beams are available, there are other tools for
determining the nature of the new gauge boson interactions.  I will
conclude these lectures by discussing one powerful technique, in
which one studies the decay of the heavy gauge boson to $\tau N$
(where $N$ is a neutral heavy lepton which may or may not be the
$\nu_\tau$).
This is a particularly useful decay mode, because the $\tau$-lepton
decays are self-analyzing. By studying the energy
distribution of the $\tau$ decay
products, one deduces information about the $\tau$ polarization,
and hence learns about the coupling of the gauge boson to $\tau N$%
\refmark{\summaryofresults,\hagiwara}.

To illustrate the method, consider the decay of a new
charged gauge boson, $W^\prime_L$, where the subscript $L$ indicates
that $W^\prime_L$ has the same couplings to fermions as the $W^\pm$
of electroweak interactions.  The following decay chain is assumed.
$$
\eqalign{%
 &W_L^\prime\to \tau N   \cr
&\hphantom{W_L^\prime\to\enskip}\ccca
    \cases{\mu \nu \bar \nu\qquad [{\rm or}\ e\nu\bar\nu]&\cr
                        \pi\nu & \cr
                        \rho\nu & \cr }
\cr}\eqn\decaychains
$$
In what follows,
I shall provide details of the computation of the differential
cross-section for $W^\prime_L$ production followed by the decay
chain
$$
W^\prime_L\to \tau N\to e\nu\bar\nu N\,,\eqn\wtotau
$$
and quote the results for the other two decay chains shown above.
In this calculation, I will make the (very good) approximation
that all final state masses are negligible compared to $W_L'$.

\REF\lackner{R.M. Barnett, H.E. Haber and K.S. Lackner, {\sl Phys. Rev.}
{\bf D29} (1984) 1381.}
The basic formula for the cross-section for the scattering
process $a+b\rta c+d$ followed by the decay $d\rta 1+2+\dots+n$
was described in section 1.7 as a trace over the product of the
production and decay density matrix elements [see Eq.~\totalcross].
Here, I
shall write out the relevant formula in complete detail\refmark\pilkuhn\
$$
\eqalign{
\sigma&={1\over4p\ls{\rm CM}\sqrt{s}}
\sum_{\lambda\lambda'} \int \cM(ab\to cd_\lambda)
\cM^*(ab\to cd_{\lambda'})\,{ds_d\over2\pi}\,d{\rm Lips}(s;p_c,p_d) \crr
&\times {\cM(d_\lambda\to1+2+\dots+n)\cM^*(d_{\lambda'}\to1+2+\dots+n)\over
(m^2-s_d)^2+m^2\Gamma^2}\, d{\rm Lips}(s_d\,;\,p_1,\dots,p_n)\,, \cr}
\eqn\bigformula
$$
where
$s=(p_a+p_b)^2$ is the CM energy squared for the process,
$m$ and $\Gamma$ are the mass and width of particle $d$,
$p\ls{\rm CM}$ is the CM-momentum of particle $a$ (or $b$) [if initial
state masses are neglected, $p\ls{\rm CM}=\sqrt{s}/2$],
$s_d=p^2_d$ is the invariant mass squared of the decaying particle $d$,
and
$$
d{\rm Lips}(s\,;\,p_1,\dots,p_n)=
(2\pi)^{4-3n}\prod^n_{i=1}\,{d^3p_i\over2E_i}\, \delta^4
(p_a+p_b-\sum_ip_i)\eqn\lips
$$
is Lorentz invariant phase space for an $n$-body final state with
total CM-energy squared equal to $s$.  Using the above formula, the
computation proceeds in the seven steps sketched out
below\refmark\lackner.

{\sl Step 1:}
Since $m_\tau\ll m_{W'_L}$, the emitted $\tau$'s are
completely left-handed. Thus, only the $\lambda=\lambda'=-1$
term in the sum
over $\lambda$, $\lambda^\prime$ in Eq.~\bigformula\ survives.
This observation simplifies this calculation immensely, and is
special to this particular problem.

{\sl Step 2:}
Use the narrow width approximation
$$
{1\over(m^2-s_d)^2+m^2\Gamma^2}\longrightarrow
{\pi\over m\Gamma}\delta (m^2-s_d)\eqn\nrwidth
$$
to integrate over $s_d$ in Eq.~\bigformula.  In this calculation,
$$
B_e\Gamma={G^2_F m^5_\tau\over 192\pi^3}\,,\eqn\bewidth
$$
where $B_e\equiv\Gamma(\tau\to e\nu\bar\nu)/\Gamma$ is the tau-lepton
branching ratio into electrons.

{\sl Step 3:}
The squared matrix elements for $a+b\rta c+d$ and $d\rta 1+2+3$
are well-known and are given below.

(i) $\bar ud\to W_L^{\prime -}\to \tau^-N$

\noindent The squared amplitude for this process,
averaged over initial
and summed over final spins and colors, is easily computed.  The result
is
$$
|\cM(\bar ud\to W_L^{\prime -}\to \tau^-N)|^2 = \left({1\over 12}\right)
{4g^4 p_d\cdot p_\nu \,p_{\bar u}\cdot p_\tau
	\over (m^2_W-s)^2+\Gamma_W^2m^2_W}\,,\eqn\udwtaun
$$
where $p_i$ is the four-momentum of particle $i$, and the
$1/12$ inside the parentheses consists of a factor of $1/4$ for the
spin-average and a color factor of $1/3$.
As remarked above, the produced $\tau^-$ is
purely left-handed (to a very good approximation), since
$m_\tau\ll m_{W^\prime}$.

\break

(ii) $\tau^-\to e^-\nu\bar\nu$

\noindent In order to apply Eq.~\bigformula, we need the squared
matrix element for the decay of a left-handed $\tau^-$.  This
computation is most easily done using the spin projection operator
method described in section 1.4.  We introduce the spin four-vector
[Eq.~\spinvec]
for a negative helicity $\tau^-$ of four-momentum $p_\tau=(E_\tau,
{\bold{\vec p_\tau}})$\foot{A capital $S$ is used here
since $s$ is already being used for the CM-energy squared.}
$$
S=-\left({|{\bold{\vec p_\tau}}|\over m_\tau}\,;\,
 {E_\tau{\bold{\hat p_\tau}}\over m_\tau}\right)\,.\eqn\spintau
$$
The calculation of the squared decay amplitude is straightforward.
Here, I simply quote the well known formula for the decay rate of a
polarized muon found in Ref.~[\bailin], which can also be used here
$$
E_\tau E_e{d\Gamma\over d^3p_e}={G_F^2\over3(2\pi)^4}
  \left[\,q^2p_e\cdot(p_\tau-m_\tau S) +
  2q\cdot(p_\tau-m_\tau S)q\cdot p_e\,\right]\,,
$$
where $p_e=(E_e;{\bold{\vec p_e}}\,)$ is the electron four-momentum,
$q\equiv p_\tau-p_e$ and $S$
is the spin four-vector of the negative-helicity $\tau^-$.

{\sl Step 4:}  Using Eq.~\bigformula\ and the results given above,
we integrate over $s_d$ and obtain
$$
\eqalign{%
\sigma={B_eg^4\over192\pi^3m^6_\tau\,s\,[(m^2_W-s)^2+\Gamma_W^2m^2_W]}
     & \int d\Omega_\tau {d^3p_e\over E_e}\, 4p_d\cdot p_\nu \,p_{\bar u}
      \cdot p_\tau\crr
 \times \Bigl[q^2p_e\cdot&(p_\tau-m_\tau S)+
 2q\cdot(p_\tau-m_\tau S)\,q\cdot p_e\Bigr]\,,\cr}
\eqn\sigtotaun
$$
where $d\Omega_\tau$ is the differential solid angle of the $\tau^-$.

{\sl Step 5:}  It is convenient to choose the rest frame of the
$W^{\prime -}_L$ to evaluate the integral above.  Explicitly,
$$
q^2p_e\cdot(p_\tau-m_\tau S)+2q\cdot(p_\tau-m_\tau S)\,q\cdot p_e =
E_e\left[m^2_\tau z+\sqrt{s}(m^2_\tau-2E_ez)(1-
{\bold{\hat p_\tau\cdot\hat p_e}})\right]\,,\eqn\restframeexpi
$$
where
$$
z\equiv{s+m_\tau^2-(s-m_\tau^2){\bold{\hat p_\tau\cdot\hat
p_e}}\over\sqrt{s}}\,,\eqn\zdef
$$
and
$$
4p_d\cdot p_\nu \,p_{\bar u}\cdot p_\tau=\quarter s^2
(1+{\bold{\hat p_d\cdot\hat p_\tau}})^2+{\cal O}(m_\tau^2)\,.
\eqn\restframeexpii
$$
Since $m_\tau\ll m\ls{W^\prime}$, we can drop the ${\cal O}(m_\tau^2)$
term in Eq.~\restframeexpii.

{\sl Step 6:} Insert the above expressions into the integral
[Eq.~\sigtotaun].
Choosing the direction of the d-quark to lie along
the $z$-axis, and the direction of the electron to lie in the $x$--$z$
plane, we write ${\bold{\hat p_d}}={\bold{\hat z}}$ and
${\bold{\hat p_e}}=(\sin\theta_e,0,\cos\theta_e)$.  Our plan is to
integrate over all possible $\tau$ directions, ${\bold{\hat p_\tau}}$,
holding the electron direction fixed.  For this purpose, it is
somewhat easier to rotate the $z$-axis to lie along the electron
direction.  That is, we take ${\bold{\hat p_e}}=(0,0,1)$ and
${\bold{\hat p_d}}=(-\sin\theta_e,0,\cos\theta_e)$.  Relative to this
choice, we integrate over ${\bold{\hat p_\tau}}=
(\sin\theta\cos\phi,\sin\theta\sin\phi,\cos\theta)$.  Then,
$$
\eqalign{
{\bold{\hat p_\tau\cdot\hat p_e}} & = \cos \theta\,,\cr
{\bold{\hat p_d\cdot\hat p_\tau}} & =\cos\theta\cos\theta_e-
\sin\theta\sin\theta_e\cos\phi\,.
\cr}\eqn\dotprods
$$
In the integration over ${\bold{\hat p_\tau}}$ at fixed electron energy,
the limits of integration depend on $E_e$.  This is most easily
obtained by noting that $(p_\tau-p_e)^2\geq 0$ implies that $z\leq
m_\tau^2/E_e$.  From the definitions of $z$ [Eq.~\zdef] and
$\cos\theta$ [Eq.~\dotprods], one sees that $y\leq\cos\theta\leq 1$,
where
$$
y \equiv {E_e(s+m^2_\tau)-m^2_\tau\sqrt{s}\over E_e(s-m^2_\tau)}\,.
\eqn\ydef
$$
It follows that
$$
\eqalign{
{d\sigma\over d\Omega_e dE_e} &= {g^4sE_e^2B_e\over768\pi^3m^6_\tau
                                  [(m_W^2-s)^2+m_W^2\Gamma^2_W]} \crr
& \times\int^{2\pi}_0 d\phi\int^1_yd\cos\theta\,
   (1+\cos\theta\cos\theta_e -\sin\theta\sin\theta_e\cos\phi)^2\crr
&\qquad\qquad \times
\left[m^2_\tau z+\sqrt{s}\,(1-\cos\theta)(m_\tau^2-
2E_ez)\right]\,.\cr}\eqn\dsigdomegade
$$

{\sl Step 7}: The above integration can be performed analytically.
Recall that we are working
in the limit of small $m_\tau$.  Although the factor of $m_\tau^6$ in
the denominator of Eq.~\dsigdomegade\ may be a cause for concern,
it is easy to see
that the leading $m_\tau$ behavior of the integral is
${\cal O}(m^6_\tau)$.  Since $y\to 1$ as $m_\tau\to 0$, one can expand
the integrand in Eq.~\dsigdomegade\ around $\cos\theta=1$.  Furthermore,
for $\cos\theta\simeq 1$,
one sees from Eq.~\zdef\ that $z={\cal O}(m_\tau^2)$.
Thus, it is sufficient to keep terms in Eq.~\dsigdomegade\ up to
$(1-\cos\theta)^2$ and drop all higher terms.
One therefore ends up with a finite value for
the differential cross-section in the limit of $m_\tau\to0$:
$$
{d\sigma\over dE_e\,d\cos\theta_e} =
{\pi\alpha^2\sqrt{s}\,(1+\cos\theta_e)^2B_e\over
36\sin^4\theta_W[(m^2_W-s)^2+\Gamma_W^2m^2_W]}
\left[1-\left({2E_e\over\sqrt{s}}\right)^3\,\right]\,,\eqn\dsigdedcos
$$
where $\alpha\equiv g^2/4\pi\sin^2\theta_W$.  Note that
the allowed range of electron energies is
$0\leq E_e\leq{1\over2}\sqrt{s}$.

Two features of this result are
immediately apparent.  First, the $(1+\cos\theta_e)^2$
angular distribution matches precisely the $(1+\cos\theta)^2$
angular distribution of the $\tau^-$ obtained in
Eq.~\udwtaun,\ in the $m_\tau=0$ limit [see Eq.~\restframeexpii].
This angular distribution can be understood using simple
helicity arguments as shown in the figure below.

\medskip
\vbox{%
\centerline{\psfig{file=ssi3-9.ps,height=2.5cm,angle=90}}
}
\medskip

\noindent The arrows above the fermion lines denote helicity.
The $W^{\prime -}_L$ couples to a left-handed $d$-quark and $\tau^-$,
and a right-handed $u$-quark and anti-neutrino.  Angular momentum
conservation favors $\theta$ near $0^\circ$ and disfavors $\theta$
near $180^\circ$.  It is interesting to note that precisely the same
angular factor would arise in $W_R^{\prime -}$ production, assuming that
the $W_R^{\prime -}$ couples to right-handed fermions and left-handed
anti-fermions.  In the figure above, the configurations shown
are both favored by angular momentum conservation, and one concludes
that the electron angular distribution will be the same for both $W_L'$
and $W_R'$ production!

Second, let us define $x\equiv 2E_e/\sqrt{s}$.  From
Eq.~\dsigdedcos, one sees that the polarization of the
$\tau^-$ is reflected in the $1-x^3$ energy distribution of the electron.
Furthermore, by repeating the above
calculation for $W_R^{\prime -}$ production,
one finds that although the final state electron
angular distribution is the same, the electron energy
distribution is proportional to $(1+2x)(1-x)^2$.  Thus, the
electron energy distribution is {\it softer} for $W_R'$ production as
compared with $W_L'$ production.
Thus, by measuring the energy distribution of the observed electron, one
gains information about the $\tau$ polarization and thereby obtains a
probe of the $W^\prime$ coupling to $\tau N$.

Other $\tau$ decay modes can be used in a similar fashion.  The
relevant calculations are similar to the one presented above, so I
shall simply summarize the results below\refmark\summaryofresults.

(i) $\bar ud\to W^{\prime -}_{L,R}\to\tau^-N$, $\tau^-\rta e^-\nu\bar\nu$
$$
{d\sigma\over dE_e\,d\cos\theta_e}=
B_eC(\theta_e)\,\times\,
  \cases{{2\over3}(1-x^3)&,\ $W'_L$\cr (1+2x)(1-x)^2&,\ $W'_R$\cr}
\eqn\enunu
$$

(ii) $\bar ud\to W^{\prime -}_{L,R}\to\tau^-N$, $\tau^-\rta \rho^-\nu$
$$\eqalign{%
{d\sigma\over dE_\rho\,d\cos\theta_\rho}
&=
{B_\rho m_\tau^2C(\theta_\rho)\over(m_\tau^2-
m^2_\rho)^2(m^2_\tau+2m^2_\rho)}             \crr
&\qquad\times\,\cases{2m_\rho^2(m^2_\tau-m^2_\rho)
+m^2_\tau(m^2_\tau-2m_\rho^2)(1-x)&,\ $W_L'$\cr
\ \ m^2_\tau\left[m_\rho^2+(m_\tau^2-2m^2_\rho)x\right] &,\ $W_R'$\cr}
\cr}\eqn\rhonu
$$

(iii) $\bar ud\to W^-_{L,R}\to\tau^-N$, $\tau^-\rta \pi^-\nu$

$$
{d\sigma\over dE_\pi\,d\cos\theta_\pi}=
B_\pi C(\theta_\pi)\,\times\,\cases{1-x&,\ \ $W_L'$\cr
  x&,\ \ $W_R'$\cr}\eqn\pinu
$$
where $m_\pi$ has been set to zero in the last computation.  In all the
above expressions, $B$ is the relevant $\tau$ branching ratio,
$$
C(\theta)\equiv {\pi\alpha^2\sqrt{s}\,
(1+\cos\theta)^2\over24\sin^4\theta_W
[(m^2_W-s)^2+\Gamma_W^2m_W^2]}\,,\eqn\ctheta
$$
and $x\equiv 2E/\sqrt{s}$, where $E$ is the energy of the final state
negatively charged particle.

There is a simple way to rederive the form of Eq.~\pinu\ for
$\bar ud\to W^{\prime -}_{L,R}\to\tau^-N$, $\tau^-\rta \pi^-\nu$.
Choose the quantization axis of the $\tau$-spin to lie along the
$z$-axis.
Let $k$ be the four-momentum of the $\pi$ in the rest frame of the $\tau$.
Then, setting $m_\pi=0$,
$k={1\over2} m_\tau\,(1;\,\sin\theta,0,\cos\theta)$.
Now, boost to a frame where the $\tau$ is moving along
${\bold{\hat z}}$ with velocity $v$. In this frame,
$$
\eqalign{
E_\pi &= \half\gamma m_\tau(1+v\cos\theta)\,,    \cr
E_\tau &=\gamma m_\tau\,,\cr}\eqn\epietau
$$
where $\gamma\equiv (1-v^2)^{1/2}$ and
$$
x\equiv{E_\pi\over
E_\tau}=\half(1+v\cos\theta)\longrightarrow\cos^2\half\theta\eqn\eratios
$$
as $v\to 1$.  But, using Eq.~\reduceddecay,
we know that the decay rate for $\tau^-(\lambda)\to\pi^-\nu$ is
$$
\Gamma(\tau^-(\lambda)\to\pi\nu)\propto|d^{1/2}_{\lambda,1/2}(\theta)|^2
=
\cases{\cos^2{\theta\over2}, & $\lambda=+{1\over2}$\crr
       \sin^2{\theta\over2}, & $\lambda=-
{1\over2}$\crr}\eqn\dhalflambda
$$
which corresponds to pion energy distributions of $x$ and $1-x$, respectively,
in agreement with the previously quoted result.

This result also has a simple physical interpretation.  The $\tau^-$
is either left or right-handed depending on whether it came from $W_L'$
or $W_R'$.  But the neutrino emitted in $\tau^-$ decay is always left-%
handed.  Thus, because the $\pi$ is spinless, conservation of angular
momentum implies that the $\pi$ is emitted preferentially forward in
the case of $W_R'$ decay and backward in the case of $W_L'$ decay.
This is illustrated in the figure below.

\medskip
\vbox{%
\centerline{\psfig{file=ssi3-10.ps,height=2.25cm,angle=90}}
}
\medskip
\REF\newwandz{J.F. Gunion and H.E. Haber, in
{\it Proceedings of the 1984 Summer Study on the Design
and Utilization of the Superconducting Super Collider},
Snowmass, Colorado, June 23-July 13, 1984, edited by R. Donaldson
and J.G. Morfin (Fermilab, Batavia, IL, 1985) pp.~150--152.}
\noindent
The arrows above the $\tau$ and $\nu$ denote helicity.  Angular
momentum conservation implies that the configurations shown above are
the ones favored.  Therefore, in the $\bar ud$ CM-frame, the energy
spectrum of the $\pi$ is harder in $W_R$ decay and softer in $W_L$
decay.

Finally, one must convolute the partonic cross-sections given above
with parton distribution functions to obtain predictions for the
energy distributions of $\tau$-lepton decay products in the laboratory
frame.  The computations of Ref.~[\newwandz] show that it may be
possible to distinguish between $W^\prime_L$ and $W^\prime_R$ on the
basis of their $\tau$-decay spectra.

\bigskip

\centerline{\fourteenbf CODA}

\REF\anomwwz{See \eg, K. Hagiwara, R.D. Peccei, D. Zeppenfeld and K. Hikasa,
{\sl Nucl. Phys.} {\bf B282} (1987) 253; E. Yehudai, {\sl Phys. Rev.}
{\bf D41} (1990) 33; {\bf D44} (1991) 3434; SLAC-383 (1991);
G.J. Gounaris, J. Layssac,
G. Moultaka and F. Renard, {\sl Int. J. Mod. Phys.} {\bf A8} (1993)
3285; G.J. Gounaris, J. Layssac and F. Renard, Montpelier preprint
PM-93-26 (1993).}
\REF\toppol{G.L. Kane, G.A. Ladinsky and C.P. Yuan, {\sl Phys. Rev.}
{\bf D45} (1992) 124; C.P. Yuan, {\sl Phys. Rev.} {\bf D45} (1992)
782; G.A. Ladinsky, {\sl Phys. Rev.} {\bf D46} (1992) 3789 [E: {\bf D47}
(1993) 3086]; C.R. Schmidt and M.E. Peskin, {\sl Phys. Rev. Lett.}
{\bf 69} (1992) 410; C.R. Schmidt, {\sl Phys. Lett.} {\bf B293}
(1992) 111; W. Bernreuther, J.P. Ma, and T. Schroder, {\sl Phys. Lett.}
{\bf B297} (1992) 318.}
In these lectures, I presented an introduction to spin formalism and
illustrated its use in a number of examples of searches for new
physics beyond the Standard Model at future colliders.  I have only
touched the surface; many important topics have been omitted.
Some of the important applications not included in these lectures
involve the search for evidence of physics beyond the Standard Model
in precision measurements of top-quarks and gauge bosons.  For
example, a crucial test of the Standard Model consists of checking the
details of the $W^+W^-Z$ and $W^+W^-\gamma$ vertices\refmark\anomwwz.
Similarly, it
may be possible to see hints of non-Standard Model effects in $t\bar
t$ couplings to $\gamma$ and/or $Z$\refmark\toppol.
In both cases, one can search
for anomalous moments, CP-violating form-factors, evidence for form-%
factors, \etc\  Using spin information to separate out definite
helicity final states could enhance a particular signal as well as
help control backgrounds.

\REF\peskinlane{E. Eichten, K. Lane and M.E. Peskin, {\sl Phys. Rev.
Lett.} {\bf 50} (1983) 811.}
There are many other precision tests of electroweak theory and QCD
which become available with polarized beams and/or spin analysis of
final state particles.  Some of these have been addressed in other
lectures presented at this summer school.  Another method for
searching for evidence of new physics is to detect the presence of
four-fermion operators, which would be remnants of new physics at a
higher energy scale\refmark\peskinlane.
There are many such operators possible.  These
could be detected at a future hadron supercollider if deviations are
seen from Standard Model predictions in two-jet and/or Drell-Yan
cross-sections (to give just two examples).  The ability to separate
out the definite helicity properties of such operators (if they exist)
would play a crucial role in trying to interpret their origin.

The discovery of new physics beyond the Standard Model hopefully
lies ahead in the not too distant future.  In the first decade of the
third millennium, powerful supercolliders will be ready to fully
explore the TeV-energy scale and reveal its secrets.  Once deviations
from the Standard Model are found, the challenge to theorists and
experimentalists will be to interpret the results and build a new
Standard Model of particle physics.  Spin techniques will surely play
an important role in unraveling the true nature of TeV-scale physics.
We have only just begun to explore its power.

\bigskip
\centerline{ACKNOWLEDGMENTS}
\medskip
I would like to thank Roberto Vega for fruitful discussions pertaining
to Lecture 1, Hitoshi Murayama and Jonathan Feng for their help in
presenting some of the material of Lecture 2, and Jack Gunion for
his assistance and collaboration in various aspects of Lecture 3.
I am also very grateful to Mayling and Lance Dixon and to Valerie and
Michael Peskin for their gracious hospitality during my stay at the
SLAC Summer School.  I am pleased to acknowledge the Institute for
Theoretical Physics in Santa Barbara for their support during the preparation
of the written version of these lectures.  Finally,
I wish to extend my special thanks
and appreciation to Nora Rogers at SCIPP and Darla Sharp at the ITP
for their assistance in the \TeX\ input
and processing.

\endpage

\refout
\bye